\documentclass[preprint,12pt]{elsarticle}

\usepackage[linesnumbered,lined,boxed,commentsnumbered,ruled,vlined]{algorithm2e}
\usepackage{algpseudocode}
\usepackage{amssymb}
\usepackage{amsthm}
\usepackage{booktabs}
\usepackage{comment}
\usepackage[symbol]{footmisc}
\usepackage[a4paper, total={6in, 8in}]{geometry}
\usepackage{glossaries}
\usepackage{multirow}
\usepackage{pgfplots}
\usepgfplotslibrary{groupplots}
\usepackage{subcaption}
\usepackage{siunitx}
\usepackage{tikz}
\usepackage[export]{adjustbox}
\usepackage{placeins}
\usepackage{gensymb}

\usepackage{lineno}

\newacronym{gsa}{GSA}{global sensitivity analysis}
\newacronym{lar}{LAR}{least angle regression}
\newacronym{mcs}{MCS}{Monte Carlo sampling}
\newacronym{mvsa}{MVSA}{multivariate sensitivity-adaptive}
\newacronym{ols}{OLS}{ordinary least squares}
\newacronym{omp}{OMP}{orthogonal matching pursuit}
\newacronym{pca}{PCA}{principal component analysis}
\newacronym{pce}{PCE}{polynomial chaos expansion}
\newacronym{pdf}{PDF}{probability density function}
\newacronym{pod}{POD}{proper orthogonal decomposition}
\newacronym{rmse}{RMSE}{root-mean-square error}
\newacronym{rom}{ROM}{reduced order model}
\newacronym{sp}{SP}{subspace pursuit}
\newacronym{td}{TD}{total-degree}
\newacronym{uq}{UQ}{uncertainty quantification}

\DeclareMathOperator*{\argmax}{\arg\,\max}
\DeclareMathOperator*{\argmin}{\arg\,\min}

\journal{XYZ}

\begin{document}

\begin{frontmatter}



\title{Multivariate sensitivity-adaptive polynomial chaos expansion for high-dimensional surrogate modeling and uncertainty quantification}

\author[aff1,aff2]{Dimitrios Loukrezis\corref{cor1}}
\ead{dimitrios.loukrezis@siemens.com}
\author[aff1]{Eric Diehl}
\author[aff2]{Herbert De Gersem}

\affiliation[aff1]{
organization={Siemens AG, Technology},
city={Munich},
country={Germany}
}

\affiliation[aff2]{
organization={Technische Universität Darmstadt, Institute for Accelerator Science and Electromagnetic Fields (TEMF)},
city={Darmstadt},
country={Germany}
}

\cortext[cor1]{Corresponding author.}


\begin{abstract}
This work develops a novel basis-adaptive method for constructing anisotropic polynomial chaos expansions of multidimensional (vector-valued, multi-output) model responses. 
The adaptive basis selection is based on multivariate sensitivity analysis metrics that can be estimated by post-processing the polynomial chaos expansion and results in a common anisotropic polynomial basis for the vector-valued response. 
This allows the application of the method to problems with up to moderately high-dimensional model inputs (in the order of tens) and up to very high-dimensional model responses (in the order of thousands). 
The method is applied to different engineering test cases for surrogate modeling and uncertainty quantification, including use cases related to electric machine and power grid modeling and simulation, and is found to produce highly accurate results with comparatively low data and computational demand.
\end{abstract}



\end{frontmatter}


\section{Introduction}
\label{sec:intro}

The \gls{pce} method is commonly employed in various forms \cite{shen2020polynomial} for surrogate modeling and \gls{uq} in scientific and engineering applications.
Therein, the \gls{pce} offers inexpensive polynomial approximations, commonly referred to as reduced order models or surrogate models, to be used in computationally demanding design tasks such as optimization and reliability analysis \cite{lim2021distribution, loukrezis2022power, suryawanshi2016reliability}, or for online model-based estimations in real time, e.g., in the context of digital twins \cite{thelen2022comprehensive, thelen2023comprehensive}. 
Additionally, uncertainty and sensitivity metrics regarding the model response can be estimated by simply post-processing the expansion's terms \cite{crestaux2009polynomial, knio2006uncertainty, sudret2008global} very efficiently, for example compared to \gls{mcs} methods \cite{zhang2021modern}.
Particularly popular are the regression-based variants of the \gls{pce}, 
where the expansion's coefficients are computed by means of least squares regression \cite{hadigol2018least, migliorati2014analysis}. 
In that case, the \gls{pce} is essentially a supervised machine learning regression model, with the additional benefit of typically requiring few training data and little computation time compared to other data-driven modeling methods \cite{torre2019data}.  

One of the bottlenecks of the \gls{pce} method is the so-called ``curse of dimensionality'', which in this case concerns the rapid growth of the polynomial basis, caused by large input dimensionality and high polynomial degrees. 
For regression-based \gls{pce} methods, this can lead to an ill-conditioned, ill-posed, or simply intractable least squares problem, since the training data demand rises along with the size of the polynomial basis \cite{migliorati2014analysis, cohen2018multivariate}. 
This issue is particularly relevant for numerous engineering applications where data acquisition is complex and/or expensive.
Remedies have been sought in sparse and/or basis-adaptive \gls{pce} methods (in the following referred to as sparse/adaptive). 
These methods construct the polynomial basis by exploiting possible anisotropies in the way the input parameters affect the model response. 
That is, depending on the model at hand, changes in certain parameters or parameter combinations can have a significant or negligible impact on the response. 
Taking into account the differences in parameter impact, an anisotropic polynomial basis can be constructed, which improves the performance of the \gls{pce} in terms of accuracy versus data demand.
Extensive reviews on sparse/adaptive, regression-based \gls{pce} methods can be found in the works of L\"uthen et al. \cite{luethen2021sparse, luethen2022automatic}. 

On the one hand, sparse PCEs are computed by solving a regularized least squares problem that induces sparsity in its solution and then omitting the zero coefficients.
Common methods for computing sparse \glspl{pce} are \gls{lar} \cite{blatman2011adaptive}, \gls{omp} \cite{jakeman2015enhancing}, and compressive sensing \cite{alemazkoor2017divide, diaz2018sparse, sargsyan2014dimensionality, tsilifis2019compressive}. 
On the other hand, basis-adaptive \glspl{pce} are adaptive approximation methods where the polynomial basis is progressively enriched with terms of high impact, while low-impact terms are neglected. 
Various approaches and criteria have been suggested in the literature for adaptive basis selection, for example, based on residual correlation \cite{blatman2011adaptive,  jakeman2015enhancing}, cross-validation error \cite{zhang2023novel}, coefficient confidence intervals \cite{abraham2017robust}, entropy principles \cite{he2020adaptive}, or variance contribution \cite{thapa2020adaptive, loukrezis2020robust}.
Considering non-trivial applications of practical value, sparse/adaptive \glspl{pce} are generally applicable for up to moderately high-dimensional inputs, typically in the order of tens or, less commonly, low hundreds.
In the case of very high-dimensional inputs, linear dimension reduction methods such as the Karhunen-Lo\`eve expansion are most commonly employed to enable the use of \glspl{pce} \cite{ghanem1990polynomial}. 
More recently, nonlinear dimension reduction methods have also been used for that purpose \cite{lataniotis2020extending, kontolati2022survey}. 

The dimensionality of the model's response is also a cause for concern regarding the computational cost of regression-based \glspl{pce}, as the least squares problem must be solved $M$ times, where $M$ denotes the dimension of the response.
Particularly considering high-dimensional responses, e.g., time-series or frequency responses \cite{mai2017surrogate, yaghoubi2017sparse}, the computational cost for computing the \gls{pce} may become unfeasible. 
This is particularly troublesome if sparse/adaptive \glspl{pce} are used, since 
most, if not all, of the available methods have been developed considering scalar model responses and rely on iterative algorithms that cannot be efficiently vectorized. 
For a multidimensional model response, these methods must be applied element-wise, often resulting in unacceptable computation times.
To render sparse/adaptive \gls{pce} methods tractable, dimension reduction techniques are used to reduce the dimensionality of the response to its minimum necessary components.
Most commonly, linear dimension reduction is used, e.g., based on \gls{pca} or \gls{pod} \cite{blatman2013sparse, bhattacharyya2020uncertainty, jacquelin2019random, hawchar2017principal, nagel2020principal}. 
Nonlinear dimension reduction has also been considered recently \cite{kontolati2022manifold, giovanis2024polynomial}. 
Note that \gls{uq} based on \glspl{pce} combined with dimension reduction necessitates the transformation of the \gls{pce} coefficients, such that they refer to the original (non-reduced) output. 
While this is straightforward for linear dimension reduction methods, it is not trivial for nonlinear methods, in which case \gls{uq} is typically performed by sampling the \gls{pce} instead of post-processing it.
In any case, if dimension reduction does not result in a sufficiently small response dimension, the computational cost may remain unacceptable.
Parallel computing can also be used to accelerate computation, assuming however that suitable hardware is available.

This paper suggests a new basis-adaptive \gls{pce} method that specifically targets multidimensional model responses.
Its main novelty lies in the basis selection criterion employed for the adaptive approximation, which utilizes the metrics of a multivariate \gls{gsa} method \cite{gamboa2014sensitivity}. 
The latter can be seen as a generalization of the well-known, variance-based Sobol sensitivity analysis \cite{sobol1993sensitivity} from scalar to functional and vector-valued model responses. 
Similar to the Sobol indices, the multivariate sensitivity indices can be estimated by post-processing a vector-valued \gls{pce} \cite{garcia2014global, sun2020global} and correspond to contributions to the aggregated variance over the multidimensional model response. 
Accordingly, the vector-valued \gls{pce} coefficients can be interpreted as multivariate sensitivity indicators. 
Basis enrichment is then performed according to the multivariate sensitivity indicators, such that the terms included in the expansion are the ones with the maximum contribution to the aggregated variance of the vector-valued model response. 
With respect to the candidate terms for basis enrichment, a forward neighbor approach is used \cite{luethen2022automatic}, such that the polynomial basis remains downward-closed \cite{cohen2018multivariate}.
We call this basis-adaptive \gls{pce} method \emph{\gls{mvsa}}.

The \gls{mvsa} \gls{pce} method is designed specifically for vector-valued model responses, contrary to sparse/adaptive \gls{pce} methods available in the literature.
Using the multivariate \gls{gsa}-based basis selection criterion, the \gls{mvsa} algorithm constructs a single anisotropic polynomial basis for the model response. 
The common polynomial basis allows the use of vectorized \gls{ols} solvers, leading to significantly reduced computation times.  
This can be an important advantage in several use cases, for example, surrogate models that must be computed online and in real time, e.g., in the context of digital twin applications, \gls{uq}-informed decision-making in short time windows, or if multiple models must be computed using different data sets, as in bootstrapping or ensemble modeling.
The common basis and the vectorized \gls{ols} solver also allow the \gls{mvsa} \gls{pce} method to be applied to very high-dimensional model responses, i.e., with sizes up to the order of thousands.
On possible downsides, the common basis obtained with the \gls{mvsa} algorithm can be sub-optimal considering all elements of the model response.
However, the trade-off is found to be more than acceptable, as illustrated by the numerical results presented in this work.

The remaining of this paper is structured as follows.
Section~\ref{sec:model} describes the general problem setting of parameter-dependent models with random inputs. 
Section~\ref{sec:gsa} recalls variance-based \gls{gsa} methods for scalar and multidimensional (vector-valued) model responses.
Section~\ref{sec:pce} discusses the \gls{pce} method, its computation by means of regression, and its connection to \gls{gsa}.
Section~\ref{sec:mvsa-pce} presents the \gls{mvsa} \gls{pce} method for the construction of anisotropic vector-valued \glspl{pce}. 
In Section~\ref{sec:num_exp}, the \gls{mvsa} \gls{pce} method is applied for surrogate modeling and \gls{uq} in different engineering test cases.
Our conclusions are discussed in Section~\ref{sec:conclusion}.

\section{Parameter-dependent model with random input data}
\label{sec:model}

In the following, we shall consider a parameter-dependent model  $f(\mathbf{x})$, such that
\begin{equation}
	\label{eq:parameter-dependent-model}	
	f \colon \mathbf{x} \to \mathbf{y}, \: \text{equivalently}, \: \mathbf{y} = f\left(\mathbf{x}\right), 
\end{equation}
where $\mathbf{x} \in \mathbb{R}^N$ is an input parameter vector and $\mathbf{y} \in \mathbb{R}^M$ the corresponding model response.
The model, which is here abstractly given as the function $f$, is considered to be deterministic, meaning that its response cannot vary given the same input.
As a minimal requirement, we will assume that $f$ is continuous and relatively smooth, such that the \gls{pce} method developed in this work can be applied as described in section~\ref{sec:mvsa-pce}. For non-smooth functions, a multi-element method would be necessary \cite{galetzka2023hp}.

Throughout this work, the model's input parameters $\mathbf{x}$ are considered to be realizations of a random vector $\mathbf{X}=(X_1,X_2,\dots,X_N)^\top$, where  $X_n,\,n=1,2,\dots,N$, are assumed to be independent random variables. 
The multivariate random variable  $\mathbf{X}$ is defined on the probability space $(\Theta,\Sigma,P)$, where $\Theta$ is the sample space, $\Sigma$ the sigma algebra of events, and $P \colon \Sigma \to [0,1]$ a probability measure.
Accordingly, $\mathbf{x} = \mathbf{X}(\theta) \in \Xi$ is a random variable realization for $\theta \in \Theta$, where $\Xi \subset \mathbb{R}^N$ is called the image space. 
The random vector is additionally characterized by the \gls{pdf} $\varrho_{\mathbf{X}}(\mathbf{x}) \colon \Xi \to \mathbb{R}_{\geq 0}$.
Under the independence assumption for the random variables $X_n$, it holds that  $\varrho_{\mathbf{X}}(\mathbf{x}) = \prod_{n=1}^N \varrho_{X_n}(x_n)$ and $\Xi = \Xi_1 \times\cdots \times \Xi_N$, where $\varrho_{X_n}(x_n)$ and $\Xi_n$ are the marginal (univariate) \glspl{pdf} and image spaces, respectively.
Due to the propagation of uncertainty through the model, its  response is now a dependent random vector $\mathbf{Y} = f(\mathbf{X})$.

\section{Global sensitivity analysis}
\label{sec:gsa}

\Gls{gsa} is broadly defined as the study of how uncertainty in the response of a model can be allocated to the considered input uncertainty sources \cite{saltelli2008global}. 
We are interested in variance-based \gls{gsa}, also referred to as the Sobol method \cite{sobol1993sensitivity, sobol2001global}.
The corresponding sensitivity metrics are known as Sobol indices.
Note that the Sobol method concerns scalar model responses only. 
Extensions were developed later to address multidimensional model responses \cite{campbell2006sensitivity, gamboa2014sensitivity}.

\subsection{Global sensitivity analysis for scalar model response}
\label{sec:gsa-scalar}

We consider a parametric model similar to the one described in Section~\ref{sec:model}, however, with a scalar response $y = f(\mathbf{x}) \in \mathbb{R}$.
The Sobol method begins with a decomposition of the scalar random response $Y = f(\mathbf{X})$, such that
\begin{equation}
	\label{eq:anova_decomp_func}
	Y = f\left(\mathbf{X}\right) = f_0 + \sum_{n=1}^N f_n\left(X_n\right) + \sum_{n<k}^N f_{nk}\left(X_n, X_k\right) + \cdots + f_{1,\dots,N}\left(X_1, \dots, X_N\right),
\end{equation}
where $f_0$ is a constant function, $f_{n}$ is a function of $X_n$ only, $f_{nk}$ is a function of $X_n$ and $X_k$, and so forth.
The terms of \eqref{eq:anova_decomp_func} are proven to be mutually orthogonal \cite{sobol2001global}, therefore, the expected value of the response is $\mathbb{E}\left[Y\right] = f_0$.
Then, the variance of the response, here denoted as $\mathbb{V}\left[Y\right]$, can be decomposed as
\begin{align}
	\label{eq:anova_decomp_var}
	\mathbb{V}\left[Y\right] = \sum_{n=1}^N \mathbb{V}\left[f_n\right] + \sum_{n<k}^N \mathbb{V}\left[f_{nk}\right] + \cdots + \mathbb{V}\left[f_{1,\dots,N}\right]
	= \sum_{n=1}^N \mathbb{V}_n + \sum_{n<k}^N \mathbb{V}_{nk} + \cdots + \mathbb{V}_{1,\dots,N},
\end{align}
where $\mathbb{V}_n$ is a partial variance attributed only to $X_n$, $\mathbb{V}_{nk}$ is a partial variance attributed to the combination of $X_n$ and $X_k$, and so forth.
First-order Sobol indices quantify the impact of the random variable $X_n$ alone, i.e., with all other parameters regarded as constant, and are given as
\begin{equation}
	\label{eq:sobol-first}
	S_n^{\text{F}} = \frac{\mathbb{V}_n}{\mathbb{V}\left[Y\right]}. 
\end{equation}
Total-effect Sobol indices quantify the impact of the random variable $X_n$ when interacting with all remaining random variables $X_k$, $k \neq n$, and are given as
\begin{equation}
	\label{eq:sobol-total}
	S_n^{\text{T}} = \frac{1}{\mathbb{V}\left[Y\right]} \left(\mathbb{V}_n + \sum_{\substack{n < k}}^N\mathbb{V}_{nk} + \cdots + \mathbb{V}_{1,\dots,N}\right).
\end{equation}
Sobol indices quantifying other interactions among the random variables, e.g., second- or third-order, can be computed in a similar manner \cite{saltelli2008global}.

\subsection{Global sensitivity analysis for vector-valued model response}
\label{sec:gsa-mv}
We now consider a multidimensional random response $\mathbf{Y} = \left(Y_1, \dots, Y_M\right)$ and the dependence $\mathbf{Y} = f(\mathbf{X})$, as described in Section~\ref{sec:model}.
In principle, the \gls{gsa} method presented in Section \ref{sec:gsa-scalar} can be applied element-wise to each scalar component $Y_m$, $m=1,\dots,M$.
However, correlations among the response's components render this approach questionable, as it may lead to redundant and difficult to interpret results \cite{lamboni2009multivariate, lamboni2011multivariate}. 
To circumvent this impasse, two multivariate \gls{gsa} methodologies have been developed, based on decomposing either the vector-valued response itself \cite{campbell2006sensitivity} or the corresponding covariance matrix  \cite{gamboa2014sensitivity}. 
Both approaches compute generalised sensitivity indices that quantify the impact of the input random variables on the full multivariate response.
In fact, the generalised sensitivity indices provided by these two multivariate \gls{gsa} approaches are equivalent \cite{garcia2014global}.

Following the work of Gamboa et al. \cite{gamboa2014sensitivity}, we consider a non-empty index set $\mathbf{u} \subset (1, \dots, N)$, its complement $\mathbf{v} = (1, \dots, N) \setminus \mathbf{u}$, and the corresponding input random variable subsets $\mathbf{X}_{\mathbf{u}} = \left\{X_n\right\}_{n \in \mathbf{u}}$, $\mathbf{X}_{\mathbf{v}} = \left\{X_n\right\}_{n \in \mathbf{v}}$.
By applying the Hoeffding decomposition \cite{van2000asymptotic}, we obtain the output decomposition
\begin{equation}
	\label{eq:hoeffding}
	\mathbf{Y} = f(\mathbf{X}) = f_0 + f_{\mathbf{u}}(\mathbf{X}_{\mathbf{u}}) + f_{\mathbf{v}}(\mathbf{X}_{\mathbf{v}}) + f_{\mathbf{u},\mathbf{v}}(\mathbf{X}_{\mathbf{u}}, \mathbf{X}_{\mathbf{v}}),
\end{equation}
where $f_0 \in \mathbb{R}^M$, $f_{\mathbf{u}}: \mathbb{R}^{\#\mathbf{u}} \rightarrow \mathbb{R}^M$, $f_{\mathbf{v}}:\mathbb{R}^{\#\mathbf{v}} \rightarrow \mathbb{R}^M$, and $f_{\mathbf{u},\mathbf{v}}:\mathbb{R}^{N} \rightarrow \mathbb{R}^M$, where $\#$ denotes the cardinality of a set.
Taking the covariance matrices in both sides of \eqref{eq:hoeffding} yields
\begin{equation}
	\label{eq:covar-decomp}
	\mathbf{C} = \mathbf{C}_{\mathbf{u}} + \mathbf{C}_{\mathbf{v}} + \mathbf{C}_{\mathbf{u}, \mathbf{v}},
\end{equation}
where $\mathbf{C}$, $\mathbf{C}_{\mathbf{u}}$, $\mathbf{C}_{\mathbf{v}}$, and $\mathbf{C}_{\mathbf{u}, \mathbf{v}}$ are the covariance matrices for $f$ (equivalently, for $\mathbf{Y}$), $f_{\mathbf{u}}$, $f_{\mathbf{v}}$, and $f_{\mathbf{u}, \mathbf{v}}$, respectively.
By considering index sets comprising single indices, index pairs, triplets, and so forth, the covariance decomposition \eqref{eq:covar-decomp} can be written as 
\begin{equation}
	\label{eq:covar-decomp-2}
	\mathbf{C} = \sum_{n=1}^N \mathbf{C}_{n} + \sum_{n<k}\mathbf{C}_{nk} + \cdots + \mathbf{C}_{1,\dots,N},
\end{equation}
which is the multidimensional analog to the variance decomposition \eqref{eq:anova_decomp_var}.
In fact, for a scalar response, the variance decomposition \eqref{eq:anova_decomp_var} is recovered from the covariance matrix decomposition \eqref{eq:covar-decomp-2}.
Based on this observation, generalised sensitivity indices for multidimensional responses are motivated as follows.
The sum of the variances of all response components $Y_m$, $m=1,\dots,M$, is equal to the trace of the covariance matrix $\mathbf{C}$, here denoted as $\mathrm{tr}(\mathbf{C})$, and corresponds to the aggregated variance of the multidimensional response.
Then, the multivariate equivalent to the first-order Sobol index \eqref{eq:sobol-first} is obtained as
\begin{equation}
	\label{eq:gen-sobol-first}
	G_n^{\text{F}} = \frac{\mathrm{tr}\left(\mathbf{C}_n\right)}{\mathrm{tr}\left(\mathbf{C}\right)},
\end{equation}
where the numerator quantifies the aggregated variance caused by the random variable $X_n$ alone.
In analogous fashion, the multivariate counterpart to the total-effect Sobol index \eqref{eq:sobol-total} is given by
\begin{equation}
	\label{eq:gen-sobol-total}
	G_n^{\text{T}} = \frac{\mathrm{tr}\left(\mathbf{C}_n\right) + \sum_{n<k} \mathrm{tr}\left(\mathbf{C}_{nk}\right) + \cdots + \mathrm{tr}\left(\mathbf{C}_{1,\dots,N}\right)}{\mathrm{tr}\left(\mathbf{C}\right)}.
\end{equation}

\section{Polynomial chaos expansion}
\label{sec:pce}

Originating from the work of Norbert Wiener on homogeneous chaos, the \gls{pce} was popularized much later as an efficient \gls{uq} method \cite{ghanem1990polynomial, xiu2002wiener}.
Recall from Section~\ref{sec:model} the (dependent) random model response $\mathbf{Y} = f(\mathbf{X})$, with $\mathbf{Y} = \left(Y_1, \dots, Y_M\right)$ and $\mathbf{X} = \left(X_1, \dots, X_N\right)$, and the fixed response $\mathbf{y} = f(\mathbf{x})$ for $\mathbf{x} = \mathbf{X}\left(\theta\right)$.
Then, a \gls{pce} is a global polynomial approximation of the form
\begin{equation}
	\label{eq:spectral_approx}
	f(\mathbf{x}) \approx \widetilde{f}(\mathbf{x}) = \sum_{k=1}^K \mathbf{c}_k \Psi_k(\mathbf{x}),
\end{equation}
where $\mathbf{c}_k \in \mathbb{R}^M$ are the expansion coefficients and $\Psi_k$ are multivariate polynomials that satisfy the orthogonality property
\begin{equation}
	\label{eq:orthogonality}
	\mathbb{E}\left[\Psi_k \Psi_l\right] = \int_{\Xi}  \Psi_k\left(\mathbf{x}\right) \Psi_l\left(\mathbf{x}\right) \varrho_{\mathbf{X}}\left(\mathbf{x}\right) \mathrm{d}\mathbf{x} =  \mathbb{E}\left[\Psi_k^2\right] \delta_{kl},
\end{equation}
where $\delta_{kl}$ is the Kronecker delta. 
Depending on the distribution of the random inputs, the polynomials are either selected according to the Wiener-Askey scheme \cite{xiu2002wiener} or constructed numerically \cite{soize2004physical, wan2006multi, oladyshkin2012data}.
In the remaining of this paper, we always consider orthonormal polynomials, such that $\mathbb{E}\left[\Psi_k^2\right] = 1$.
As mentioned in Section~\ref{sec:model}, the input random variables are assumed to be independent, hence, the joint \gls{pdf} is given as $\varrho_{\mathbf{X}} \left(\mathbf{x}\right) = \prod_{n=1}^N \varrho_{X_n}\left(x_n\right)$.
Note that it is possible to construct \glspl{pce} for dependent input random variables also \cite{feinberg2018multivariate, jakeman2019polynomial, rahman2018polynomial}.
Denoting with $\psi_n^{k_n}$ univariate polynomials of degree $k_n$ which are orthogonal, respectively, orthonormal with respect to the corresponding marginal \glspl{pdf} $\varrho_{X_n}$, and using the multi-index notation $\mathbf{k} = \left(k_1, \dots, k_N\right)$, multivariate orthogonal polynomials can be constructed as 
\begin{equation}
	\label{eq:pce-polys}
	\Psi_\mathbf{k}(\mathbf{x}) = \prod_{n=1}^N \psi_n^{k_n} (x_n).
\end{equation}

The \gls{pce} \eqref{eq:spectral_approx} can now be equivalently written as
\begin{equation}
	\label{eq:spectral_approx_multi_index}
	f(\mathbf{x}) \approx \widetilde{f}(\mathbf{x}) = \sum_{\mathbf{k} \in \Lambda} \mathbf{c}_{\mathbf{k}} \Psi_{\mathbf{k}}(\mathbf{x}),
\end{equation}
where the multi-indices $\mathbf{k}$ in \eqref{eq:spectral_approx_multi_index} are uniquely associated to the single indices $k$ in \eqref{eq:spectral_approx} and $\Lambda$ is a multi-index set with cardinality $\#\Lambda = K$. 

It is typical to use a multi-index set $\Lambda$ corresponding to a \gls{td} basis, where $\Lambda = \left\{\mathbf{k} : \lVert \mathbf{k} \rVert_1 \leq p\right\}$ for maximum univariate polynomial degree $p \in \mathbb{Z}_{\geq 0}$. 
Another common option is the hyperbolic truncation basis, where $\Lambda = \left\{\mathbf{k} : \lVert \mathbf{k} \rVert_q \leq p\right\}$, with $q \in \left(0, 1\right)$ and $ \lVert \mathbf{k} \rVert_q = \left(\sum_{n=1}^N \left(k_n\right)^q\right)^{1/q}$.
Note that both bases are isotropic, meaning that the same univariate basis terms are considered for all inputs $X_n$, $n=1,\dots,N$. 
As a result, the basis grows rapidly for an increasing maximum polynomial degree $p$ and, more crucially, for a high-dimensional input parameter vector $\mathbf{X}$, which is a manifestation of the so-called curse of dimensionality.

\subsection{Regression-based computation of polynomial chaos expansion coefficients}
\label{sec:pce-coeff}
In this work, the coefficients of the \gls{pce} are computed by means of least-squares regression \cite{hadigol2018least, migliorati2014analysis}.
Alternative options include pseudo-spectral projection \cite{constantine2012sparse, conrad2013adaptive, winokur2016sparse} and, less commonly, interpolation \cite{buzzard2013efficient, galetzka2023hp}.
Collecting the \gls{pce} coefficients $\mathbf{c}_k \in \mathbb{R}^M$, $k=1, \dots, K$, into a matrix $\boldsymbol{\Gamma} \in \mathbb{R}^{K \times M}$, such that the coefficient $\mathbf{c}_k$ is the $k$-th row of $\boldsymbol{\Gamma}$, the coefficients are computed by solving the least squares regression problem
\begin{align}
	\label{eq:regression}
	\underset {\boldsymbol{\Gamma} \in \mathbb{R}^{K \times M}}{\arg\min} \left\{\frac{1}{Q}\sum_{q=1}^{Q} \left( \mathbf{y}^{(q)} - \sum_{k=1}^K \mathbf{c}_k \Psi_k\left(\mathbf{x}^{(q)}\right) \right)^2\right\},
\end{align}
where $\left\{\mathbf{x}^{(q)}, \mathbf{y}^{(q)} = f\left(\mathbf{x}^{(q)}\right)\right\}_{q=1}^{Q}$ is a set of input parameter realizations along with the corresponding model responses, called the experimental design or the training data set.
In matrix format, the regression problem \eqref{eq:regression} is equivalently written as 
\begin{align}
	\label{eq:regression-matrix}
	\underset {\boldsymbol{\Gamma} \in \mathbb{R}^{K \times M}}{\arg\min} \left\{\frac{1}{Q} \lVert \mathbf{B} - \mathbf{D} \boldsymbol{\Gamma}  \rVert_2 \right\},
\end{align}
where $\mathbf{D} \in \mathbb{R}^{Q \times K}$ with $d_{qk} = \Psi_k\left(\mathbf{x}^{(q)}\right)$ is the least squares system matrix, also called the design matrix, and $\mathbf{B} \in \mathbb{R}^{Q \times M}$ with $b_{qm} = y_m^{(q)}$ the right-hand side.

For the regression problem to be well posed and uniquely solvable with an \gls{ols} solver, the size of the experimental design must be at least equal to the size of the polynomial basis, i.e., $Q \geq K$. 
For a well conditioned \gls{ols} problem, $Q$ is typically chosen to be 2--5 times larger than $K$, depending on the problem at hand \cite{cohen2018multivariate, migliorati2014analysis}.
This can lead to severe computational problems if the polynomial basis is too large, e.g., for a \gls{td} basis with high maximum polynomial degree $p$ and input dimensionality. 
This limitation can be overcome by methods that penalize \eqref{eq:regression}--\eqref{eq:regression-matrix} to compute sparse solutions \cite{luethen2021sparse}, for example using the least absolute shrinkage and selection operator (LASSO) method \cite{tibshirani1996regression} or compressive sensing \cite{donoho2006compressed}.
Sparse solutions are typically computed using iterative algorithms such as \gls{lar} \cite{blatman2011adaptive}, \gls{omp} \cite{jakeman2015enhancing}, or \gls{sp} \cite{diaz2018sparse}.
Alternatively, basis-adaptive algorithms can be used  \cite{luethen2022automatic}, where the basis is sequentially enriched with new terms, up to a size that does not lead to conditioning issues.
A limitation here is that the sparse/adaptive \gls{pce} algorithms available in the literature consider scalar model responses almost exclusively, therefore, must be applied element-wise for each of the $M$ components of a vector-valued model response. 
This can lead to an undesirable computational cost with respect to the computation of the \gls{pce} coefficients if the response is high-dimensional. 

\subsection{Uncertainty quantification and sensitivity analysis based on polynomial chaos expansion}
\label{sec:pce-gsa}

For a \gls{pce} approximation in the form of \eqref{eq:spectral_approx_multi_index}, it is straightforward to show that the mean and variance of the response can be estimated as
\begin{subequations}
	\label{eq:pce-moments}
	\begin{align}
		\mathbb{E}\left[\mathbf{Y}\right] &\approx \mathbf{c}_{\mathbf{0}}, \label{eq:pce-mean}\\
		\mathbb{V}\left[\mathbf{Y}\right] &\approx \sum_{\mathbf{k} \in \Lambda \setminus \mathbf{0}} \mathbf{c}_{\mathbf{k}}^2,\label{eq:pce-variance}
	\end{align}
\end{subequations}
where the zeroth multi-index is denoted as $\mathbf{0} = \left(0,\dots,0\right)$ and the \gls{pce} basis is considered to be orthonormal \cite{ghanem1990polynomial}.

For a scalar response $Y = f\left(\mathbf{X}\right)$, the \gls{pce} takes the form of the Sobol decomposition \eqref{eq:anova_decomp_func} by appropriately ordering its terms.
Hence, Sobol sensitivity indices can be estimated by simply post-processing the \gls{pce} \cite{crestaux2009polynomial, sudret2008global}. 
Indeed, the \gls{pce} coefficients $c_{\mathbf{k}}$ for  $\mathbf{k} \in \Lambda \setminus \mathbf{0}$ can be interpreted as partial variances due to specific random variable interactions defined by the multi-indices $\mathbf{k}$. 
The multi-indices corresponding to partial variances caused by $X_n$, either individually (first-order) or in combination with all other random variables (total-effect), can then be collected into the multi-index sets $\Lambda_n^\text{F} \subset \Lambda$ and $\Lambda_n^\text{T} \subset \Lambda$, respectively defined as
\begin{subequations}
	\begin{align}
		\Lambda_n^{\text{F}} &= \{\mathbf{k} \in \Lambda \; : \; k_n \neq 0 \:\: \text{and} \:\: k_l = 0, l = 1,\dots,N, l \neq n\}, \\
		\Lambda_n^{\text{T}} &= \{\mathbf{k} \in \Lambda \; : \; k_n \neq 0\}.
	\end{align}
\end{subequations}
The first-order and total-effect Sobol indices are then estimated as 
\begin{subequations}
	\label{eq:pce-sobol}
	\begin{align}
		S_n^{\text{F}} &\approx \frac{\sum_{\mathbf{k} \in \Lambda_n^{\text{F}}} c_{\mathbf{k}}^2}{\sum_{\mathbf{k} \in \Lambda \setminus \mathbf{0}} c_{\mathbf{k}}^2}, \\
		S_n^{\text{T}} &\approx \frac{\sum_{\mathbf{k} \in \Lambda_n^{\text{T}}} c_{\mathbf{k}}^2}{\sum_{\mathbf{k} \in \Lambda \setminus \mathbf{0}} c_{\mathbf{k}}^2}.
	\end{align}
\end{subequations}

For a multidimensional response $\mathbf{Y} = f\left(\mathbf{X}\right)$, multivariate \gls{gsa} requires the computation of the traces of the covariance matrices with respect to the components $Y_m$, which depend on specific combinations of the input random variables, as shown in Section \ref{sec:gsa-mv}.
The traces are equal to the sum of the variances of all response components dependent on the specific input random variable combinations. Similar to the case of a scalar response, the partial variances are easily computed from the \gls{pce} coefficients \cite{garcia2014global}. 
Then, the generalised sensitivity indices $G_n^{\text{F}}$ and $G_n^{\text{T}}$ defined in \eqref{eq:gen-sobol-first} and \eqref{eq:gen-sobol-total}, respectively, can be estimated as 
\begin{subequations}
	\label{eq:pce-gen-sobol}
	\begin{align}
		G_n^{\text{F}} &\approx \frac{\sum_{m=1}^M \left(\sum_{\mathbf{k} \in \Lambda_{n}^{\text{F}}} c_{m,\mathbf{k}}^2\right)}{\sum_{m=1}^M \left(\sum_{\mathbf{k} \in \Lambda \setminus \mathbf{0}} c_{m,\mathbf{k}}^2\right)},\\
		G_n^{\text{T}} &\approx \frac{\sum_{m=1}^M \left(\sum_{\mathbf{k} \in \Lambda_{n}^{\text{T}}} c_{m,\mathbf{k}}^2\right)}{\sum_{m=1}^M \left(\sum_{\mathbf{k} \in \Lambda \setminus \mathbf{0}} c_{m,\mathbf{k}}^2\right)}.
	\end{align}
\end{subequations}
Accordingly, a vector-valued \gls{pce} coefficient $\mathbf{c}_{\mathbf{k}} = \left(c_{m,\mathbf{k}}\right)_{1 \leq m \leq M}$ yields a contribution to the aggregated variance over the multidimensional model response, that is specific to the multi-index $\mathbf{k}$.

Note that of a common multi-index set $\Lambda$ is assumed in formulas \eqref{eq:pce-variance}-\eqref{eq:pce-gen-sobol}, implying a common \gls{pce} basis for all response elements.
This is not necessarily true, for example, if a sparse/adaptive \gls{pce} method is applied element-wise over the response, or for isotropic \glspl{pce} with different maximum polynomial degrees per response element. 
In that case, $\Lambda = \bigcup_{m=1}^M \Lambda_m$ is to be used, where if $\mathbf{k} \in \Lambda$ and $\mathbf{k} \not \in \Lambda_m$, then $c_{m,\mathbf{k}} = 0$.

\section{Multivariate sensitivity-adaptive polynomial chaos expansion}
\label{sec:mvsa-pce}

In the \gls{mvsa} \gls{pce} method, the polynomial basis is expanded sequentially, such that the newly added expansion terms are the ones corresponding to the maximum contribution to the aggregated variance of the multidimensional model response (see Sections~\ref{sec:gsa-mv} and \ref{sec:pce-gsa}). 
In that way, a common multi-index set $\Lambda$ and polynomial basis are constructed for the vector-valued response.
The candidate terms are selected based on a forward neighbor approach \cite{jakeman2015enhancing, luethen2022automatic}, such that the multi-index set remains downward-closed \cite{cohen2018multivariate}. 
The \gls{pce} coefficients are computed with an \gls{ols} solver. 
The sequential basis expansion is terminated if the \gls{ols} problem becomes ill-posed or ill-conditioned.
The main steps and building blocks of the \gls{mvsa} \gls{pce} method are explained in detail in the following. 
The complete procedure is also described in Algorithm~\ref{algo:mvsa-pce}.

\begin{algorithm}[t!]
	\caption{Multivariate sensitivity-adaptive polynomial chaos expansion}
	\label{algo:mvsa-pce}
	
	\footnotesize{
		\SetAlgoLined
		\KwData{Training data set $\left\{\mathbf{x}^{(q)}, \mathbf{y}^{(q)} = f\left(\mathbf{x}^{(q)}\right)\right\}_{q=1}^{Q}$ with $\mathbf{x}^{(q)} \in \mathbb{R}^N$ and $\mathbf{y}^{(q)} \in \mathbb{R}^M$, initial downward-closed multi-index set $\Lambda_{\text{init}}$ with $\#\Lambda_{\text{init}} < Q$, maximum condition number $\kappa$.}
		\KwResult{Multi-index set $\Lambda$ and corresponding \gls{pce}.}
		
		Set $\Lambda = \Lambda_{\text{init}}$.
		
		\While(\tcp*[f]{adaptive basis expansion}){True}{ 
			
			Compute the admissible multi-index set $\Lambda^+_{\text{adm}}$ as in \eqref{eq:admissible_index_set}.
			
			Construct the extended multi-index set $\Lambda_{\text{ext}} = \Lambda \cup \Lambda^+_{\text{adm}}$.
			
			\If(\tcp*[f]{underdetermined \gls{ols} problem}){$\#\Lambda_{\mathrm{ext}} > Q$}{
				Exit while-loop.
			}
			
			Solve the \gls{ols} problem \eqref{eq:regression}, equivalently \eqref{eq:regression-matrix}, for the extended multi-index set $\Lambda_{\text{ext}}$.
			
			\If(\tcp*[f]{ill-conditioned \gls{ols} problem}){$\mathrm{cond}\left(\mathbf{D}_\mathrm{ext}\right) > \kappa$}{
				Exit while-loop.
			}
			
			Find the multi-index $\mathbf{k}^* = \argmax_{\mathbf{k} \in \Lambda^+_{\text{adm}}} \eta_{\mathbf{k}}$, where $\eta_{\mathbf{k}} = \sum_{m=1}^M c_{m,\mathbf{k}}^2$ are the sensitivity indicators defined in \eqref{eq:eta-def}.
			
			Update the multi-index set $\Lambda$, such that $\Lambda \leftarrow \Lambda \cup \left\{\mathbf{k}^*\right\}$.
		}
		
		Set $\Lambda = \Lambda_\mathrm{ext}$.
		
		\While(\tcp*[f]{basis pruning}){$\mathrm{cond}\left(\mathbf{D}\right) > \kappa$ $\mathrm{or}$ $\#\Lambda > Q$}{
			
			Solve the \gls{ols} problem \eqref{eq:regression}, equivalently \eqref{eq:regression-matrix}, for the multi-index set $\Lambda$.
			
			Find the multi-index $\mathbf{k}^\dagger = \argmin_{\mathbf{k} \in \Lambda} \eta_{\mathbf{k}}$.
			
			Update the multi-index set $\Lambda$, such that $\Lambda \leftarrow \Lambda \setminus \left\{\mathbf{k}^\dagger\right\}$.
		}
		Solve the \gls{ols} problem \eqref{eq:regression}, equivalently \eqref{eq:regression-matrix}, for the final multi-index set $\Lambda$. 
	}
\end{algorithm}

\subsection{Downward closed multi-index set}
\label{sec:dc-set}
In the following, we will require the multi-index set $\Lambda$ defining the basis of the \gls{pce} to be downward-closed, i.e., to satisfy the property
\begin{equation}
	\forall \mathbf{k} \in \Lambda \Rightarrow \mathbf{k} - \mathbf{e}_n \in \Lambda, n=1,\dots,N, k_n \neq 0,
\end{equation}
where $\mathbf{e}_n = \left(\delta_{n \nu}\right)_{1 \leq \nu \leq N}$ is the $n$th unit vector and $\delta_{n \nu}$ denotes the Kronecker delta. 
Downward closed multi-index sets are also called monotone or lower sets.
The corresponding downward-closed polynomial space $\mathbb{P}_{\Lambda} = \text{span}\left\{\Psi_{\mathbf{k}}(\mathbf{x}), \mathbf{k} \in \Lambda \right\}$ satisfies desirable properties such as differentiation in any variable and invariance by a change of basis \cite{cohen2018multivariate}.

\subsection{Forward neighbors and admissible multi-indices}
Given a multi-index set $\Lambda$, the set of forward neighbors is defined as
\begin{equation}
	\Lambda^+ = \left\{\mathbf{k} + \mathbf{e}_n \not\in \Lambda \:|\: n=1,\dots,N, \mathbf{k} \in \Lambda\right\}.
\end{equation}

Assuming that the multi-index set $\Lambda$ is downward-closed, it can easily be observed that $\Lambda \cup \Lambda^+$ is not necessarily downward-closed, i.e., there may exist multi-indices $\mathbf{k} \in \Lambda^+$ for which $\Lambda \cup \left\{\mathbf{k}\right\}$ is not a downward-closed set.
To retain the downward-closedness property, the set of admissible forward neighbors is defined as 
	\begin{equation}
		\Lambda^+_{\mathrm{adm}} = \left\{ \mathbf{k} \in \Lambda^+  \:|\: \mathbf{k} - \mathbf{e}_n \in \Lambda, n=1,\dots,N, k_n \neq 0\right\}.
		\label{eq:admissible_index_set}
	\end{equation}
That is, for any admissible multi-index $\mathbf{k} \in \Lambda^+_{\mathrm{adm}}$, the set $\Lambda \cup \left\{\mathbf{k}\right\}$ is downward-closed. 
Equivalently, $\Lambda \cup \Lambda^+_{\mathrm{adm}}$ is downward-closed.

\subsection{Adaptive basis expansion}
\label{sec:adaptive-basis-expansion}
Assuming that the polynomial basis of a currently available \gls{pce} is based on a downward-closed multi-index set $\Lambda$, we require that the polynomial basis will be expanded by \gls{pce} terms corresponding to admissible multi-indices $\mathbf{k} \in \Lambda^+_{\mathrm{adm}}$.
If a downward-closed \gls{pce} basis is not readily available, the adaptive basis expansion starts with the zero multi-index, i.e., $\Lambda = \left\{ \mathbf{0} \right\}$. 

In each basis expansion step, the set of admissible neighbors $\Lambda_{\text{adm}}^+$ is computed and an extended multi-index set $\Lambda_{\text{ext}} = \Lambda \cup \Lambda_{\text{adm}}^+$ is formed.
Using the polynomial basis corresponding to $\Lambda_{\text{ext}}$ and the available training data $\left\{\mathbf{x}^{(q)}, \mathbf{y}^{(q)} = f\left(\mathbf{x}^{(q)}\right)\right\}_{q=1}^{Q}$, the \gls{ols} problem \eqref{eq:regression}, equivalently \eqref{eq:regression-matrix}, is solved to compute the vector-valued coefficients $\mathbf{c}_{\mathbf{k}}$, $\mathbf{k} \in \Lambda_{\text{ext}}$. 
As previously shown in Section~\ref{sec:pce-gsa},  the value of the sum 
\begin{equation}
	\eta_{\mathbf{k}} = \sum_{m=1}^M c_{m,\mathbf{k}}^2,
	\label{eq:eta-def}
\end{equation}
quantifies the contribution of the $\mathbf{k}$-indexed polynomial term to the aggregated variance of the $M$-dimensional model response, see equation \eqref{eq:pce-variance}. 
Looking at the \gls{pce}-based estimation of the multivariate sensitivity indices given in equations \eqref{eq:pce-gen-sobol}, $\eta_{\mathbf{k}}$ can be interpreted as an indicator of the sensitivity of the multidimensional response to the $\mathbf{k}$-indexed polynomial term. 
Therefore, in every step of the basis expansion process, the multi-index set $\Lambda$ is expanded as
\begin{equation}
	\Lambda \leftarrow \Lambda \cup \left\{\mathbf{k}^*\right\}, \:\: \text{where} \:\: \mathbf{k}^* = \argmax_{\mathbf{k} \in \Lambda_{\text{adm}}^+} \eta_{\mathbf{k}},
\end{equation}
such that the currently available downward-closed multi-index set is expanded with the admissible multi-index that corresponds to the maximum variance contribution, equivalently, to the maximum multivariate sensitivity indicator.
For that reason, we call the basis expansion procedure \emph{multivariate sensitivity-adaptive (\gls{mvsa})}.

\subsection{Termination criteria and basis pruning}

The adaptive basis expansion procedure described above continues as long as two conditions are respected. 
The first is that the \gls{ols} system cannot become underdetermined when computing the \gls{pce} coefficients.
The second is related to the conditioning of the \gls{ols} problem.
For that reason, a limit value $\kappa\in\mathbb{R}$ is set for the condition number of the \gls{ols} system matrix, also called the design matrix (see Section \ref{sec:pce-coeff}).
The latter is essentially a hyperparameter and its optimal value differs for each considered problem. 
Additionally, the value of $\kappa$ affects the accuracy of the obtained \gls{pce} and its computation time.
In the numerical experiments presented in Section~\ref{sec:num_exp}, the value $\kappa = 100$ is chosen, following the results of our own prior work \cite{loukrezis2020robust} regarding the sensitivity of surrogate modeling performance with to this hyperparameter. 
This proved to be a good practical choice for all test cases examined in the present work.

Denoting with $\mathbf{D}_{\mathrm{ext}}$ the design matrix corresponding to the extended multi-index set $\Lambda_{\mathrm{ext}}$, the adaptive \gls{pce} basis expansion proceeds as follows. 
As long as the conditions $\#\Lambda_{\text{ext}} \leq Q$ and $\mathrm{cond}\left(\mathbf{D}_\mathrm{ext}\right) \leq \kappa$ are satisfied, the polynomial basis is expanded as described in Section~\ref{sec:adaptive-basis-expansion}. 
If either condition is violated, the adaptive basis expansion stops and a \gls{pce} is computed using the multi-index set $\Lambda = \Lambda_\mathrm{ext}$.
Of course, this choice violates at least one of the considered \gls{ols} stability criteria, most commonly the one related to the condition number of the design matrix.
To correct this, a basis pruning procedure is followed, where polynomial terms of comparatively reduced contribution to the \gls{pce} are removed from $\Lambda$ until the \gls{ols} stability criteria are satisfied.
In each step of the basis pruning procedure, the multi-index that corresponds to the minimum sensitivity indicator $\eta_{\mathbf{k}}$ is identified and removed from the multi-index set, such that 
\begin{equation}
	\label{eq:coeff-removal}
	\Lambda \leftarrow \Lambda \setminus \left\{\mathbf{k}^\dagger\right\}, \:\: \text{where} \:\: \mathbf{k}^\dagger = \argmin_{\mathbf{k} \in \Lambda} \eta_{\mathbf{k}}.
\end{equation}
The corresponding polynomial term is then removed from the \gls{pce} basis.
The removal of multi-indices and polynomial terms continues until the conditions $\#\Lambda \leq Q$ and $\mathrm{cond}\left(\mathbf{D}\right) \leq \kappa$ are satisfied, where $\mathbf{D}$ is the design matrix corresponding to the multi-index set $\Lambda$.

Note that the final multi-index set resulting from the basis pruning procedure is not necessarily downward-closed. 
The downward-closedness constraint is crucial during the adaptive basis expansion, as it prevents the admissible multi-index set from becoming intractable due to the curse of dimensionality.
Once the adaptive basis expansion has been terminated, the goal is to retain the most significant basis terms, which are here selected according to the (multivariate) sensitivity of the vector-valued response.
The removal of insignificant multi-indices can result in a final multi-index set that is not downward-closed.

\section{Numerical experiments}
\label{sec:num_exp}

In the following, the \gls{mvsa} \gls{pce} method is applied for surrogate modeling and \gls{uq} to different engineering problems, each featuring a high-dimensional model response and moderately high-dimensional random inputs. 
In all considered test cases, the \gls{mvsa} \gls{pce} is computed for training data sets of increasing size and then compared against other methods with respect to its approximation accuracy, uncertainty estimation accuracy, and computational efficiency.
The latter concerns both accuracy versus training data demand and computation time.
We note the following remarks with respect to the presented numerical results:

\paragraph{Surrogate model accuracy}
The approximation accuracy of a surrogate model is evaluated using the \gls{rmse} metric, which is computed using a test data set. 
Naturally, the test data are not included in the training data set. 

\paragraph{Uncertainty estimates}
The estimated uncertainty metrics, in particular, expected values and standard deviations of the vector-valued responses, are compared against reference values obtained with \gls{mcs}. The performance of \gls{mcs} is not affected from the input or output dimensionality of the model, hence this choice. However, \gls{mcs} converges slowly and therefore needs a large number of model evaluations for accurate uncertainty estimates. 

\paragraph{Computation time} Where applicable, we report the time needed to compute the \glspl{pce} as a further efficiency metric next to accuracy versus training data demand.  
Despite commonly regarded as negligible next to the cost of data acquisition, a faster computation time - in our case, often several orders of magnitude faster - can be an important advantage in various use cases.
One such use case concerns online surrogate modeling, which is a common element of digital twin applications. 
Therein, data-driven regression models must often be computed using sensor data obtained during the operation of the physical asset. 
These models must be made available as quickly as possible, so that they can be used in real time during operation, especially since changes in the operating conditions might render a data-driven model invalid, in which case a new model must be computed.
Another use case concerns the use of a \gls{pce} method for providing \gls{uq} metrics, e.g., variance-based confidence intervals, to be used for a reliability- or safety-related decision within a short time window. 
Other relevant use cases arise when multiple data-driven models must be computed using the same method, as is done in, e.g., $k$-fold cross-validation, bootstrapping, or ensemble modeling.
Then, the computation time for a single model accumulates, possibly amounting to an undesired overall computational cost. 
In all aforementioned use cases, computation time can be a deciding factor regarding the suitability of a surrogate modeling or \gls{uq} method.

\paragraph{Training and test data sets} The training and test data sets are generated by randomly sampling the considered models according to the joint probability distribution of the input parameters. We consider a small-data training regime, as is the usual case in real-world engineering applications. 
The size of the training data set $Q$ depends on the specific test case.
The size of the test data set is $Q'=10^3$, unless stated otherwise. 
All numerical experiments are repeated for ten seeds, equivalently, for ten different training and test data sets.

\paragraph{Comparisons among \gls{pce} methods}
The \gls{mvsa} \gls{pce} is compared  against \gls{td} \glspl{pce} and, if feasible, sparse/adaptive \glspl{pce}.
For the latter, we opt for the degree- and $q$-norm-adaptive ($p/q$-adaptive) \gls{lar} \gls{pce} method \cite{blatman2011adaptive}, which was consistently found to be the best option compared to alternatives, for example, $p/q$-adaptive \gls{omp} \cite{jakeman2015enhancing} and \gls{sp} \cite{diaz2018sparse} \glspl{pce}. 
Note that the $p/q$-adaptive \gls{lar} \gls{pce} method must be applied element-wise for a vector-valued response, see for example \cite{blatman2013sparse, mai2017surrogate, yaghoubi2017sparse}.

\paragraph{Implementation specifics}
For the \gls{mvsa} \gls{pce} method, we use an in-house \texttt{Python} implementation\footnote{\texttt{https://github.com/dlouk/mvsa-pce}}, which is partially based on the \texttt{OpenTURNS} library \cite{baudin2016openturns}. 
The \texttt{OpenTURNS} library is also used for computing \gls{td} \glspl{pce}.
Note that the employed \gls{ols} solver can resolve underdetermined systems -- often arising when \gls{td} \glspl{pce} are computed -- by returning the minimal $2$-norm solution among multiple minimizing solutions.
The $p/q$-adaptive \gls{lar} \gls{pce} method is implemented using the \texttt{MATLAB}-based \texttt{UQLab} software \cite{marelli2014uqlab}.
Also note that our in-house implementation is at disadvantage concerning its performance in terms of computation time, since \texttt{MATLAB} outperforms \texttt{\texttt{Python}} significantly in terms of computation speed.
Still, the \gls{mvsa} algorithm computes multidimensional \glspl{pce} much faster, as the numerical results show. 
We expect that this advantage will be even more pronounced if the \gls{mvsa} \gls{pce} method is implemented with a more performant programming language.
All computations have been performed using the 64-bit double-precision floating-point format on a standard x64-based Windows machine equipped with an Intel\textsuperscript{\textregistered} Core\textsuperscript{\texttrademark} i5-1245U processor (1.6 GHz, 10 cores, 12 logical processors) and 32 GB RAM.

\subsection{Deflection of simply supported beam}
\label{sec:beam}

We first consider the model of simply supported beam under uniform load, the input dimensions of which are artificially increased.
Using this simple toy problem, our main goal is to illustrate the limitations of \gls{td} and sparse/adaptive \glspl{pce} that must be applied element-wise, especially for cases where high-dimensional inputs and outputs need to be addressed.

\begin{figure}[t!]
	\centering
	\begin{subfigure}[b]{0.49\textwidth}
		\centering
		\includegraphics[width=1\textwidth]{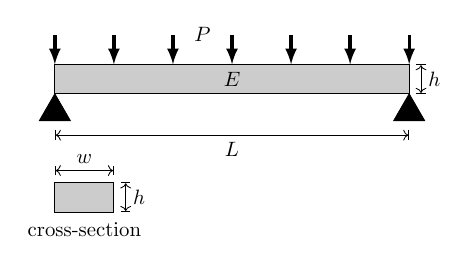}
		\caption{Simply supported beam under uniform load.}
		\label{fig:beam_sketch}
	\end{subfigure}
	\hfill
	\begin{subfigure}[b]{0.49\textwidth}
		\centering
		\includegraphics[width=1\textwidth]{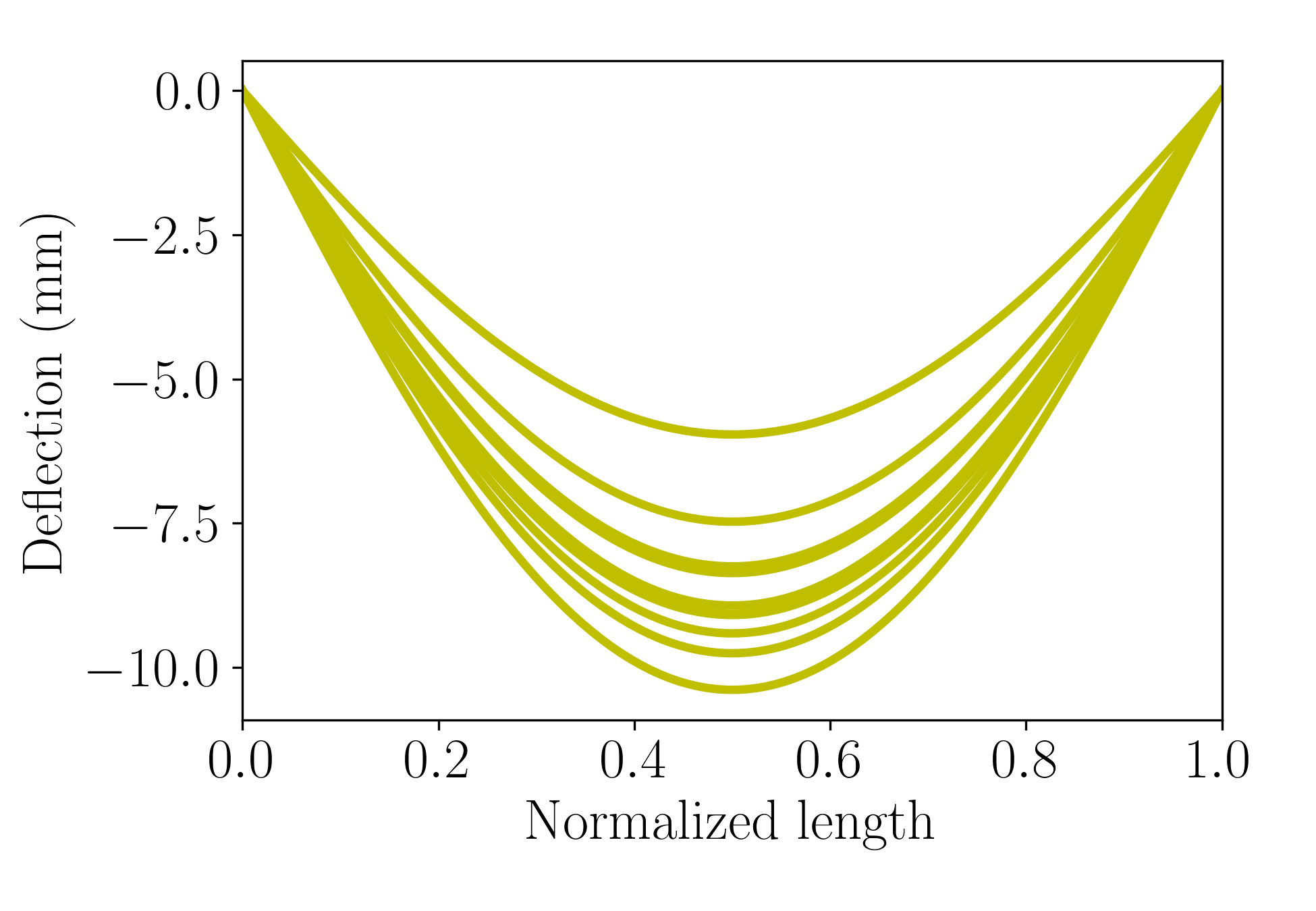}
		\caption{Deflection over normalized beam length.}
		\label{fig:beam_deflection}
	\end{subfigure}
	\caption{(a) Sketch of the simply supported beam model. (b) Beam deflection for $10$ realizations of the simply supported beam model parameters.}
	\label{fig:beam_illustration}
\end{figure} 

A sketch illustration of the simply supported beam model is shown in Figure~\ref{fig:beam_sketch}. 
The beam's geometry is defined by its width $w$, height $h$, and length $L$.
The Young's modulus of the beam is denoted with $E$. 
Under the uniform load $P$, the beam's deflection at a longitudinal coordinate $\ell_m$ along its length is given by
\begin{equation}
	\label{eq:beam_deflection}
	\delta\left(\ell_m\right) = \frac{P \, \ell_m \left( L^3 - 2 \, \ell_m^2 \, L + \ell_m^3 \right)}{2 \, E \, w \, h^3},
\end{equation} 
where the coordinates $\ell_m$, $m=1, \dots, M$, are uniformly distributed along the length of the beam excluding the points of support, such that $\ell_m = m L / (M+1)$. 
These five parameters are assumed to be random variables following a log-normal distribution, see Table~\ref{tab:beam-params}.
The beam's deflection for different realizations of the input model parameters is shown in Figure~\ref{fig:beam_deflection}.

\begin{table}[b!]
	\small
	\caption{Parameters of the simply supported beam model. All parameters follow a log-normal distribution with the given values for mean and standard deviation.}
	\centering
	\begin{tabular}{l c c c c c}
		\toprule
		Parameter & Units & Notation & Mean & St. dev.  \\[0.5ex] 
		\toprule
		Width & m & $w$ & $0.15$ & $0.0075$ \\ [0.5ex]
		Height & m & $h$ & $0.3$ & $0.015$ \\ [0.5ex]
		Length & m & $L$ & $5$ & $0.05$ \\[0.5ex]
		Young's modulus & Pa & $E$ & $3 \cdot 10^{10}$ & $4.5 \cdot 10^{9}$ \\[0.5ex]
		Load & N/m & $P$ & $10^{4}$ & $2 \cdot 10^{3}$\\[0.5ex] 
		Dummy & -- & $d_1$--$d_{15}$ & $10$ & $1$ \\[0.5ex] 
		\bottomrule
	\end{tabular}
	\label{tab:beam-params}
\end{table}

Additionally, we consider $15$ dummy parameters, denoted as $d_i$, $i=1,\dots,15$, each of which also follows a log-normal distribution, see Table~\ref{tab:beam-params}.
These parameters have no impact on the response of the model, hence, their distribution is irrelevant.
Nevertheless, the artificially increased input dimensionality still affects the performance of the considered \gls{pce} methods, particularly that of the \gls{td} \glspl{pce}, for which the curse of (input) dimensionality can be detrimental. 

In the following, three model response dimensions and three training data set sizes are considered, namely, $M \in \left\{10, 100, 1000\right\}$ and $Q \in \left\{50, 100, 150\right\}$. 
The beam deflection model \eqref{eq:beam_deflection} is approximated by \gls{mvsa}, $p/q$-adaptive \gls{lar}, and \gls{td} \glspl{pce}, the latter for maximum polynomial degrees $p=2$ and $p=3$.
Note that the \glspl{pce} are computed with respect to $\log(\mathbf{X})$ (normally distributed) instead of $\mathbf{X}$ (log-normally distributed), where $\mathbf{X} = \left(w,h,L,E,P,d_1, d_2,\dots,d_{15}\right)$ denotes the input random vector. 
Hence, according to the Wiener-Askey scheme \cite{xiu2002wiener}, the polynomial basis consists of Hermite polynomials.

\begin{figure}[t!]
	\centering
	\begin{subfigure}[b]{0.65\textwidth}
		\centering
		\fbox{\includegraphics[width=1\textwidth]{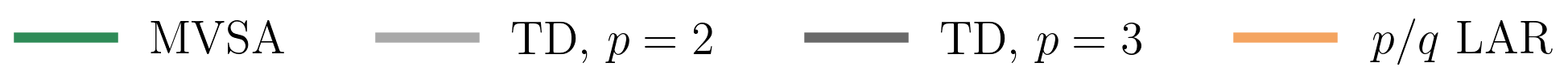}}
	\end{subfigure}
	\\
	\begin{subfigure}[b]{0.328\textwidth}
		\centering
		\includegraphics[width=\textwidth]{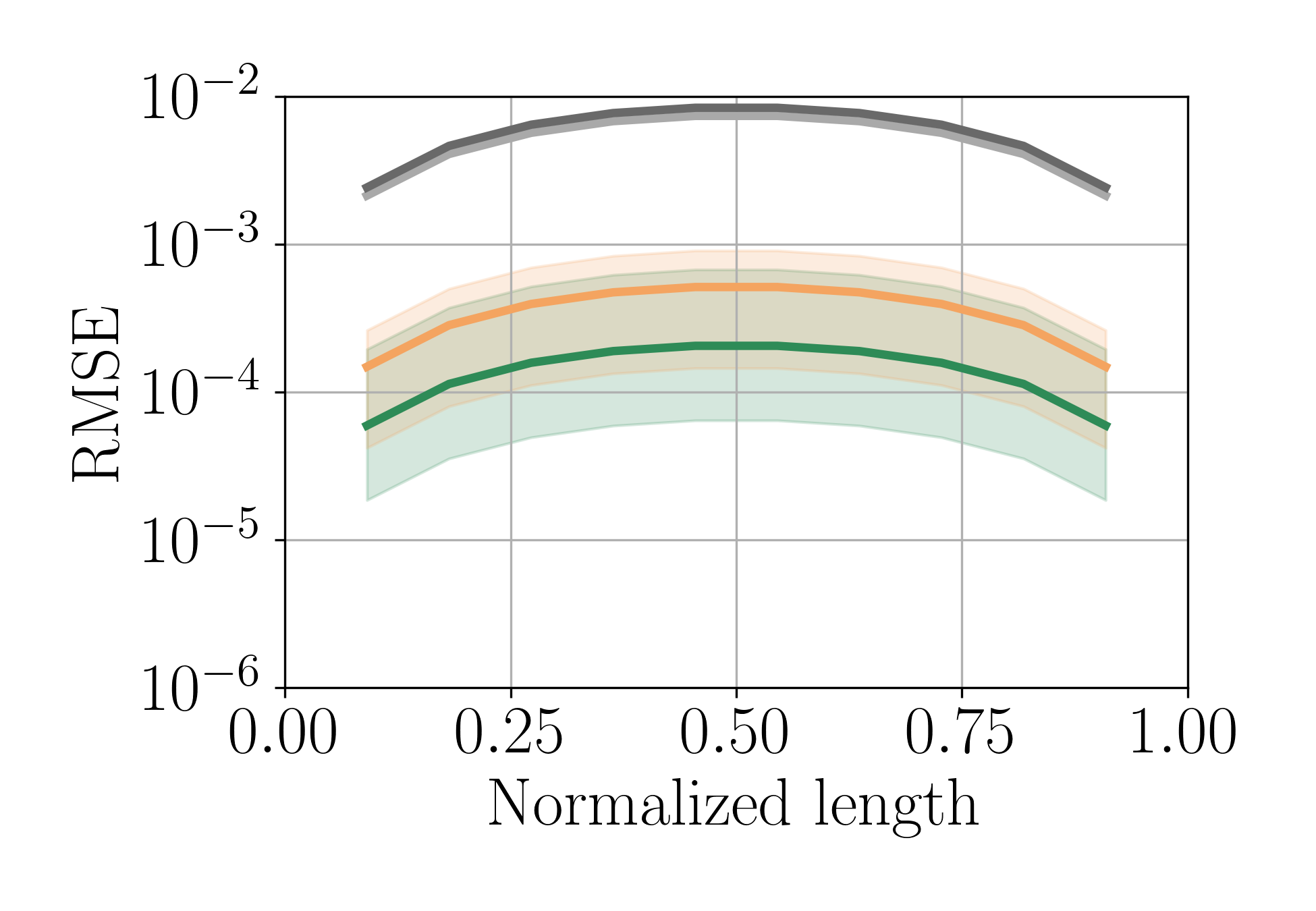}
		\caption{$M=10$, $Q=50$.}
	\end{subfigure}
	\begin{subfigure}[b]{0.328\textwidth}
		\centering
		\includegraphics[width=\textwidth]{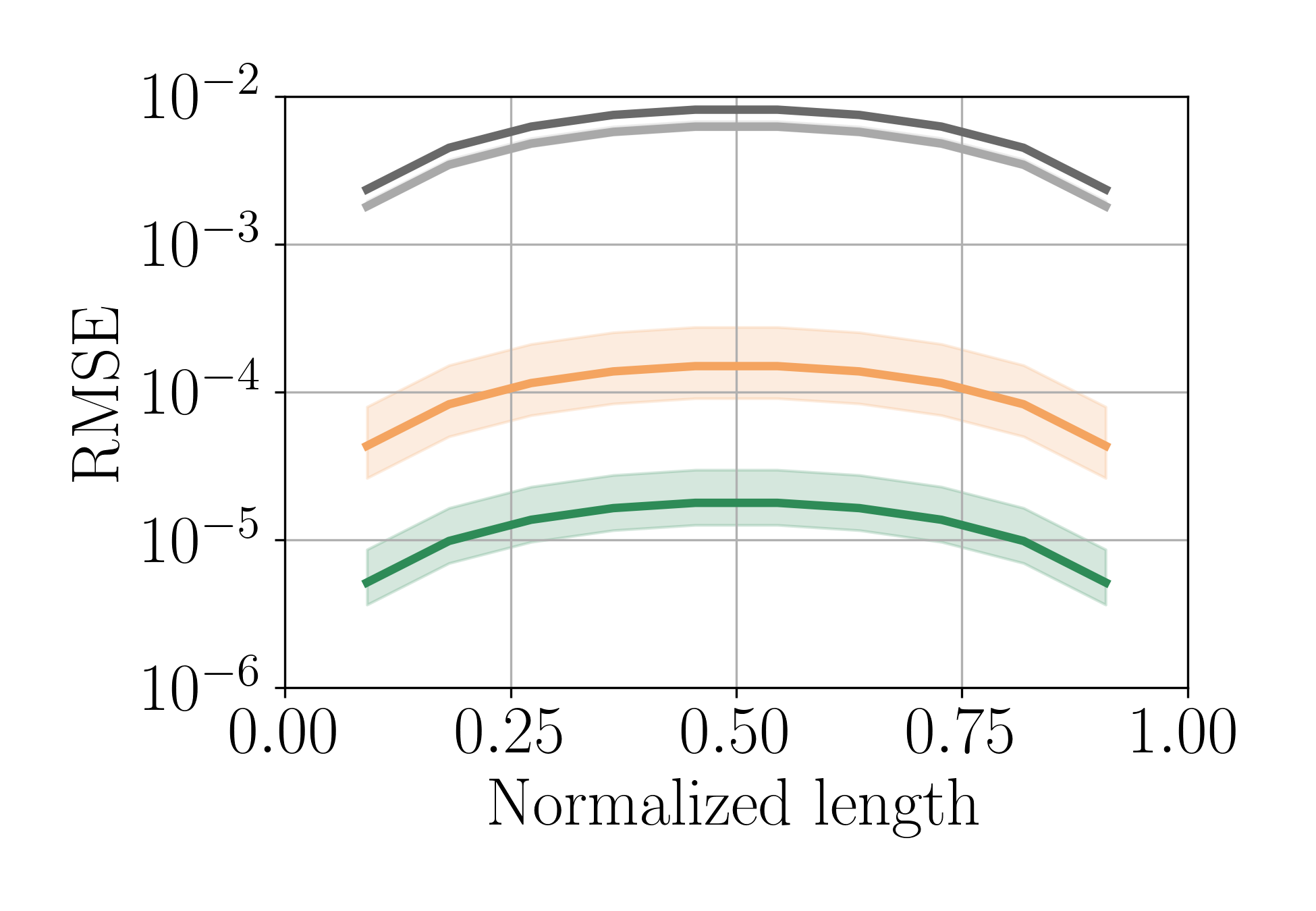}
		\caption{$M=10$, $Q=100$.}
	\end{subfigure}
	\begin{subfigure}[b]{0.328\textwidth}
		\centering
		\includegraphics[width=\textwidth]{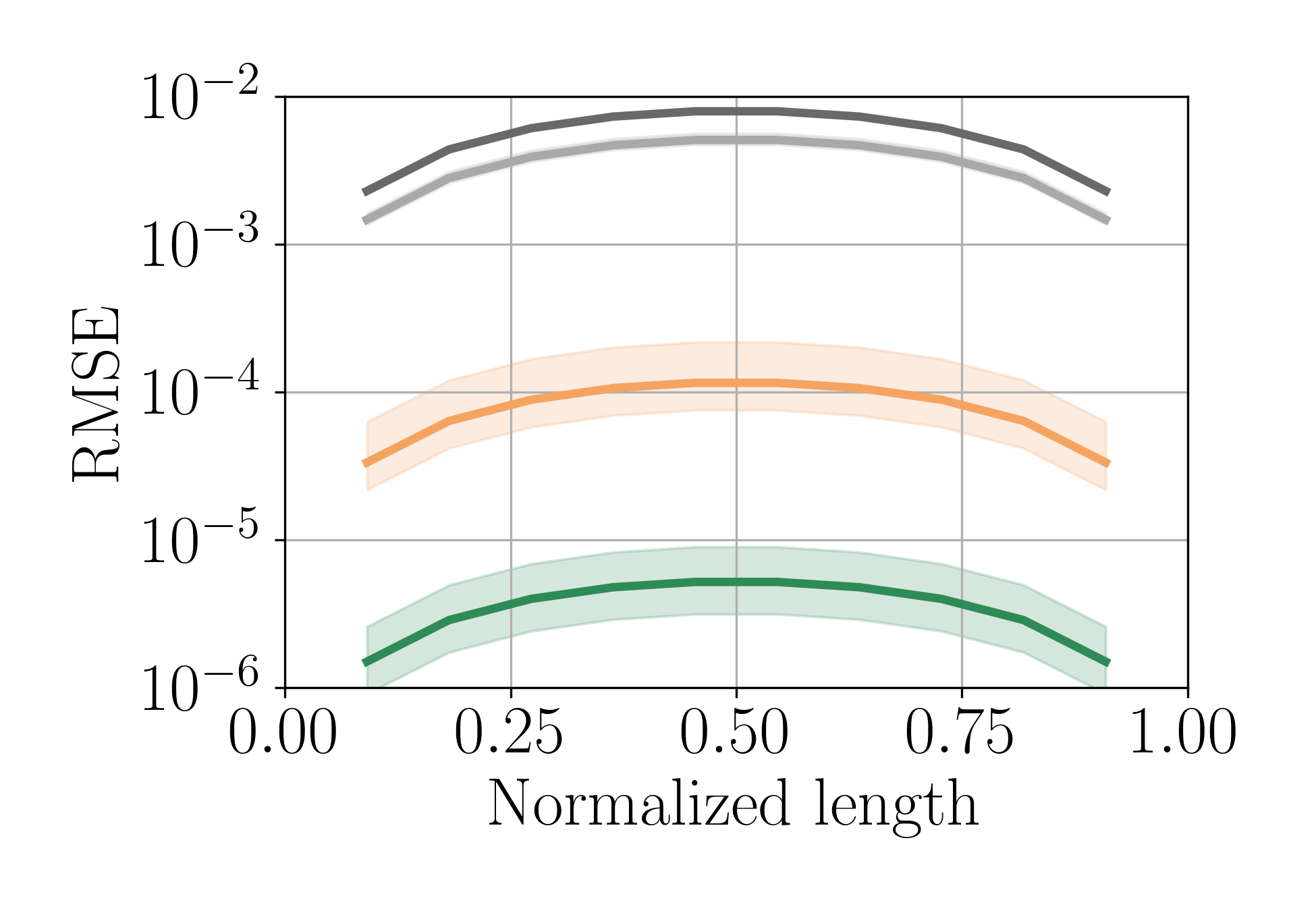}
		\caption{$M=10$, $Q=150$.}
	\end{subfigure}
	\\
	\begin{subfigure}[b]{0.328\textwidth}
		\centering
		\includegraphics[width=\textwidth]{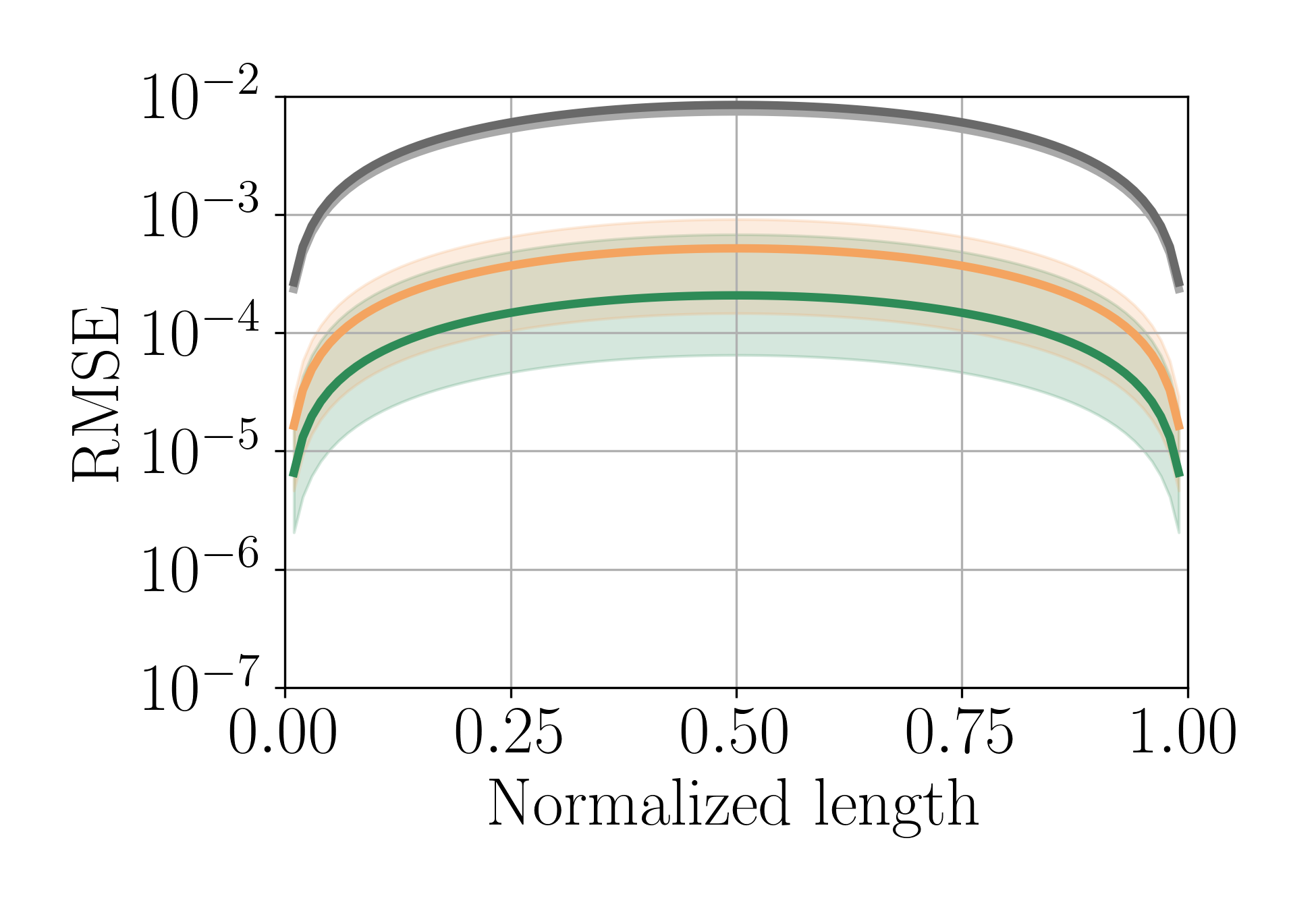}
		\caption{$M=100$, $Q=50$.}
	\end{subfigure}
	\begin{subfigure}[b]{0.328\textwidth}
		\centering
		\includegraphics[width=\textwidth]{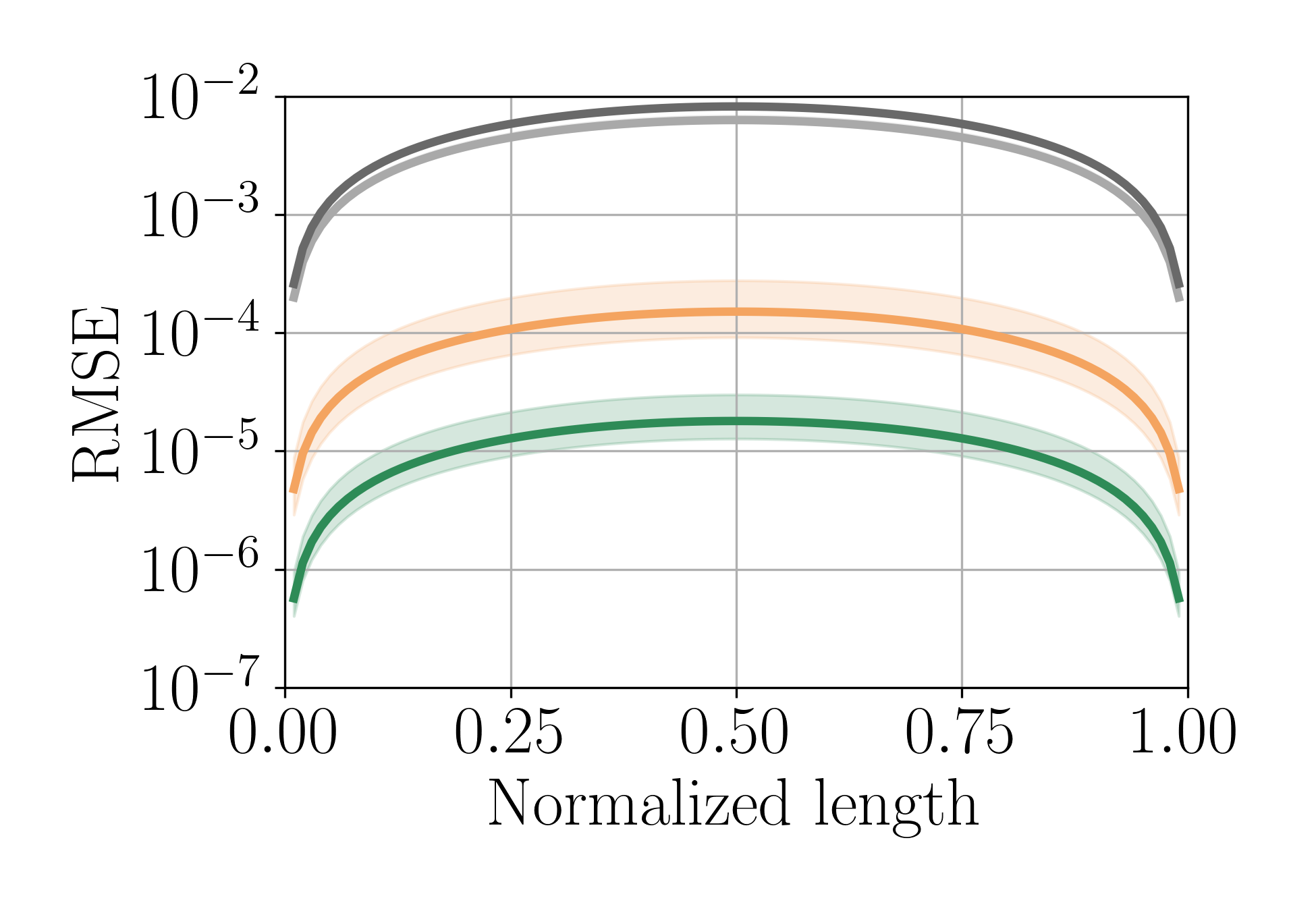}
		\caption{$M=100$, $Q=100$.}
	\end{subfigure}
	\begin{subfigure}[b]{0.328\textwidth}
		\centering
		\includegraphics[width=\textwidth]{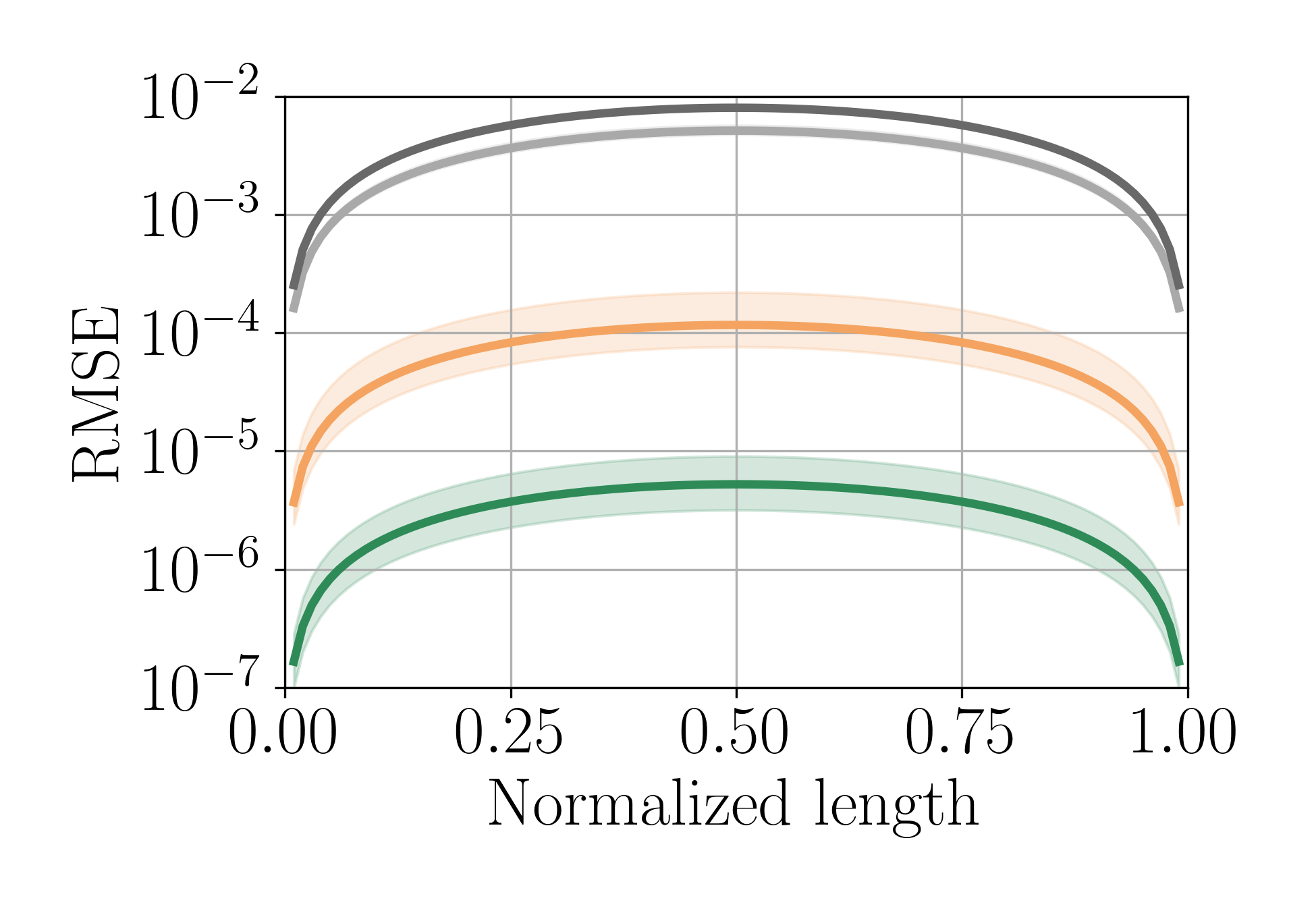}
		\caption{$M=100$, $Q=150$.}
	\end{subfigure}
	\\
	\begin{subfigure}[b]{0.328\textwidth}
		\centering
		\includegraphics[width=\textwidth]{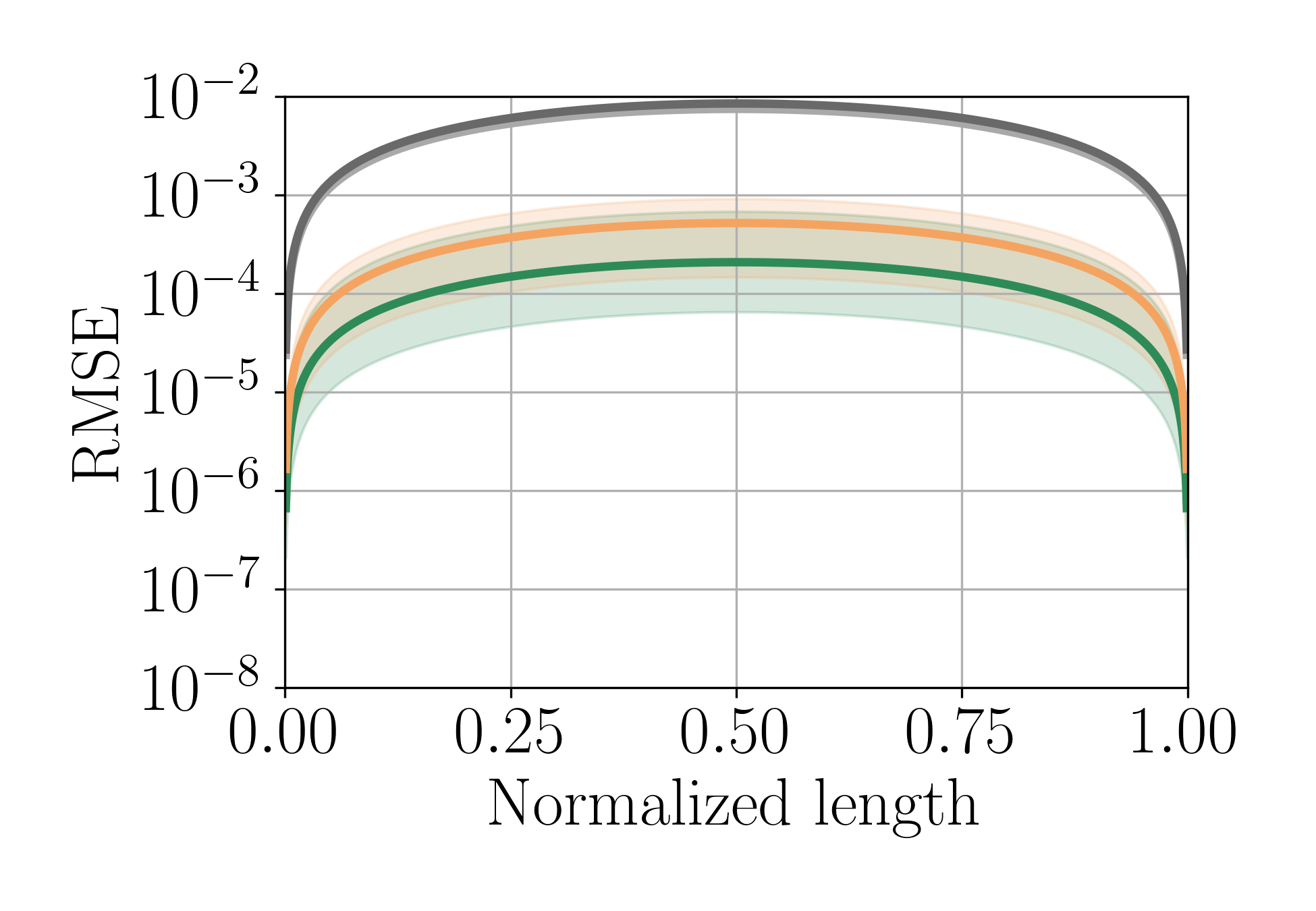}
		\caption{$M=1000$, $Q=50$.}
	\end{subfigure}
	\begin{subfigure}[b]{0.328\textwidth}
		\centering
		\includegraphics[width=\textwidth]{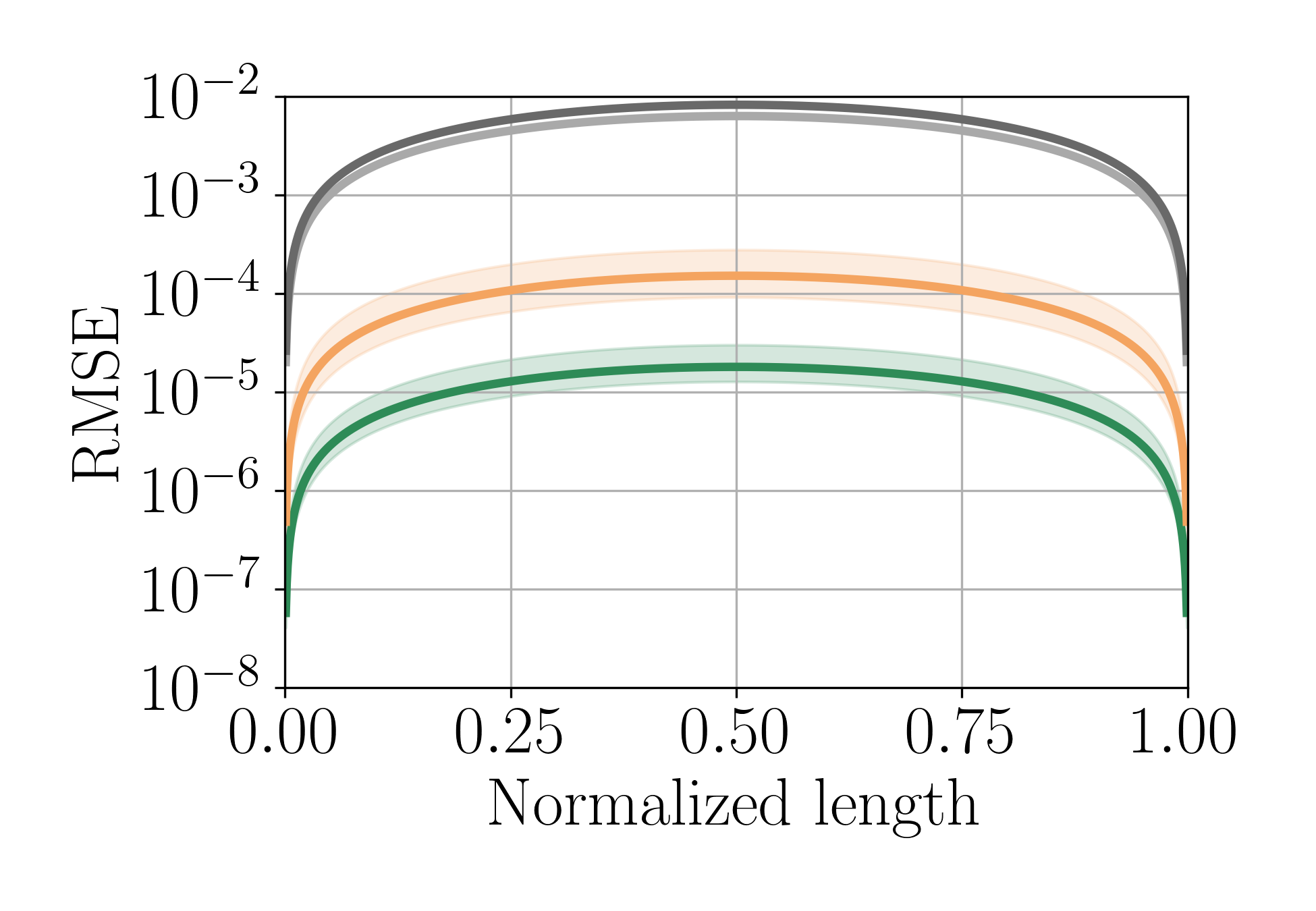}
		\caption{$M=1000$, $Q=100$.}
	\end{subfigure}
	\begin{subfigure}[b]{0.328\textwidth}
		\centering
		\includegraphics[width=\textwidth]{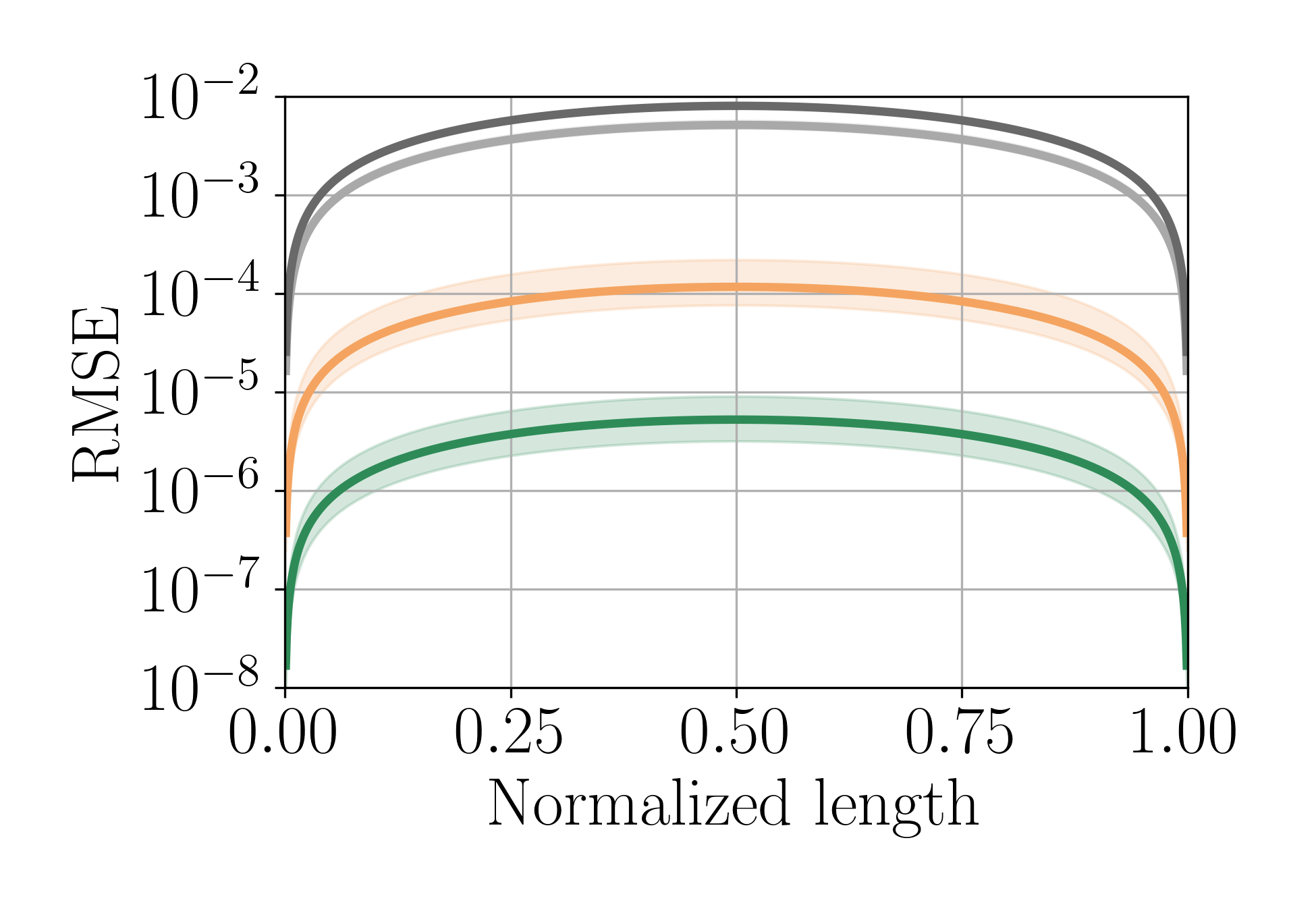}
		\caption{$M=1000$, $Q=150$.}
	\end{subfigure}
	\caption{Simply supported beam: \Gls{rmse} over model response of the different \glspl{pce} for response dimension $M \in \left\{10, 100, 1000\right\}$ and training data set size $Q \in \left\{50, 100, 150\right\}$. The solid lines show average error values over $10$ different training and test data sets. The shaded areas represent the difference between minimum and maximum errors over these $10$ data sets.}
	\label{fig:beam_vector_rmse}
\end{figure}

Figure~\ref{fig:beam_vector_rmse} shows the \gls{rmse} over the beam's length for all \gls{pce} surrogate models and combinations of response dimension $M$ and training data set size $Q$. 
As is obvious, the resolution of the response and, accordingly, of the vector-valued \gls{rmse} depends on the response dimension $M$.
For all \glspl{pce}, the \gls{rmse} is minimum near the points of support and maximum at the middle of the beam's length, as would be expected from the deflection variations observed in Figure~\ref{fig:beam_deflection}.
The \gls{td} \glspl{pce} result in comparatively low approximation accuracy, with little to no improvement as the size of the training data set increases.
This is attributed to the high-dimensional input random vector ($N=20$ inputs), despite the fact that the $15$ dummy parameters have no impact on the response at all.
The $p/q$-adaptive \gls{lar} and \gls{mvsa} \glspl{pce} result in much more accurate surrogate models, which also improve significantly when more training data become available, exactly due to their ability to discern important from negligible input parameter contributions to the approximation.
In this test case, the \gls{mvsa} \gls{pce} is more accurate than the $p/q$-adaptive \gls{lar} and the \gls{td} \glspl{pce}  
by one and three orders of magnitude, respectively, for the same training data set size. 
Note that these observations regarding surrogate model accuracy remain consistent for all three response dimensions $M \in \left\{10, 100, 1000\right\}$.

\begin{figure}[t!]
	\centering
	\begin{subfigure}[b]{0.75\textwidth}
		\centering
		\fbox{\includegraphics[width=1\textwidth]{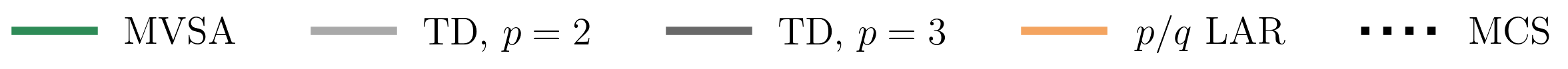}}
	\end{subfigure}
	\\
	\begin{subfigure}[b]{0.328\textwidth}
		\centering
		\includegraphics[width=\textwidth]{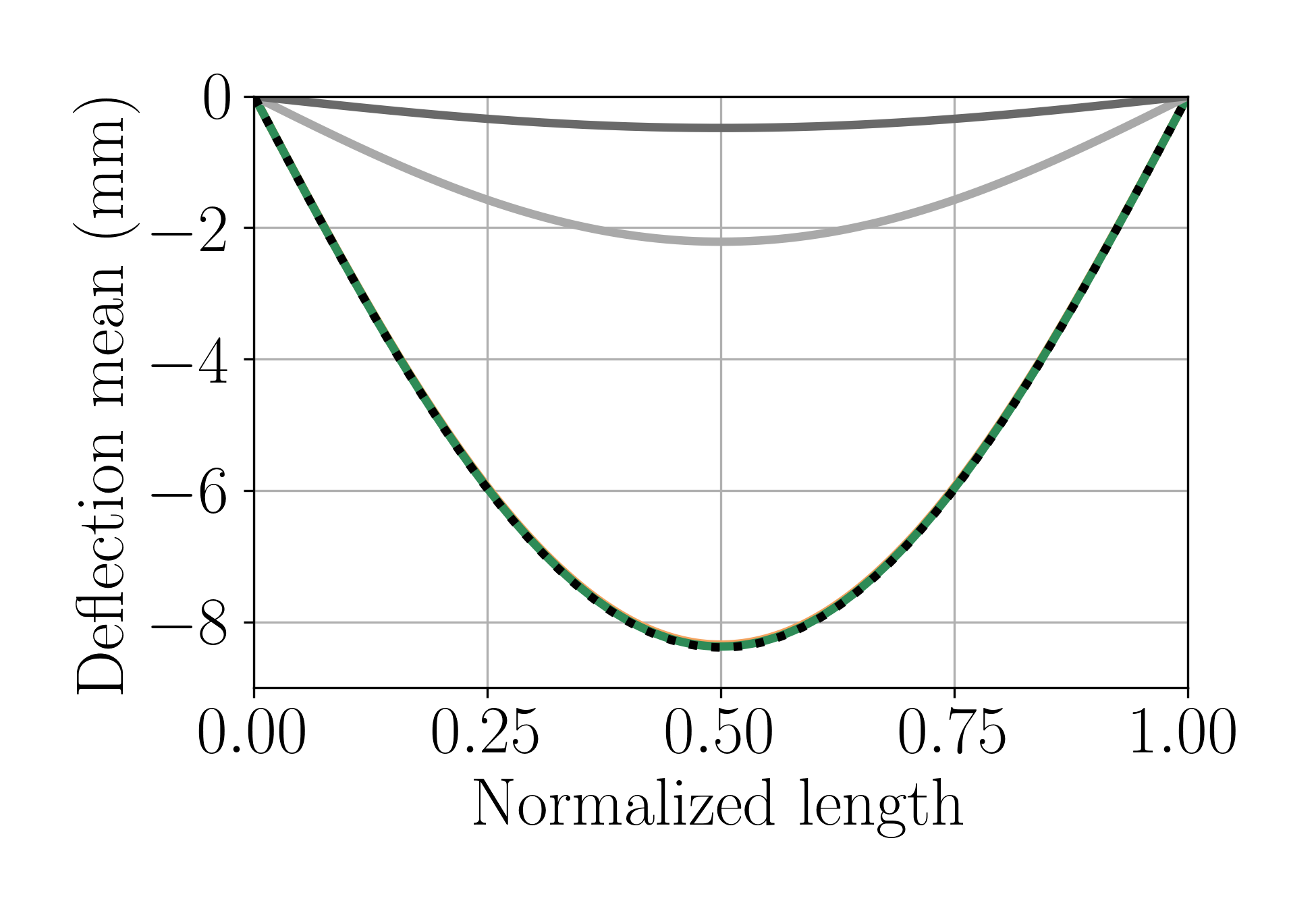}
		\caption{Mean, $Q=50$.}
	\end{subfigure}
	\begin{subfigure}[b]{0.328\textwidth}
		\centering
		\includegraphics[width=\textwidth]{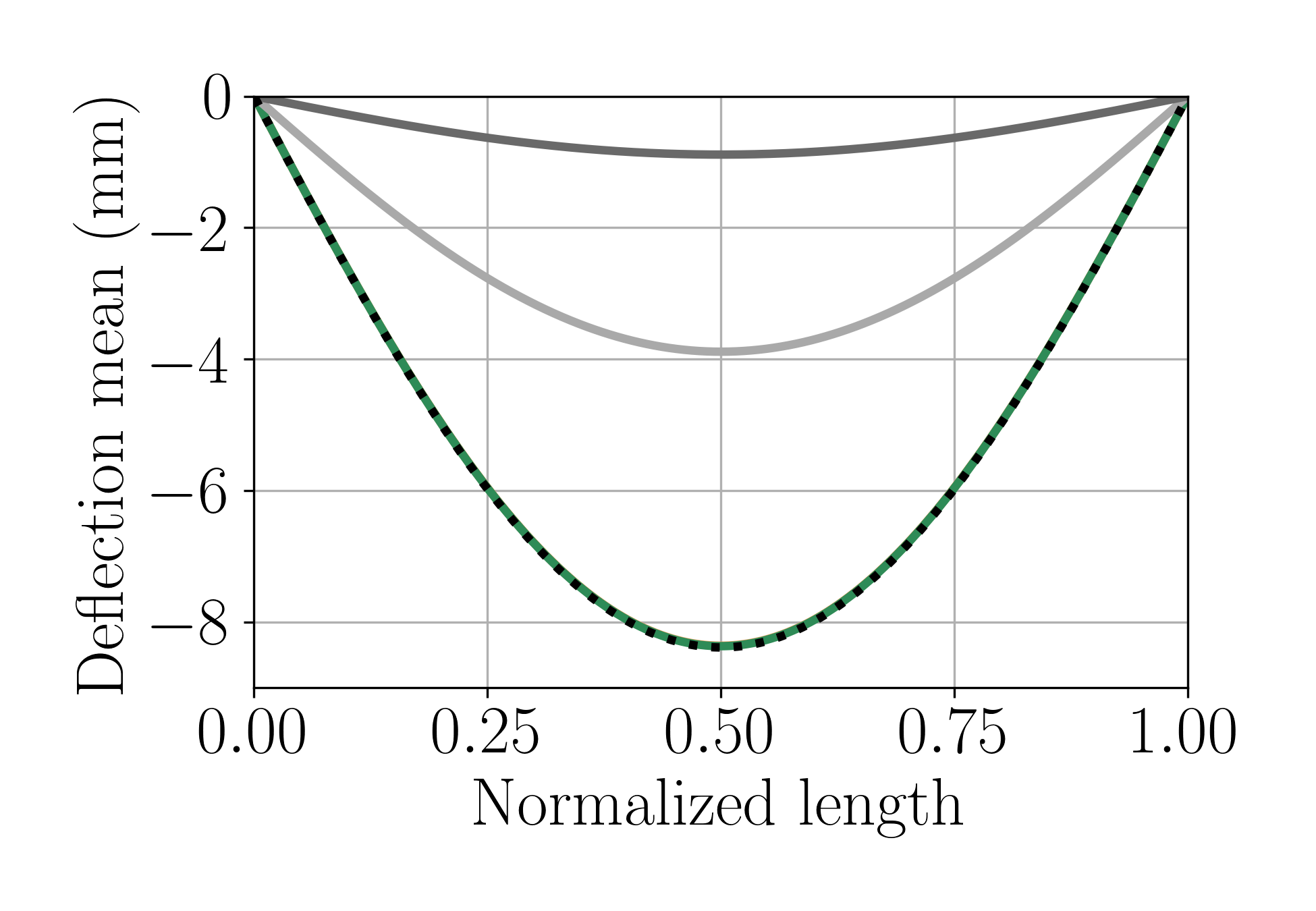}
		\caption{Mean, $Q=100$.}
	\end{subfigure}
	\begin{subfigure}[b]{0.328\textwidth}
		\centering
		\includegraphics[width=\textwidth]{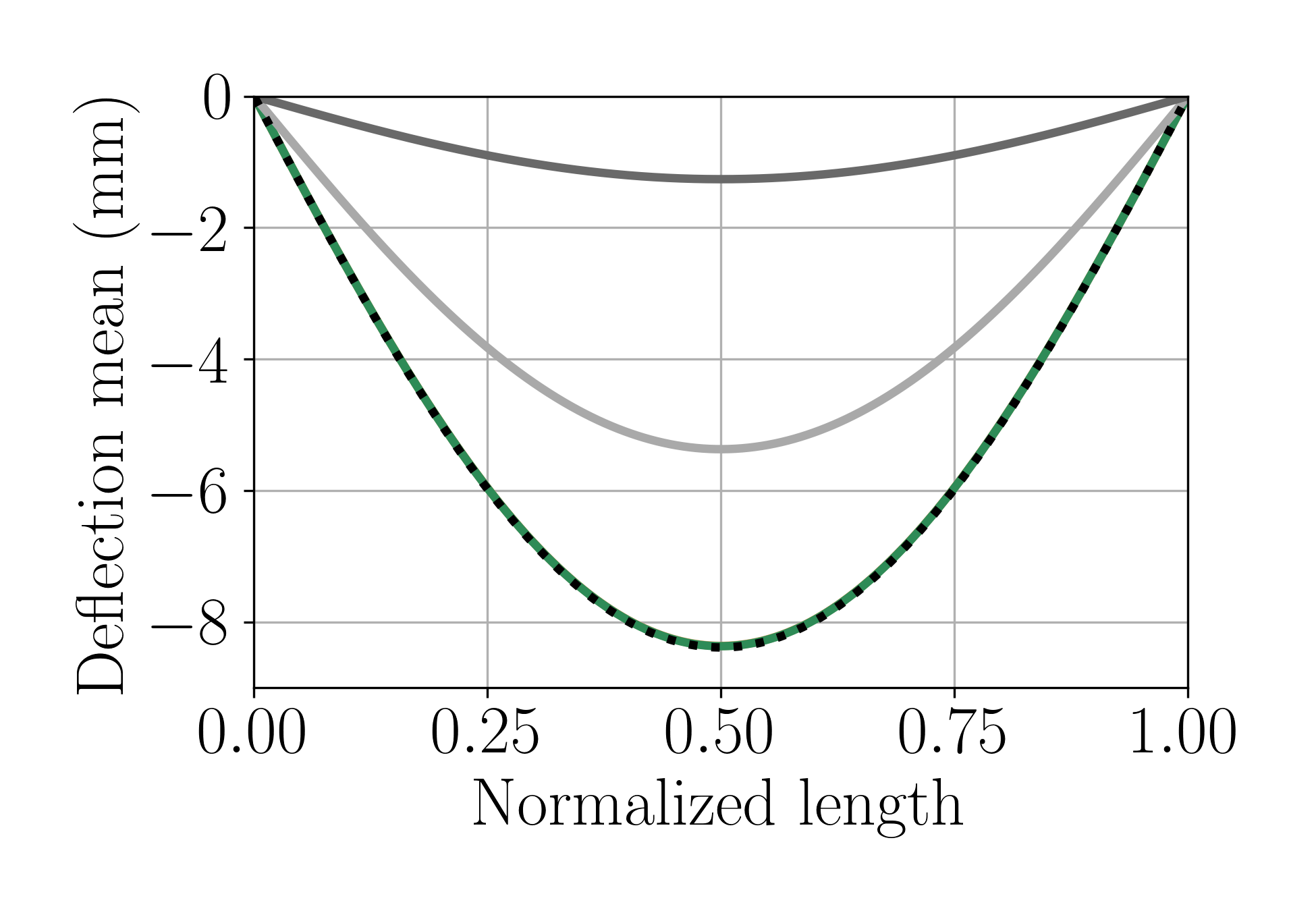}
		\caption{Mean, $Q=150$.}
	\end{subfigure}
	\\
	\begin{subfigure}[b]{0.328\textwidth}
		\centering
		\includegraphics[width=\textwidth]{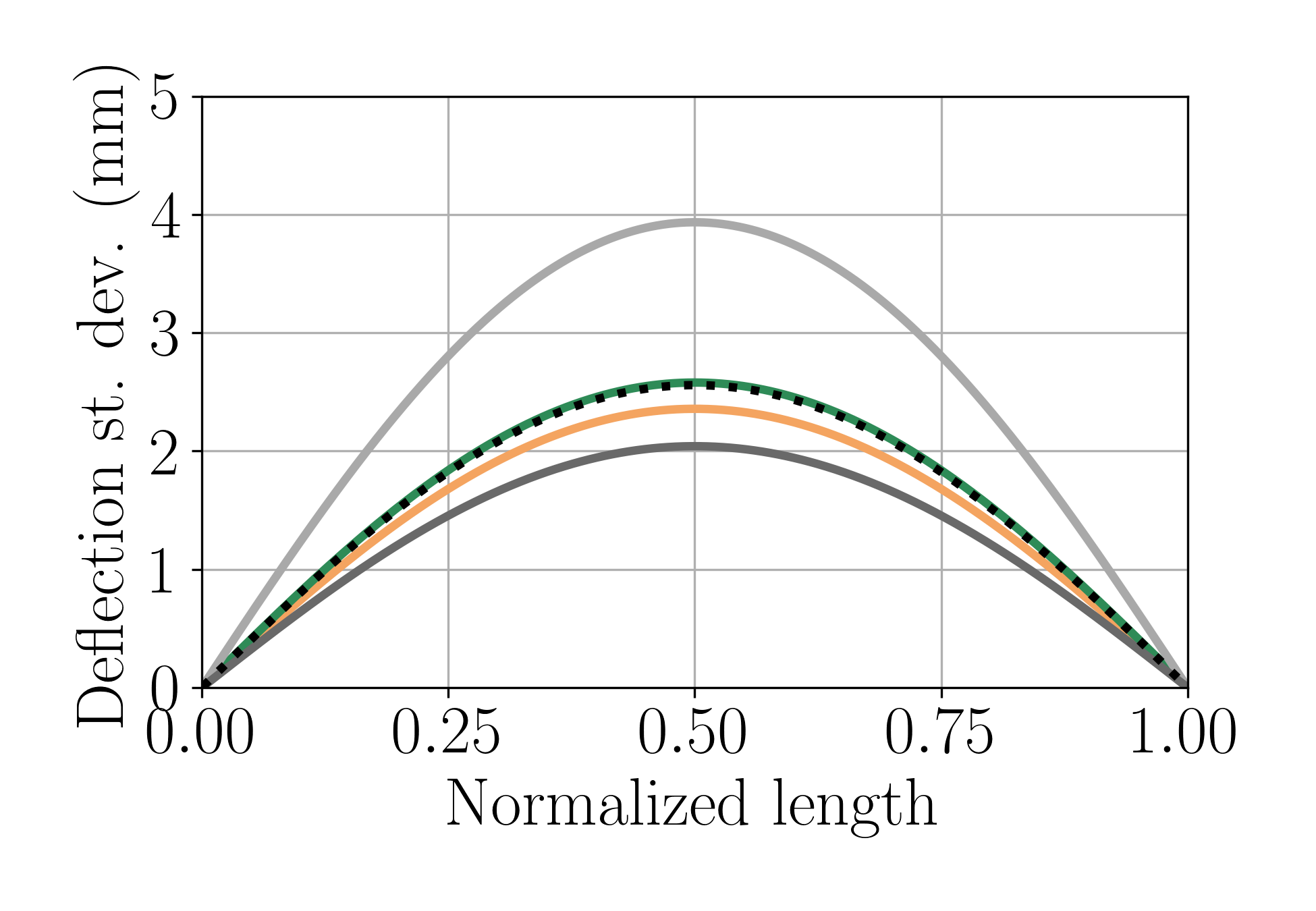}
		\caption{Standard deviation, $Q=50$.}
	\end{subfigure}
	\begin{subfigure}[b]{0.328\textwidth}
		\centering
		\includegraphics[width=\textwidth]{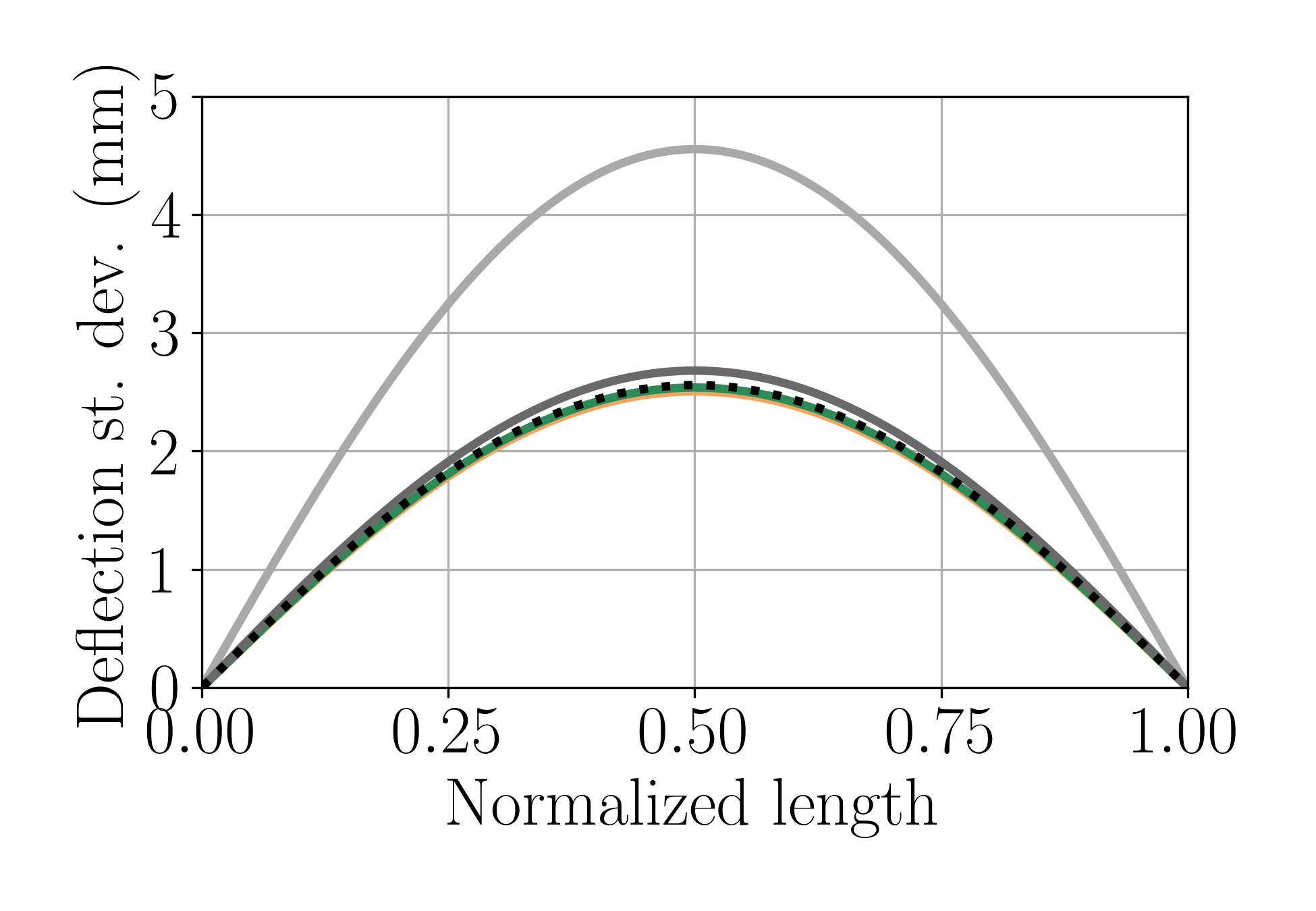}
		\caption{Standard deviation, $Q=100$.}
	\end{subfigure}
	\begin{subfigure}[b]{0.328\textwidth}
		\centering
		\includegraphics[width=\textwidth]{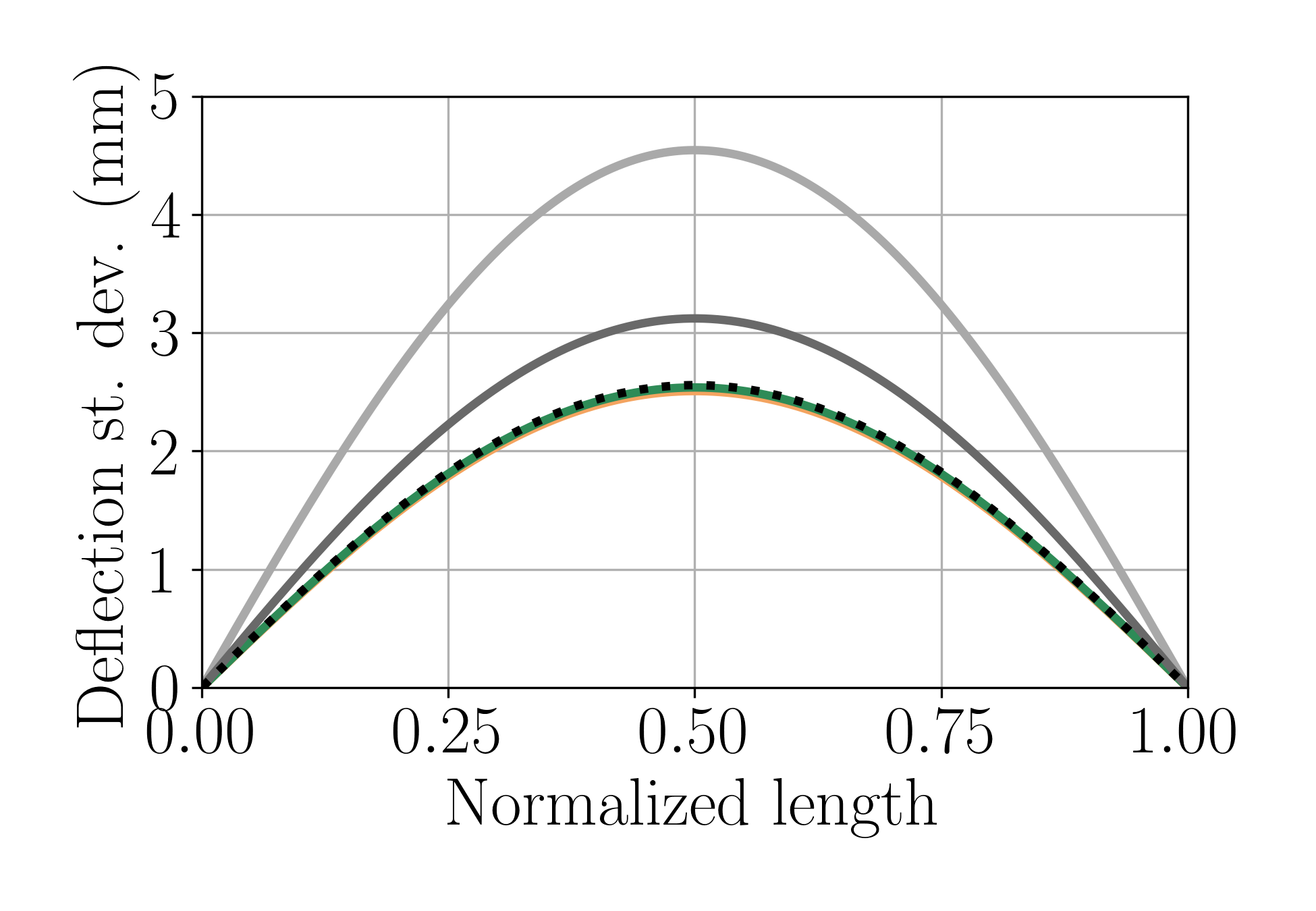}
		\caption{Standard deviation, $Q=150$.}
	\end{subfigure}
	\caption{Simply supported beam: Mean and standard deviation of the model's response with dimension $M=1000$, estimated with the different \glspl{pce} for training data set size $Q \in \left\{50, 100, 150\right\}$. The reference mean and standard deviation are computed via \gls{mcs} with $10^5$ random samples.}
	\label{fig:beam_moments}
\end{figure}

Figure~\ref{fig:beam_moments} shows mean and standard deviation estimates obtained with each \gls{pce} method, along with \gls{mcs}-based references computed with $10^5$ random samples.
In this case, only the response dimension $M=1000$ is considered, such that the resolution of the response is the maximum one.
Looking at the mean deflection, the estimates given by the \gls{mvsa} and $p/q$-adaptive \gls{lar} \glspl{pce} are indistinguishable from one another, with both methods needing only $Q=50$ training data points for an estimate identical to the \gls{mcs} reference.
In the case of the standard deviation, the \gls{mvsa} \gls{pce} is again as accurate as the \gls{mcs} reference for only $Q=50$ training data points, while the $p/q$-adaptive \gls{lar} \gls{pce} is obviously less accurate. 
For $Q=100$ and $Q=150$, both methods yield standard deviation estimates identical to the reference.
The mean and standard deviation estimates of the \gls{td} \glspl{pce} are mostly very inaccurate, as expected due to the curse of (input) dimensionality.

\begin{figure}[t!]
	\centering
	\begin{subfigure}[b]{0.65\textwidth}
		\centering
		\fbox{\includegraphics[width=1\textwidth]{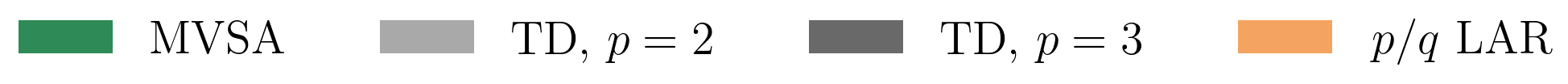}}
	\end{subfigure}
	\\
	\begin{subfigure}[b]{0.328\textwidth}
		\centering
		\includegraphics[width=\textwidth]{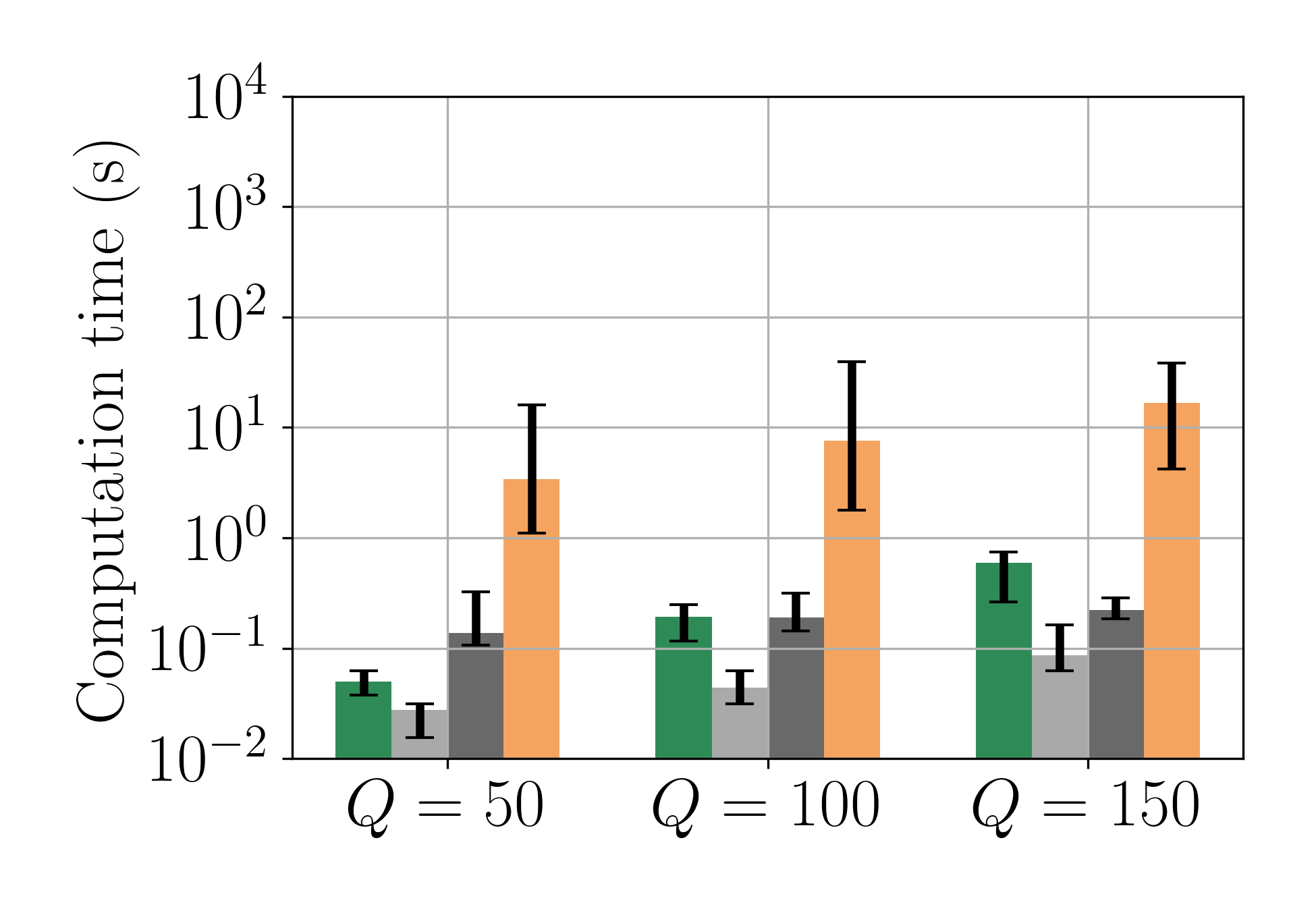}
		\caption{$M=10$.}
	\end{subfigure}
	\begin{subfigure}[b]{0.328\textwidth}
		\centering
		\includegraphics[width=\textwidth]{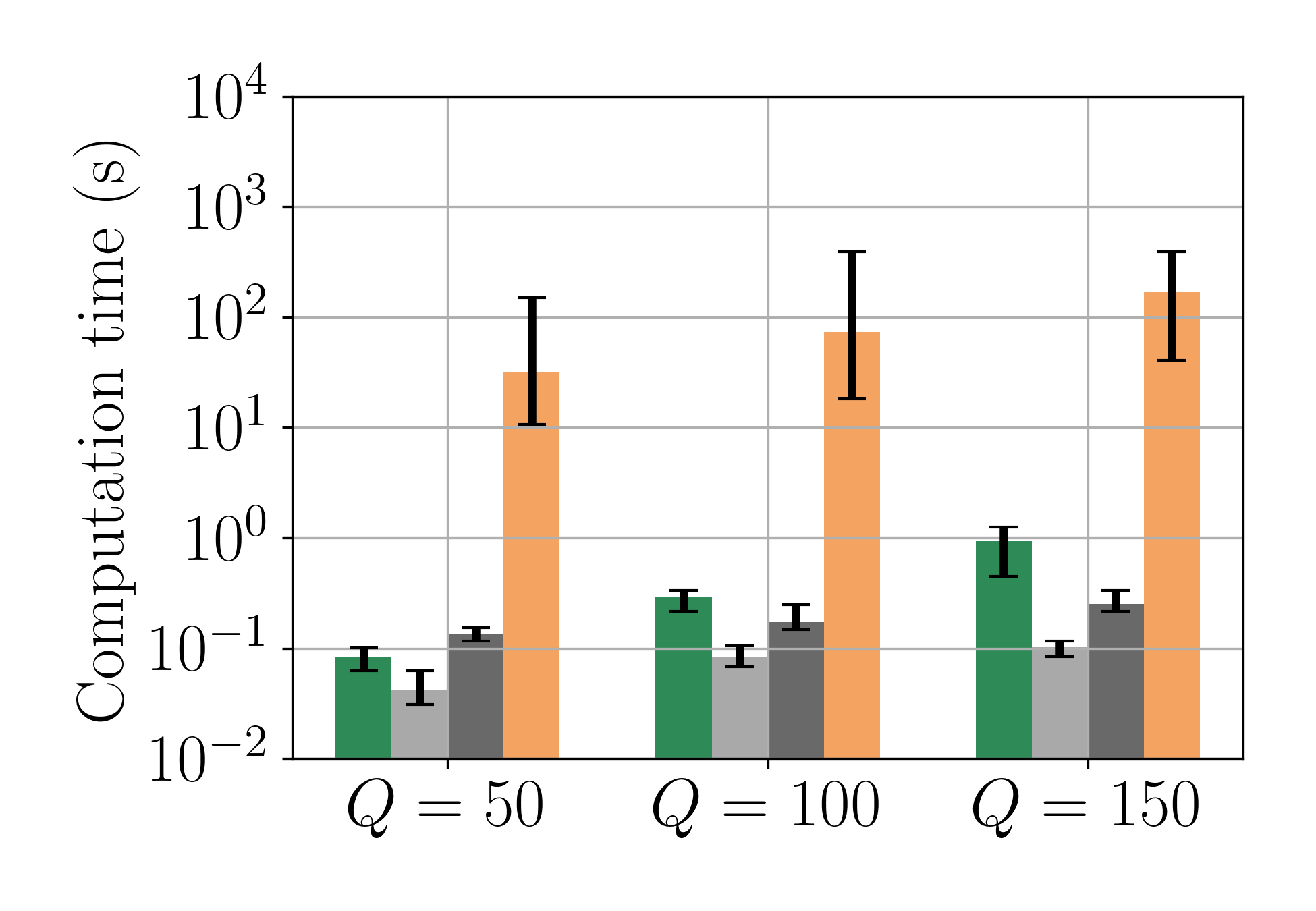}
		\caption{$M=100$.}
	\end{subfigure}
	\begin{subfigure}[b]{0.328\textwidth}
		\centering
		\includegraphics[width=\textwidth]{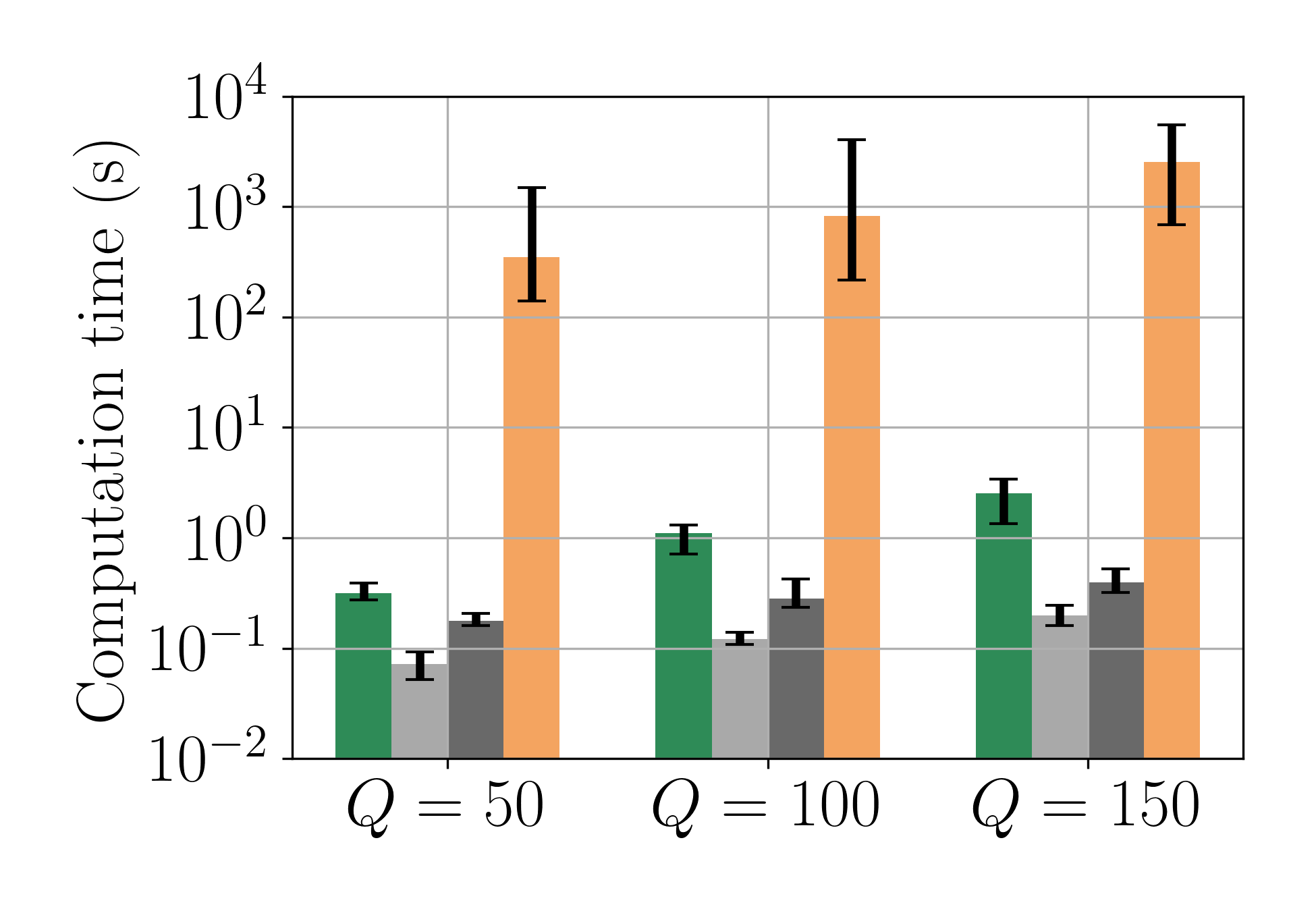}
		\caption{$M=1000$.}
	\end{subfigure}
	\caption{Simply supported beam: Computation time of the different \glspl{pce} for response dimension $M \in \left\{10 ,100, 1000\right\}$ and training data set size $Q \in \left\{50 ,100, 150\right\}$. The colored bars show the average computation time over $10$ different training data sets. The black error bars show the difference between minimum and maximum computation time over these $10$ data sets.}
	\label{fig:beam_computation_time}
\end{figure}

Figure~\ref{fig:beam_computation_time} shows the time needed to compute each \gls{pce} for all combinations of response dimension $M$ and training data set size $Q$. 
This figure illustrates clearly the limitations of applying  sprase/adaptive \glspl{pce} element-wise, particularly if the model's response is high-dimensional.
As can be observed, the $p/q$-adaptive \gls{lar} \gls{pce} results in much longer computation times than the other \gls{pce} methods, reaching more than $40$ minutes (on average) for $M=1000$ and $Q=150$.
Contrarily, the \gls{mvsa} \gls{pce} needs at most $3.5$ seconds to be computed, while also resulting in a higher surrogate modeling and uncertainty estimation accuracy in this test case.
The \gls{td} \glspl{pce} are computed even faster due to their fixed polynomial bases, however, this comes at no benefit due to their comparatively very low accuracy.

\begin{table}[t!]
	\small
	\caption{Simply supported beam: Maximum total and univariate degrees of the \gls{mvsa} and the $p/q$-adaptive \gls{lar} \glspl{pce}.}
	\centering
	\begin{tabular}{c c c c c}
		\toprule
		\multirow{2}{6em}{Training data set size $Q$}  & \multicolumn{2}{c}{Maximum total degree} & \multicolumn{2}{c}{Maximum univariate degree}  \\[1ex] 
		\cmidrule(lr){2-3}
		\cmidrule(lr){4-5}
		&  \gls{mvsa} & $p/q$ \gls{lar}  & \gls{mvsa}  & $p/q$ \gls{lar} \\[1em]
		\toprule
		50  & 3 & 3 & 3 & 3\\[0.5ex]
		100 & 4 & 3 & 4 & 3\\[0.5ex]
		150 & 5 & 3 & 4 & 3\\[0.5ex]
		\bottomrule
	\end{tabular}
	\label{tab:beam-max-degrees}
\end{table}

Last, Table~\ref{tab:beam-max-degrees} shows the maximum total degree, i.e., $\max_{\mathbf{k} \in \Lambda}\left|\mathbf{k}\right|_1$, and the maximum univariate degree of the polynomial bases resulting from the \gls{mvsa} and the $p/q$-adaptive \gls{lar} \gls{pce} methods. 
Note that in the case of the  $p/q$-adaptive \gls{lar} \gls{pce}, the maximum degrees over all polynomial bases (i.e., one basis per output) are presented. 
Interestingly, the \gls{mvsa} \gls{pce} results in higher total and univariate maximum degrees compared to the $p/q$-adaptive \gls{lar} \gls{pce}, which could to some extent explain its better performance. 
However, this is not a general phenomenon, as we will see in the next section.

\subsection{Start-up torque of induction motor}
\label{sec:e-machine}

\begin{figure}[t!]
	\centering
	\begin{subfigure}[t]{0.49\textwidth}
		\centering
		\includegraphics[scale=0.75]{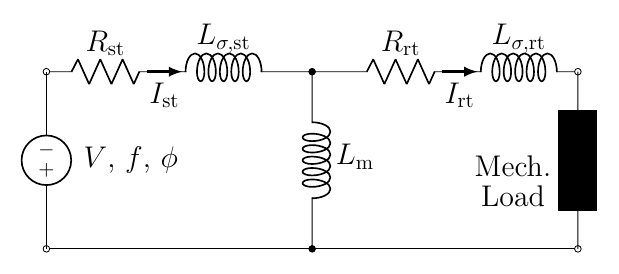}
		\caption{Equivalent circuit of induction motor.}
		\label{fig:scim_sketch}
	\end{subfigure}
	\hfill
	\begin{subfigure}[t]{0.49\textwidth}
		\centering
		\includegraphics[width=1\textwidth]{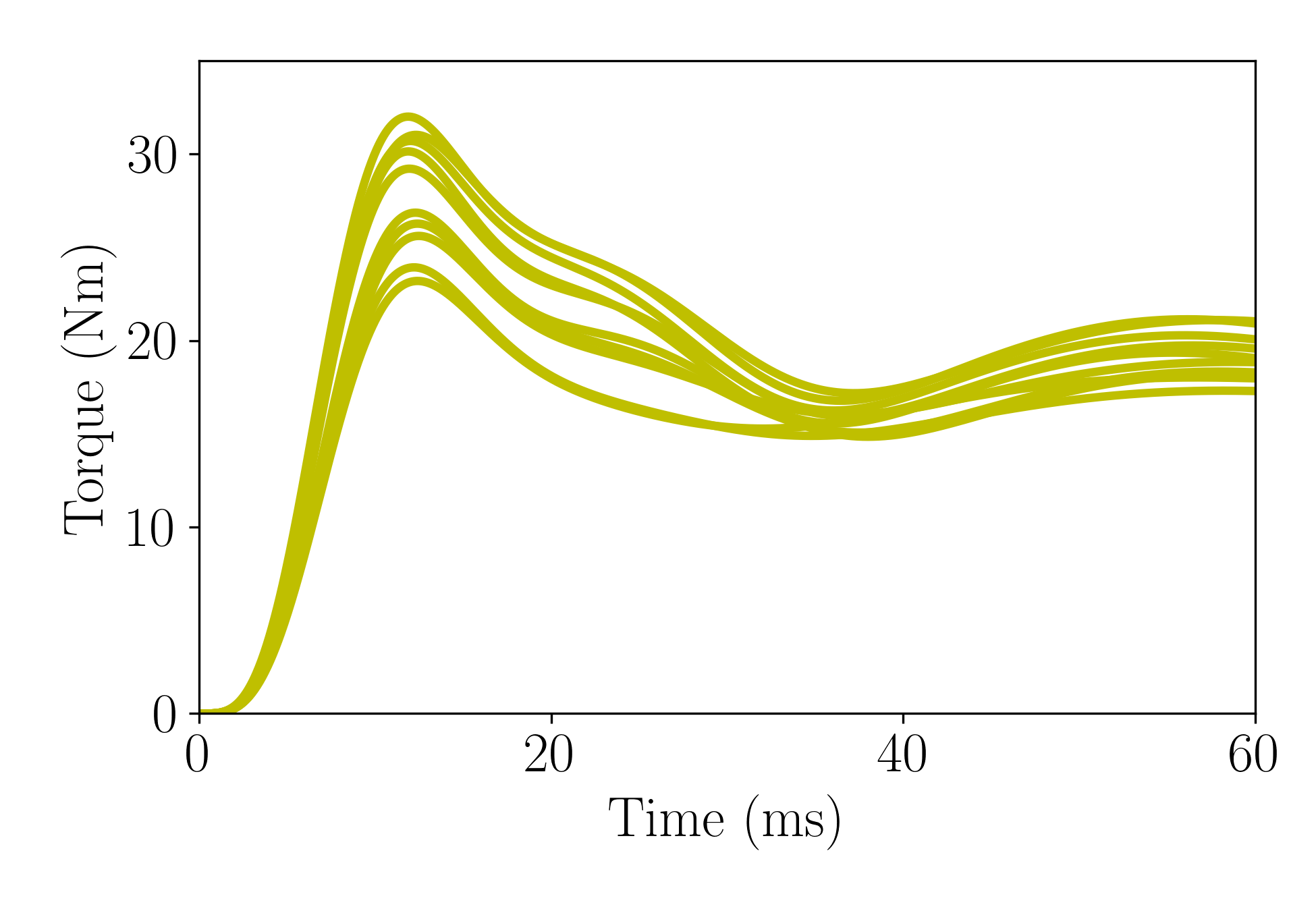}
		\caption{Electromagnetic torque during motor start-up.}
		\label{fig:scim_torque}
	\end{subfigure}
	\caption{(a) Equivalent circuit model of an induction motor. (b) Electromagnetic torque generated during the start-up of the induction motor for $10$ realizations of the model parameters.}
	\label{fig:scim_illustration}
\end{figure}

As second test case, we consider an induction motor connected to a power grid and to a mechanical load.
The power grid has phase-voltage amplitude $V=210$~V, frequency $f=50$~Hz, and initial phase $\phi = 0 \degree$.
The motor consists of a stator, where a winding is organized in three phases and $n_{\mathrm{pp}}=2$ pole pairs, and a rotor that rotates at the mechanical speed $\omega_{\mathrm{me}}$. 
The torque of the mechanical load is given by
\begin{equation}
	T_{\mathrm{ld}} \left(\omega_{\mathrm{me}}\right) = \mathrm{sign}\left(\omega_{\mathrm{me}}\right) \left(\gamma \omega_{\mathrm{me}}^2 + \beta \left| \omega_{\mathrm{me}} \right| + \alpha \right),
\end{equation}
where the coefficients $\alpha$, $\beta$, and $\gamma$ are used for modeling static, sliding, and air friction, respectively. 
The mechanical equation of motion then reads
\begin{equation}
	\frac{\mathrm{d}\omega_{\mathrm{me}}}{\mathrm{d}t} = \frac{T_{\mathrm{em}} - T_{\mathrm{ld}} \left(\omega_{\mathrm{me}}\right)}{J_{\mathrm{rt}} + J_{\mathrm{ld}}},
	\label{eq:eq-motion}
\end{equation}
where $J_{\mathrm{rt}}$ and $J_{\mathrm{ld}}$ denote the moment of inertia of the rotor and the load, respectively, and $T_{\mathrm{em}}$ is the electromagnetic torque. 
Using the per-phase equivalent circuit model of the motor depicted in Figure~\ref{fig:scim_sketch}, the electromagnetic torque is given by
\begin{equation} 
	T_{\mathrm{em}} = \frac{3 R_{\mathrm{rt}} \left| I_{\mathrm{rt}}\right|^2}{\omega_{\mathrm{syn}} - \omega_{\mathrm{me}}},
\end{equation}
where $\omega_{\mathrm{syn}} = 2 \pi f / n_{\mathrm{pp}}$ is the synchronous speed of the electromagnetic field generated by the stator winding within the air gap of the motor. 
The rotor resistance $R_{\mathrm{rt}}$ is assumed to be known, while the rotor current $I_{\mathrm{rt}}$ can be derived from the equivalent circuit's voltage and current equations \cite{pyrhonen2013design}.
To simulate the start-up phase of the induction machine, where the rotor is initially at standstill ($\omega_{\mathrm{me}}=0$), the mechanical equation of motion \eqref{eq:eq-motion} is coupled to the electrical ordinary differential equation
\begin{equation}
	v(t) = R i(t) + L\frac{\mathrm{d}i(t)}{\mathrm{d}t},
\end{equation}
which is derived separately for the three phases in the stator and the rotor, thus resulting in a system of 6 equations \cite{pyrhonen2013design}.

\begin{table}[b!]
	\small
	\caption{Parameters of the induction motor model. Grid parameters vary uniformly, namely, the phase-voltage amplitude $V$ within $\left[85\%, 100\%\right]$ of its nominal value, the frequency $f$ within $\left[49, 51\right]$ Hz, and the initial phase $\phi$ within $\left[0, 15\right]$ degrees. The remaining parameters follow normal distributions with their nominal value as mean and  $5\%$ of the nominal value as standard deviation.}
	\centering
	\begin{tabular}{l c c c c c}
		\toprule
		Parameter & Units & Notation & Nominal value \\[0.5ex] 
		\toprule
		Stator resistance & $\Omega$ &$R_{\mathrm{st}}$ &  $2.9338$\\[0.5ex]
		Rotor resistance & $\Omega$ & $R_{\mathrm{rt}}$ & $1.355$ \\[0.5ex]
		Magnetization inductance & H & $L_{\mathrm{m}}$ & $143.75 \cdot 10^{-3}$ \\[0.5ex]
		Stator leakage inductance & H & $L_{\sigma, \mathrm{st}}$ & $5.87 \cdot 10^{-3}$ \\[0.5ex]
		Rotor leakage inductance & H & $L_{\sigma,\mathrm{rt}}$ & $5.87 \cdot 10^{-3}$ \\[0.5ex]
		Moment of inertia of rotor & kg$\,$m$^2$ & $J_\mathrm{rt}$ & $1.1 \cdot 10^{-3}$ \\[0.5ex]
		Moment of inertia of load & kg$\,$m$^2$ & $J_\mathrm{ld}$ & $10^{-3}$ \\[0.5ex] 
		Constant load torque coefficient & N$\,$m & $\alpha$ & $10^{-3}$ \\[0.5ex] 
		Linear load torque coefficient & N$\,$m$\,$s & $\beta$ & $10^{-3}$ \\[0.5ex] 
		Quadratic load torque coefficient & N$\,$m$\,$s$^2$ & $\gamma$ & $10^{-3}$ \\[0.5ex] 
		Phase-voltage amplitude & V & $V$ & $210$ \\[0.5ex] 
		Frequency & Hz & $f$ & $50$ \\[0.5ex] 
		Initial phase & $\degree$ & $\phi$ & $0$ \\[0.5ex] 
		\bottomrule
	\end{tabular}
	\label{tab:scim-params}
\end{table}

The simulation model is implemented using the \texttt{gym-motor} software \cite{balakrishna2021gym} and comprises the $N=13$ input parameters listed in Table~\ref{tab:scim-params}.
The grid parameters, i.e., phase-voltage amplitude, frequency, and initial angle, vary uniformly within given ranges, while
the equivalent circuit and mechanical load parameters follow normal distributions with mean equal to their nominal value and standard deviation equal to $5\%$ of the nominal value, see Table~\ref{tab:scim-params}. 
Figure~\ref{fig:scim_torque} shows the electromagnetic torque of the induction motor for different parameter realizations, where each torque trajectory contains $M=1201$ time steps.

The full response dimension is prohibitive for the element-wise application of the $p/q$-adaptive \gls{lar} \gls{pce}, therefore a dimension reduction of the response by means of the \gls{pod} method is attempted. 
The number of \gls{pod} modes to be retained is computed using the optimal shrinkage of singular values approach of Gavish and Donoho
\cite{gavish2014optimal, gavish2017optimal, falini2022review}, in order to avoid heuristics that often lead to inaccurate approximations, e.g., based on energy or variance preservation considerations (see \cite{jacquelin2019random}).
This results in $M' \ll M$ \gls{pod} modes, where typically $M'$ varies depending on the available training data set.
In this test case, $M' \in \left[48, 51\right]$ for $Q \geq 100$ and $M' = 24$ for $Q=50$.
Contrarily, the \gls{mvsa} and \gls{td} \glspl{pce} are applied to the full model response.

\begin{figure}[t!]
	\centering
	\begin{subfigure}[b]{0.65\textwidth}
		\centering
		\fbox{\includegraphics[width=1\textwidth]{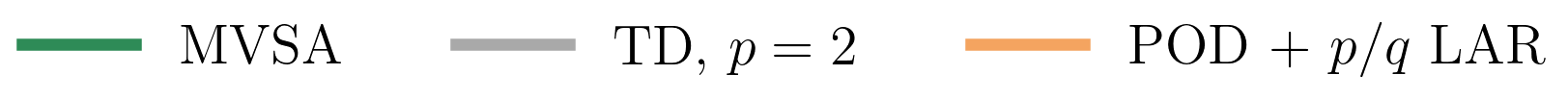}}
	\end{subfigure}
	\\
	\begin{subfigure}[b]{0.328\textwidth}
		\centering
		\includegraphics[width=\textwidth]{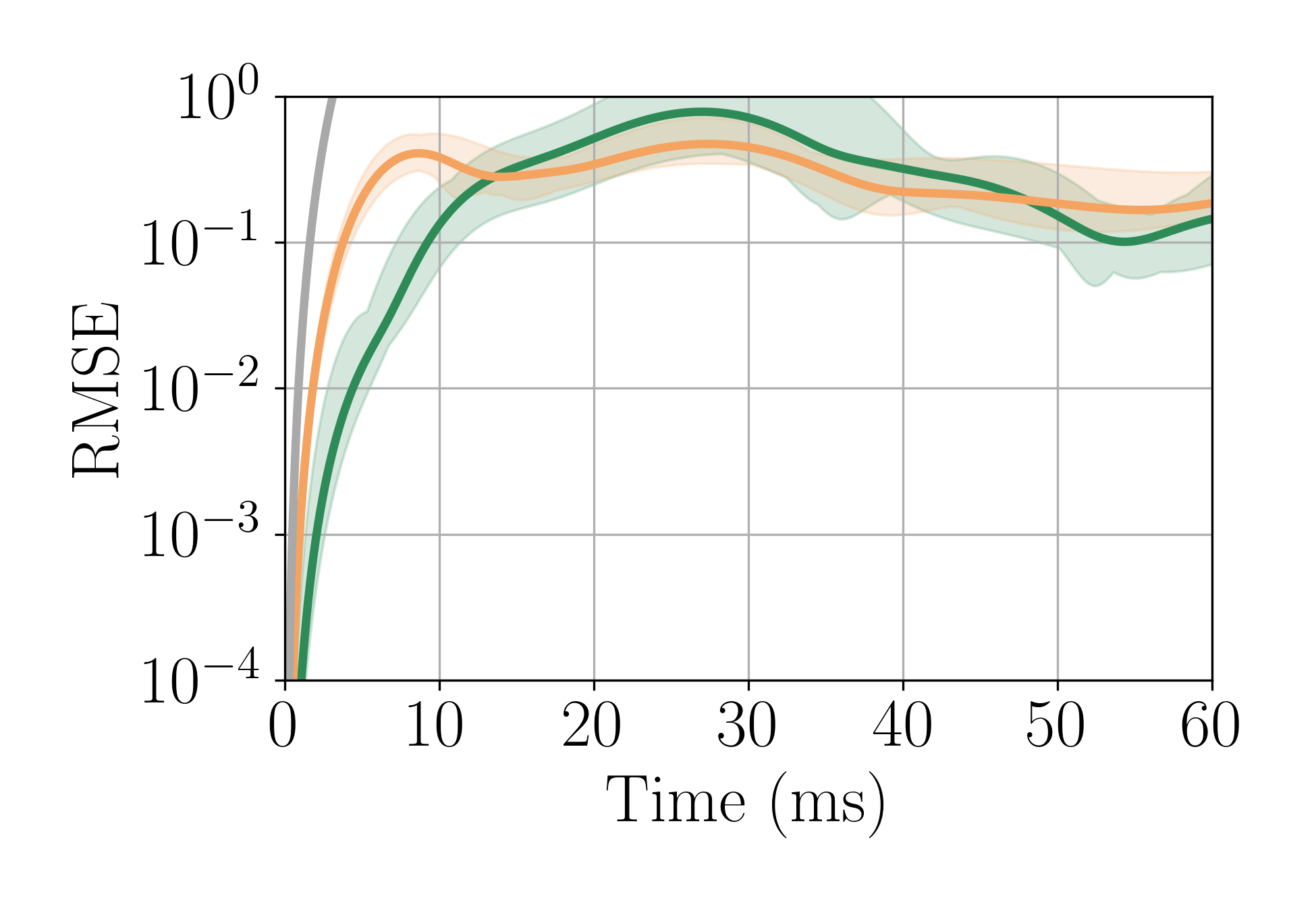}
		\caption{\gls{rmse}, $Q=50$.}
		\label{fig:scim_vector_rmse_Q50}
	\end{subfigure}
	\begin{subfigure}[b]{0.328\textwidth}
		\centering
		\includegraphics[width=\textwidth]{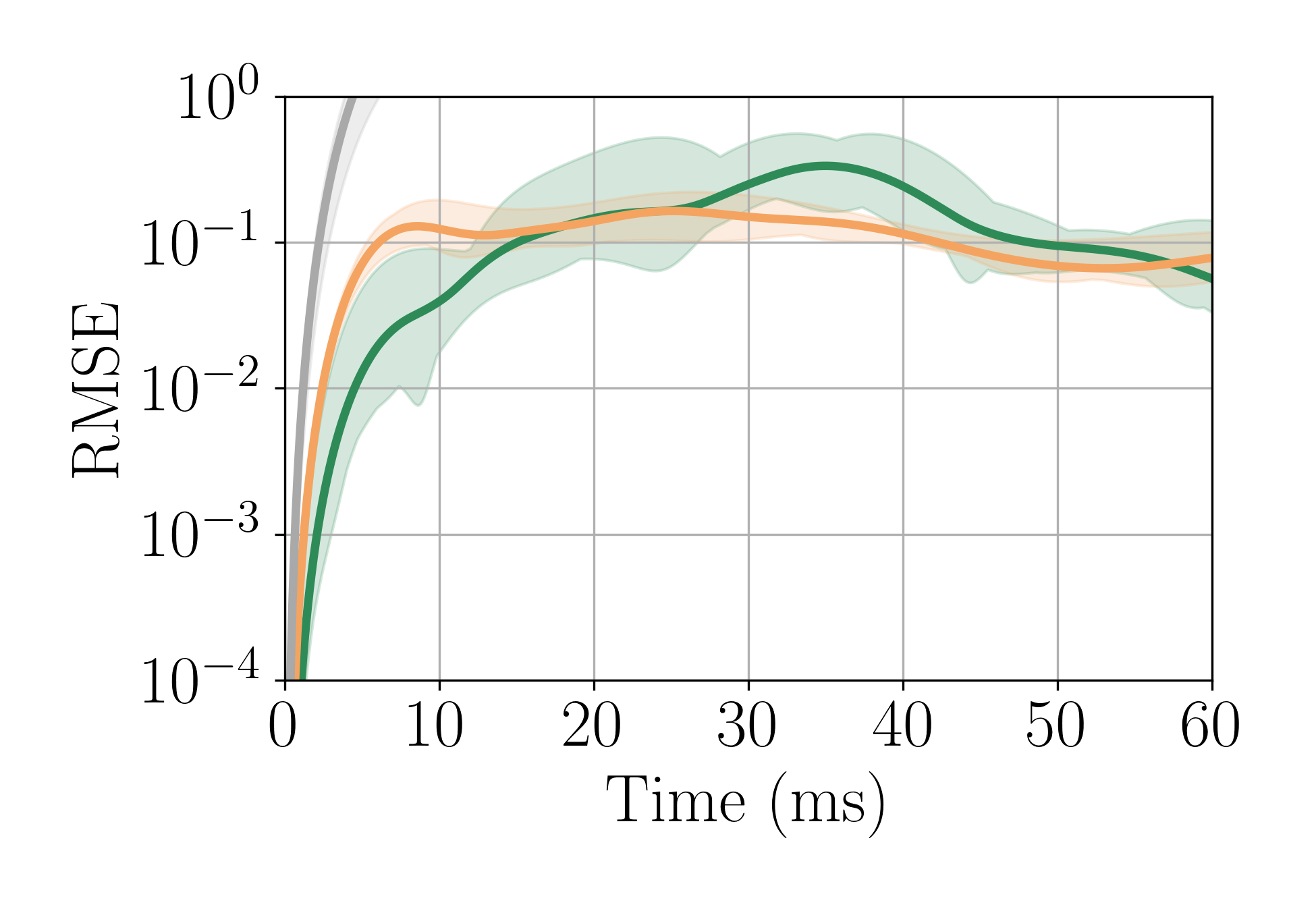}
		\caption{\gls{rmse}, $Q=100$.}
		\label{fig:scim_vector_rmse_Q100}
	\end{subfigure}
	\begin{subfigure}[b]{0.328\textwidth}
		\centering
		\includegraphics[width=\textwidth]{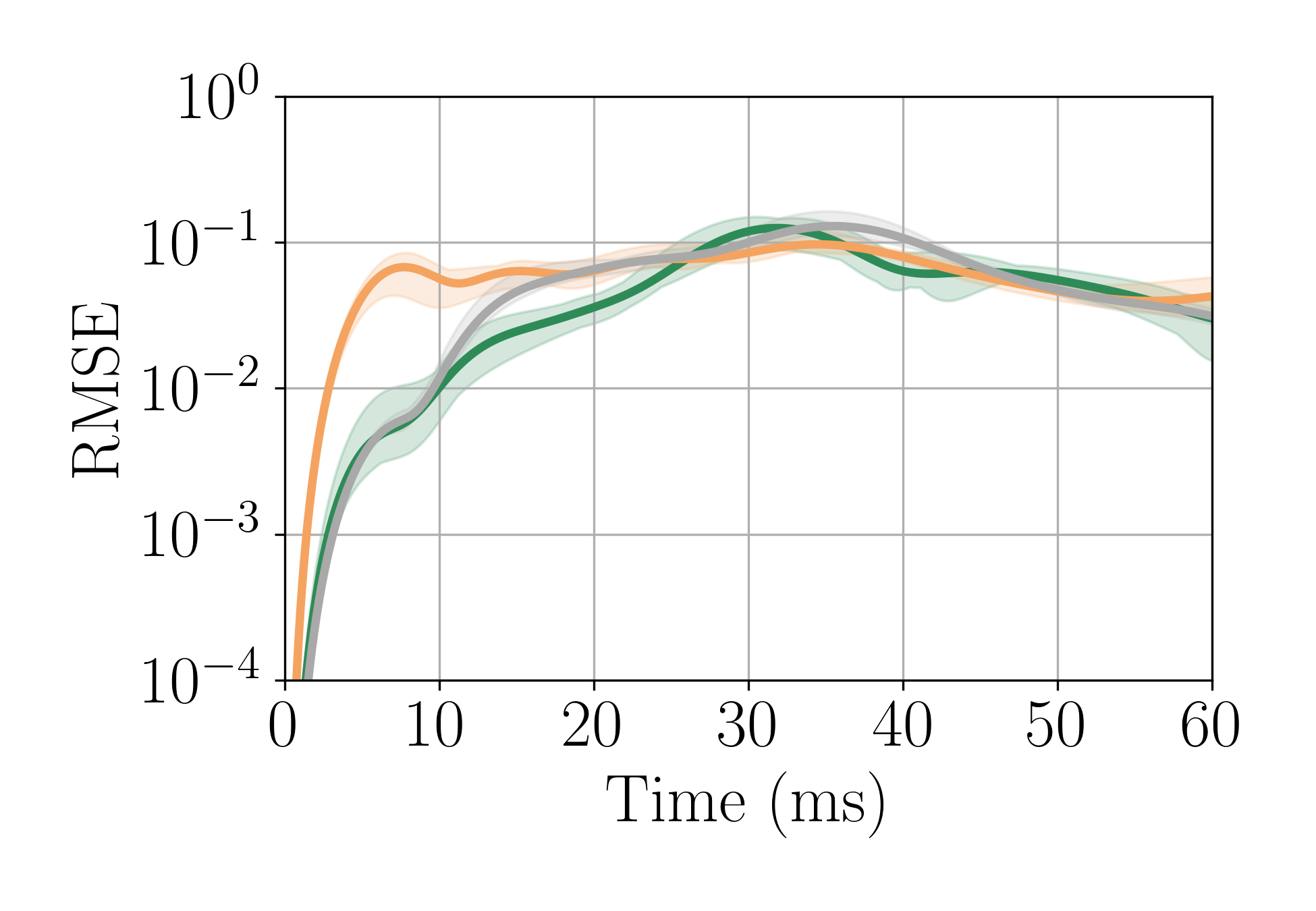}
		\caption{\gls{rmse}, $Q=200$.}
		\label{fig:scim_vector_rmse_Q200}
	\end{subfigure}
	\\
	\begin{subfigure}[b]{0.328\textwidth}
		\centering
		\includegraphics[width=\textwidth]{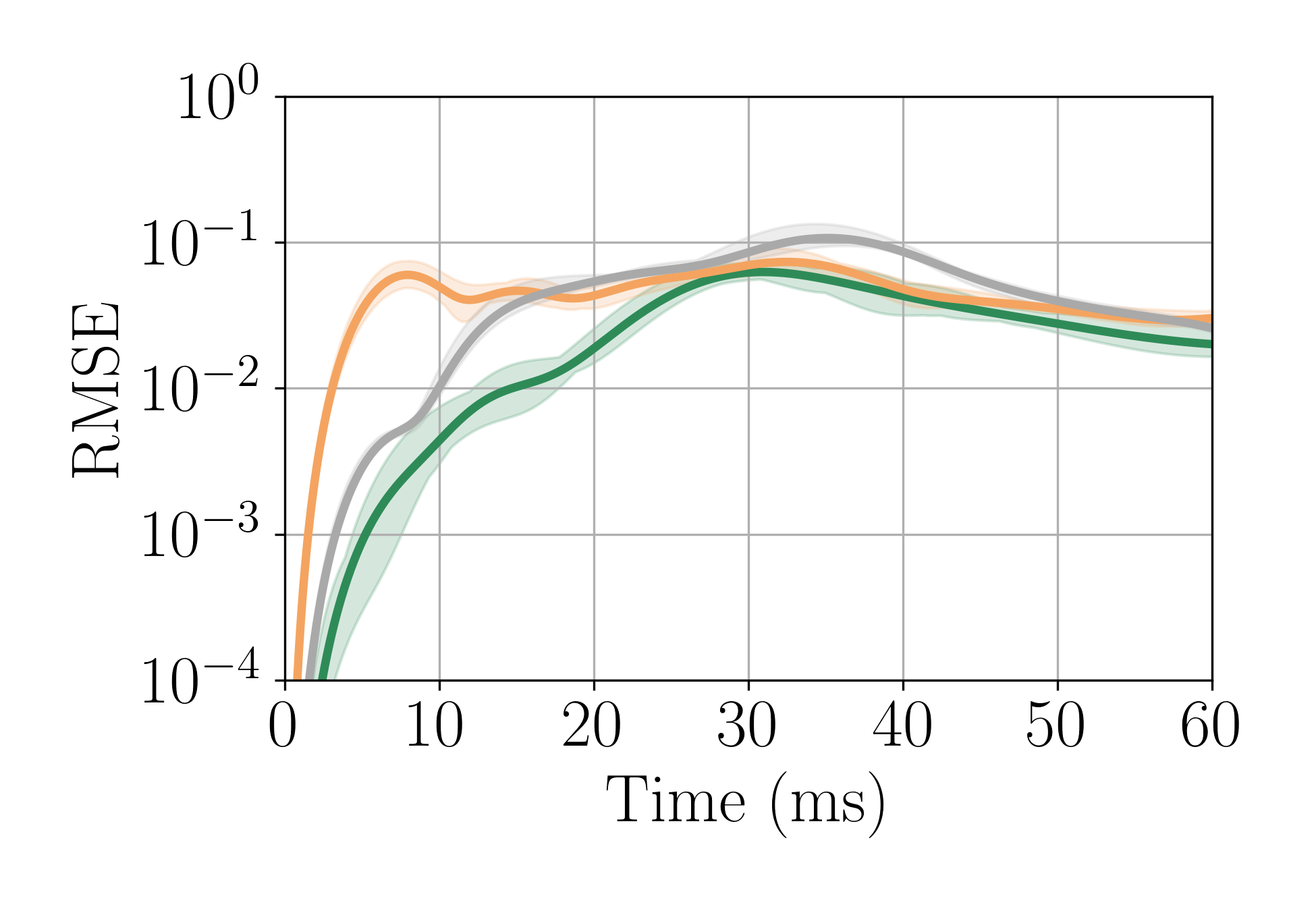}
		\caption{\gls{rmse}, $Q=300$.}
		\label{fig:scim_vector_rmse_Q300}
	\end{subfigure}
	\begin{subfigure}[b]{0.328\textwidth}
		\centering
		\includegraphics[width=\textwidth]{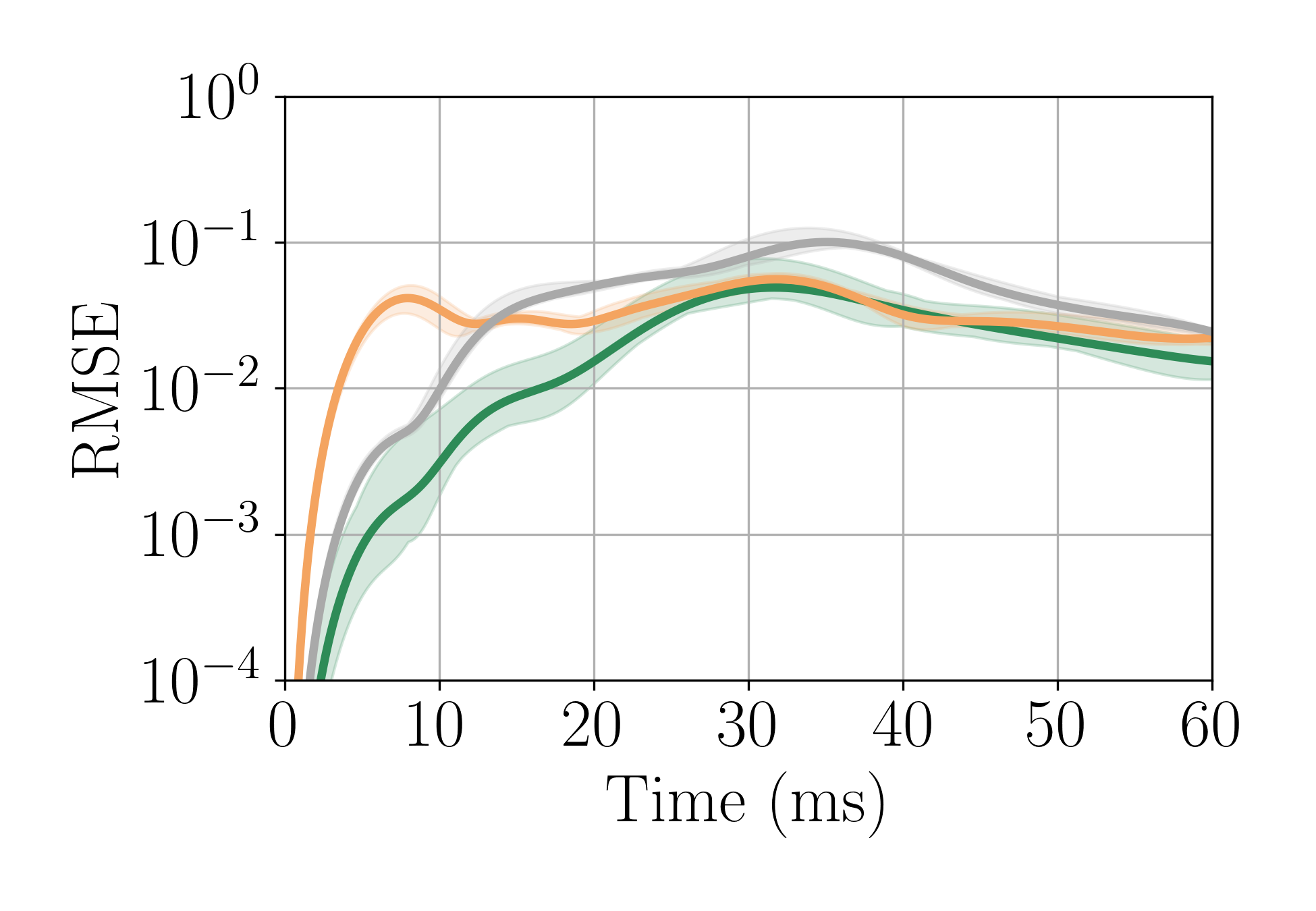}
		\caption{\gls{rmse}, $Q=400$.}
		\label{fig:scim_vector_rmse_Q400}
	\end{subfigure}
	\begin{subfigure}[b]{0.328\textwidth}
		\centering
		\includegraphics[width=\textwidth]{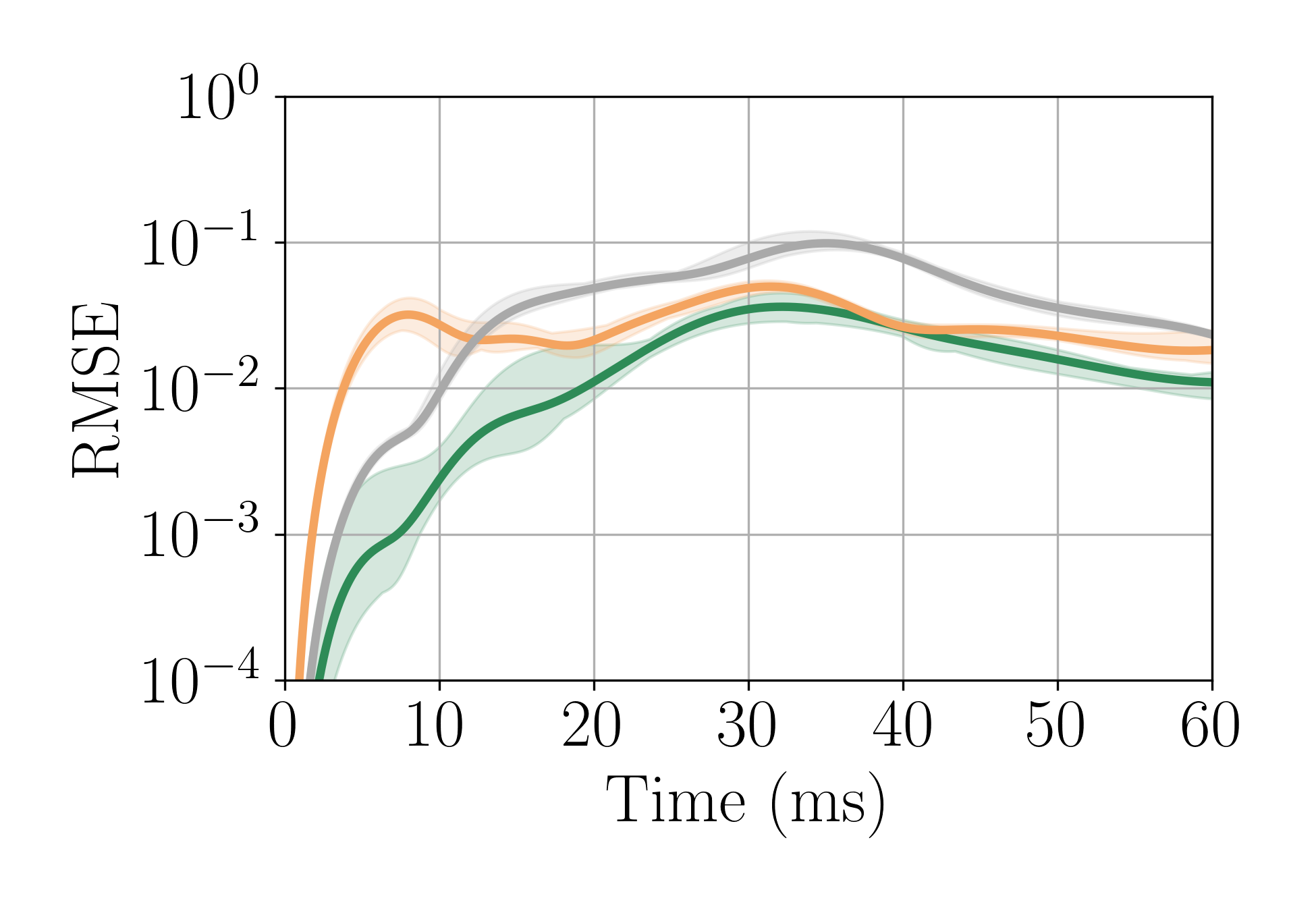}
		\caption{\gls{rmse}, $Q=500$.}
		\label{fig:scim_vector_rmse_Q500}
	\end{subfigure}
	\caption{Induction motor: \Gls{rmse} over model response of the different \glspl{pce} for training data set size $Q=50,\dots,500$. The \gls{mvsa} and \gls{td} \glspl{pce} are applied to the full response ($M=1201$). The $p/q$-adaptive \gls{lar} \gls{pce} is applied to a reduced response dimension ($M' = 24$ for $Q=50$ and $M' \in \left[48, 51\right]$ for $Q \geq 100$) based on \gls{pod}. The solid lines show average error values over $10$ different training and test data sets. The shaded areas represent the difference between minimum and maximum errors over these $10$ data sets.}
	\label{fig:scim_vector_rmse}
\end{figure}

For each \gls{pce} method and for training data sets of size  $Q=50, \dots, 500$, Figure~\ref{fig:scim_vector_rmse} presents the \gls{rmse} over the model's response.
For $Q=50,100$ training data points, the \gls{mvsa} and \gls{pod}-based $p/q$-adaptive \gls{lar} \glspl{pce} are significantly more accurate than the \gls{td} \gls{pce}, which is here considered for a maximum polynomial degree $p=2$ only, as this is the only choice with acceptable accuracy results. 
For $Q=200$, the \gls{td} \gls{pce} shows comparable accuracy to the other two \glspl{pce}, however, it does not improve further as the training data set size increases and is eventually outperformed.
For $Q \geq 200$ training data points, the \gls{mvsa} \gls{pce} is consistently more accurate than the \gls{td} \gls{pce} and comparably or more accurate than the \gls{pod}-based $p/q$-adaptive \gls{lar} \gls{pce} over the full time range.
Interestingly, the latter is consistently less accurate than the other two \glspl{pce} in the time range $0-15$~ms. 

\begin{figure}[t!]
	\centering
	\begin{subfigure}[b]{0.75\textwidth}
		\centering
		\fbox{\includegraphics[width=1\textwidth]{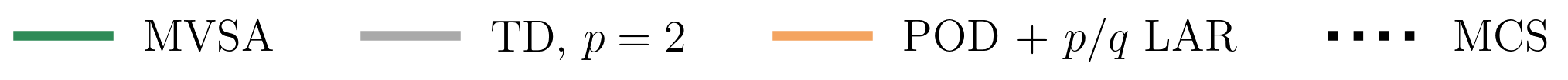}}
	\end{subfigure}
	\\
	\begin{subfigure}[b]{0.328\textwidth}
		\centering
		\includegraphics[width=\textwidth]{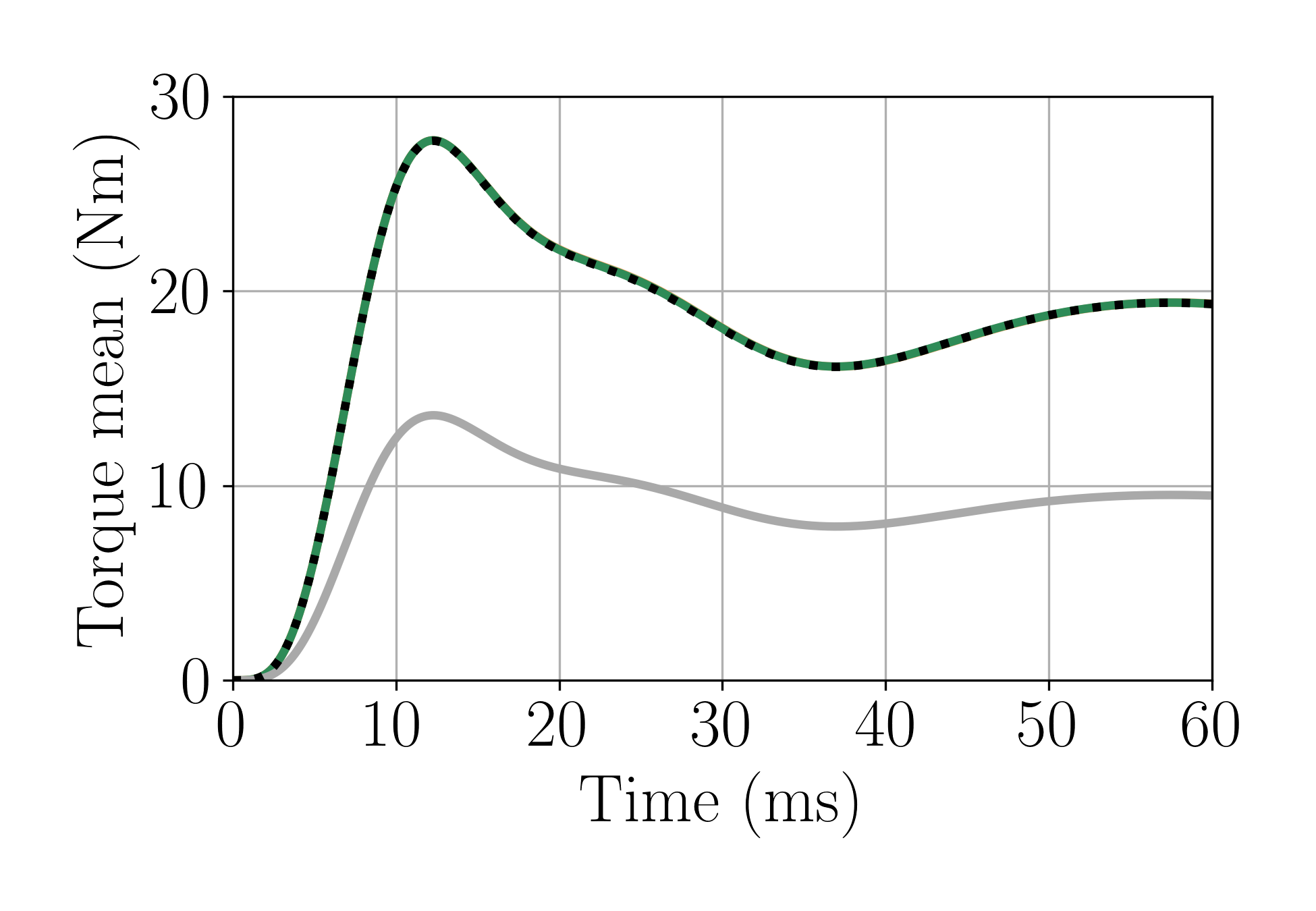}
		\caption{Mean, $Q=50$.}
	\end{subfigure}
	\begin{subfigure}[b]{0.328\textwidth}
		\centering
		\includegraphics[width=\textwidth]{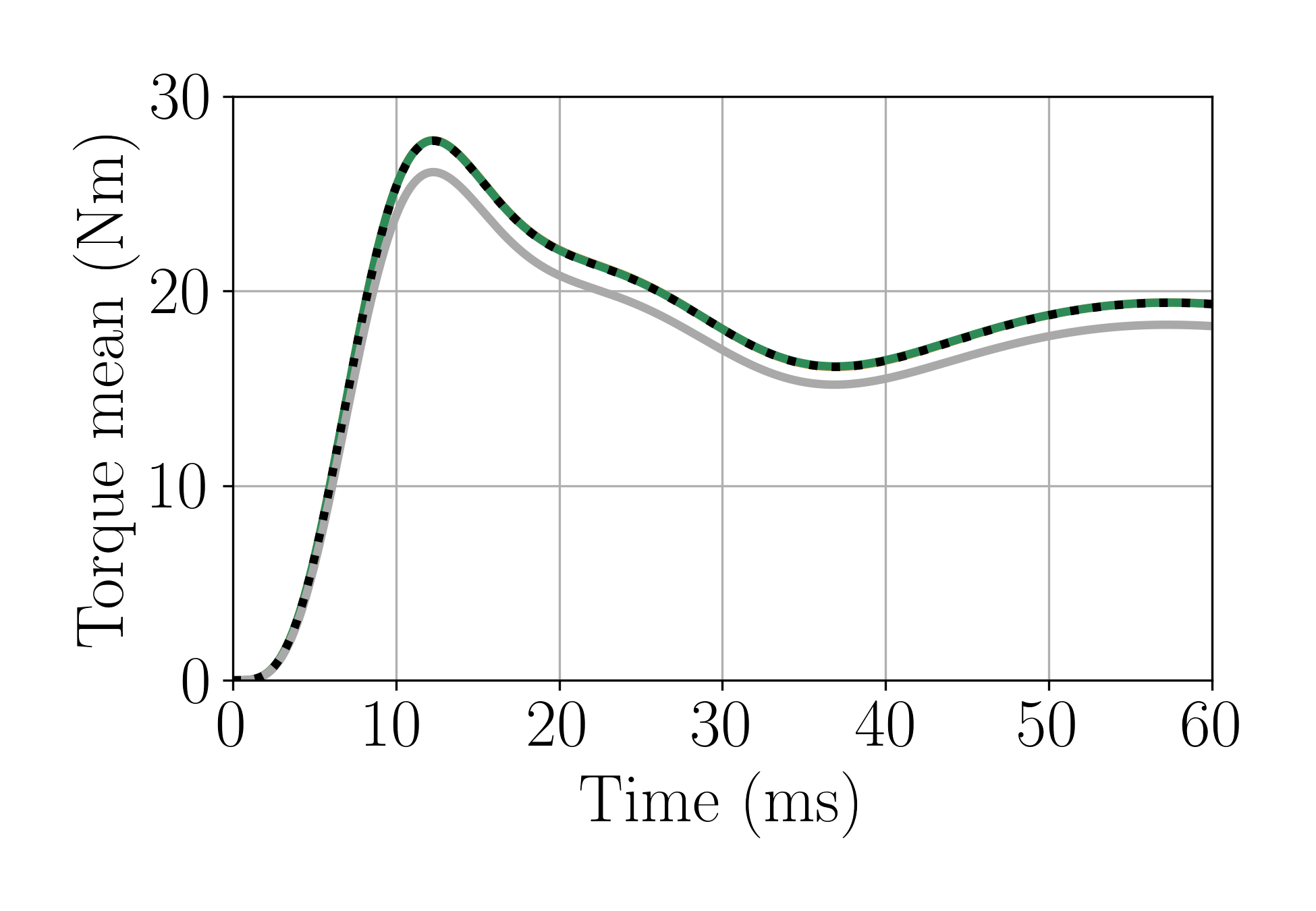}
		\caption{Mean, $Q=100$.}
	\end{subfigure}
	\begin{subfigure}[b]{0.328\textwidth}
		\centering
		\includegraphics[width=\textwidth]{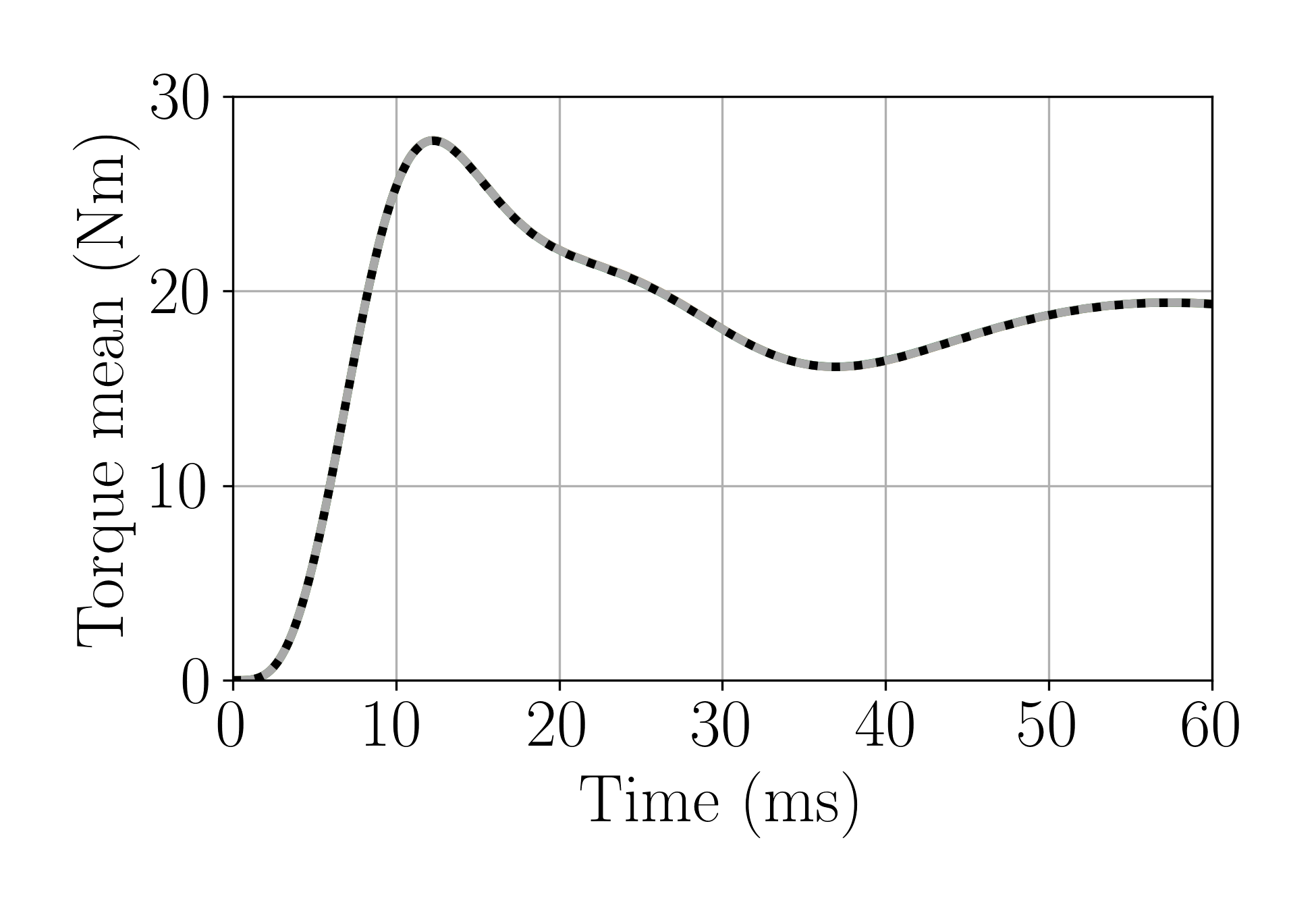}
		\caption{Mean, $Q=150$.}
	\end{subfigure}
	\\
	\begin{subfigure}[b]{0.328\textwidth}
		\centering
		\includegraphics[width=\textwidth]{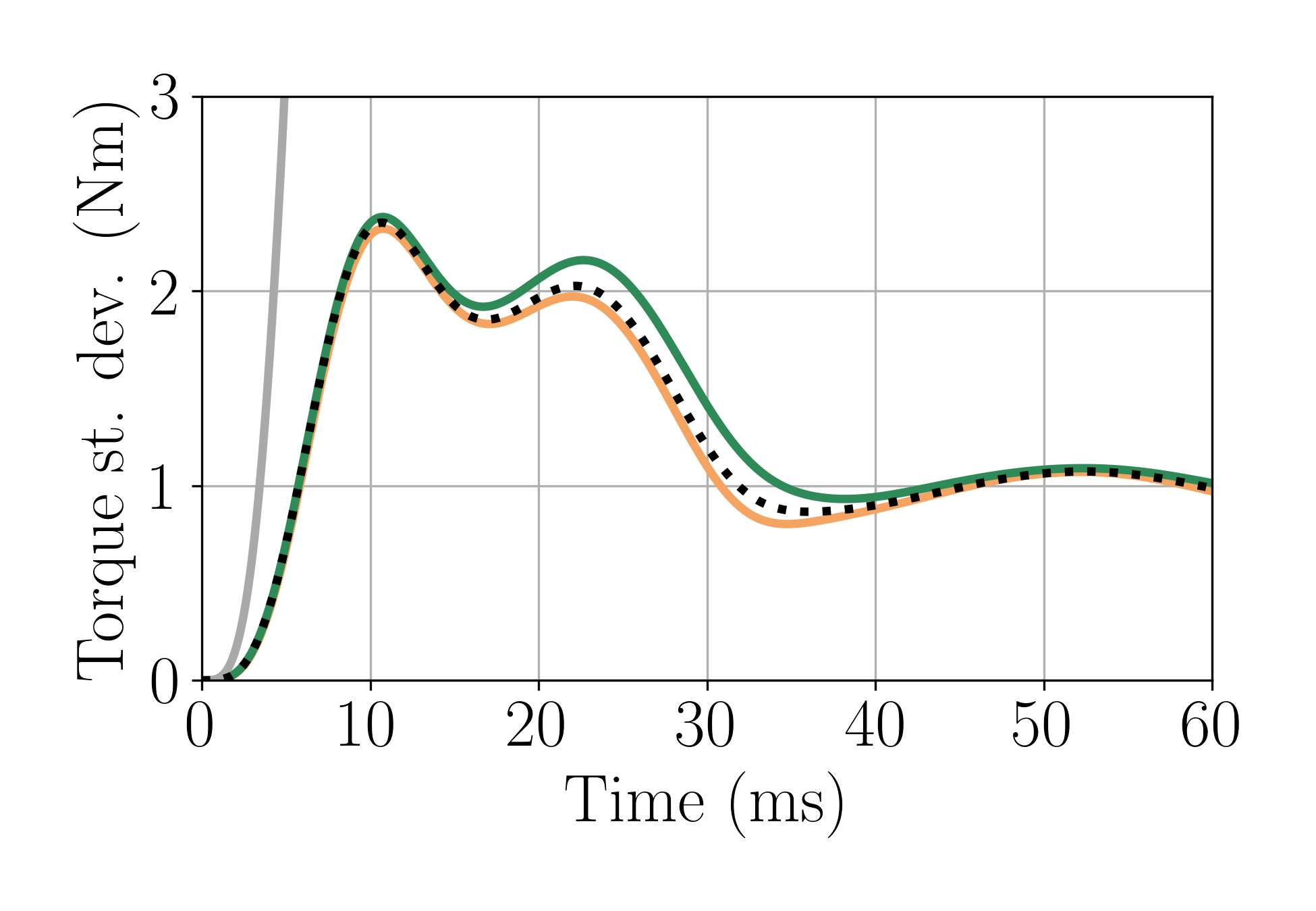}
		\caption{Standard deviation, $Q=50$.}
	\end{subfigure}
	\begin{subfigure}[b]{0.328\textwidth}
		\centering
		\includegraphics[width=\textwidth]{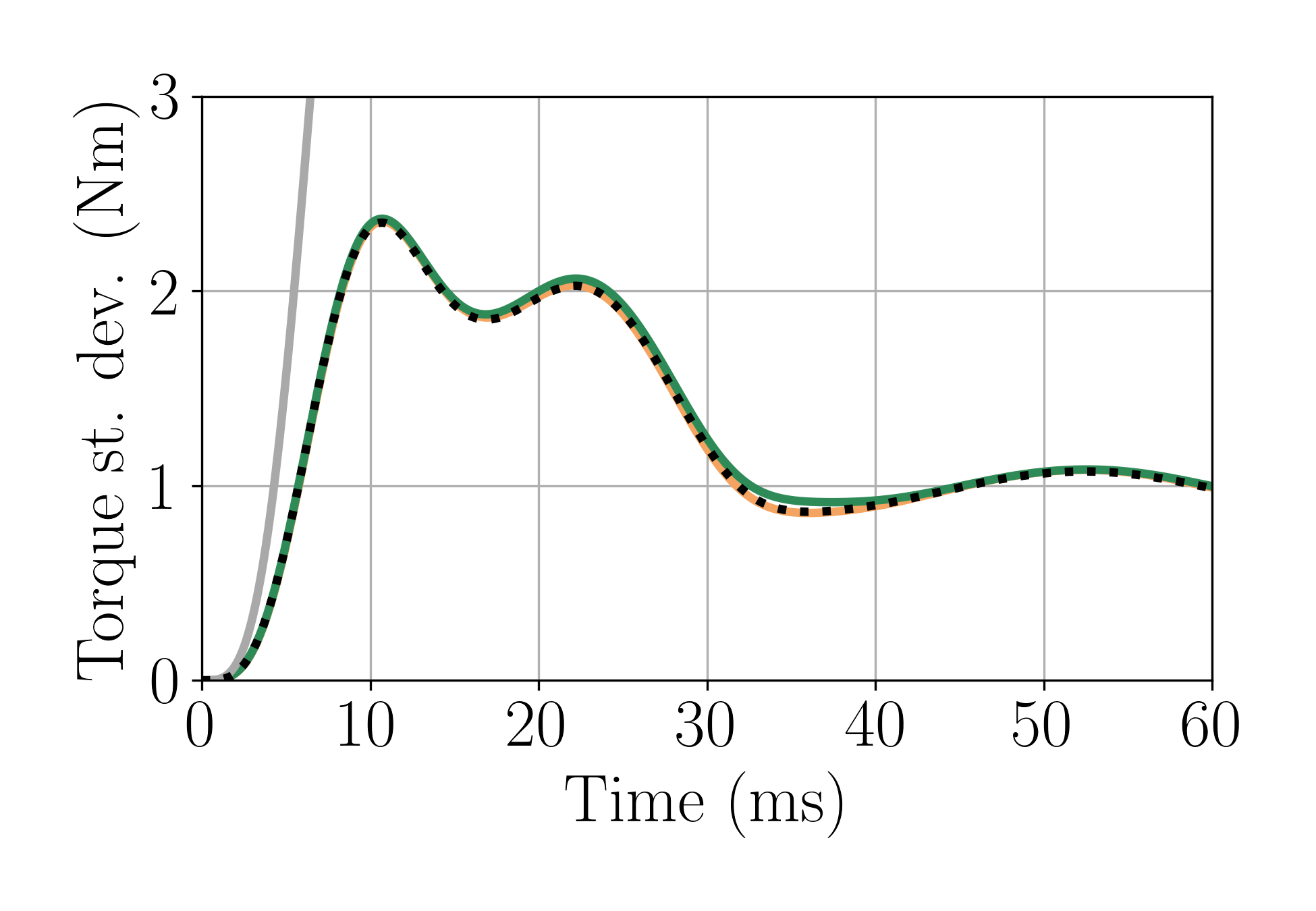}
		\caption{Standard deviation, $Q=100$.}
	\end{subfigure}
	\begin{subfigure}[b]{0.328\textwidth}
		\centering
		\includegraphics[width=\textwidth]{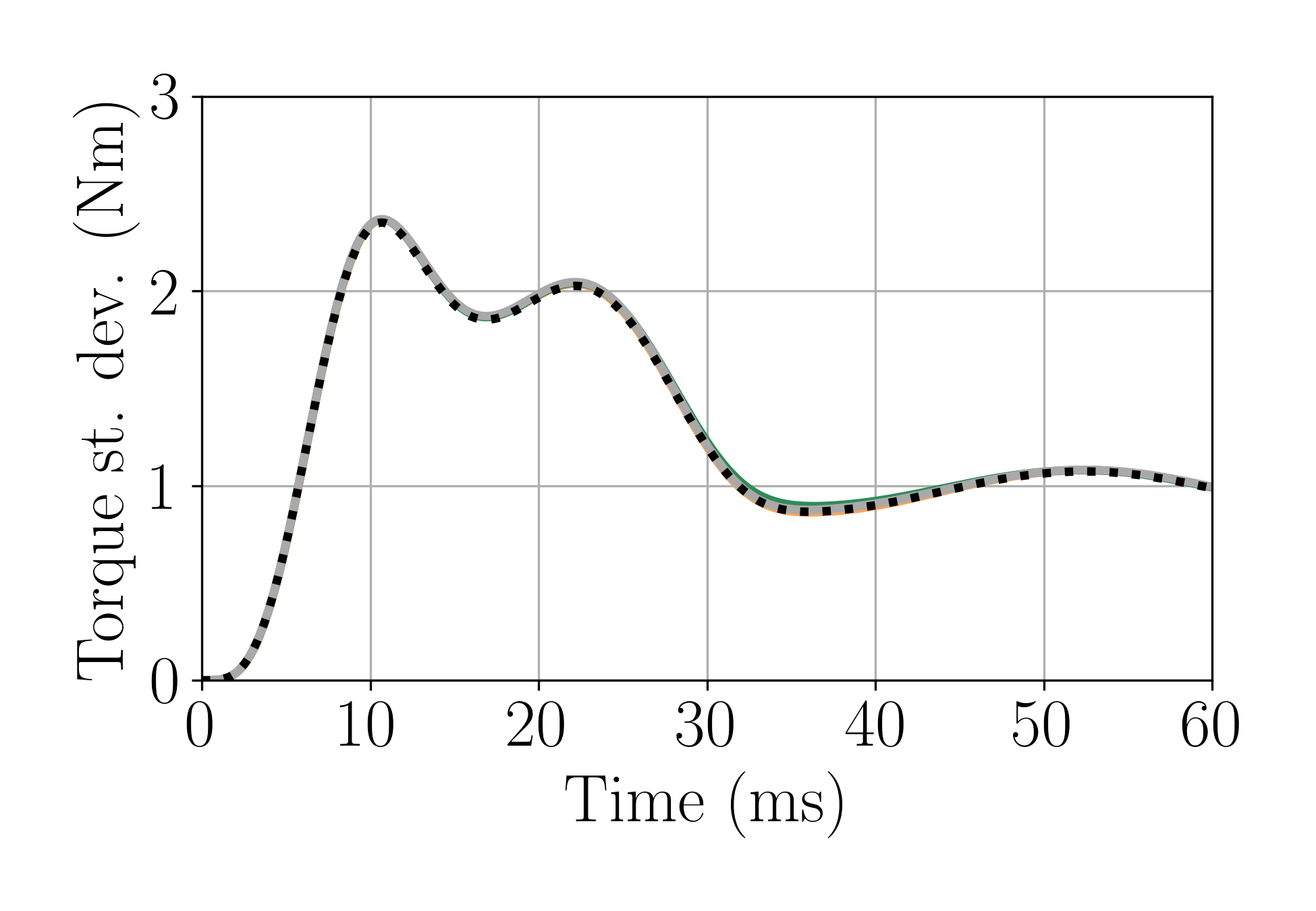}
		\caption{Standard deviation, $Q=150$.}
	\end{subfigure}
	\caption{Induction motor: Mean and standard deviation of the model's response, estimated with the different \glspl{pce} for training data set size $Q \in \left\{50, 100, 150\right\}$. The reference mean and standard deviation are computed via \gls{mcs} with $15\cdot10^3$ random samples. The \gls{mvsa} and \gls{td} \glspl{pce} are applied for the full response ($M=1201$). The $p/q$-adaptive \gls{lar} \gls{pce} is applied for a reduced dimension ($M' = 24$ for $Q=50$ and $M' \in \left[48, 51\right]$ for $Q \geq 100$) based on \gls{pod}.}
	\label{fig:scim_moments}
\end{figure}

Mean and standard deviation estimates in comparison with \gls{mcs} references are presented in Figure~\ref{fig:scim_moments}.
Already for $Q=50$ training data points, the \gls{mvsa} and \gls{pod}-based \gls{lar} \glspl{pce} yield mean estimates identical to the reference. 
The latter is slightly more accurate than the former in its standard deviation estimate for $Q=50$, while the estimates of both \glspl{pce} are almost indistinguishable to the reference for $Q=100$ and $Q=150$.
The \gls{td} \gls{pce} also results in very accurate moment estimates for $Q=150$, however, it is significantly less accurate for smaller training data set sizes, especially concerning the standard deviation.

\begin{figure}[t!]
	\centering
	\begin{subfigure}[b]{0.7\textwidth}
		\centering
		\fbox{\includegraphics[width=1\textwidth]{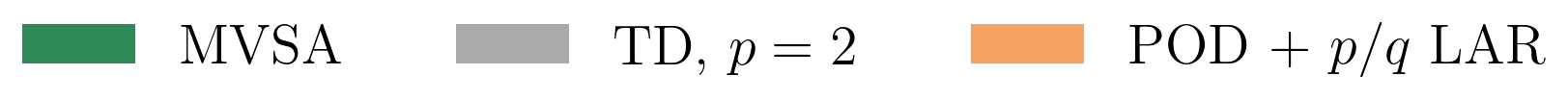}}
	\end{subfigure}
	\\
	\begin{subfigure}[b]{0.8\textwidth}
		\centering
		\includegraphics[width=\textwidth]{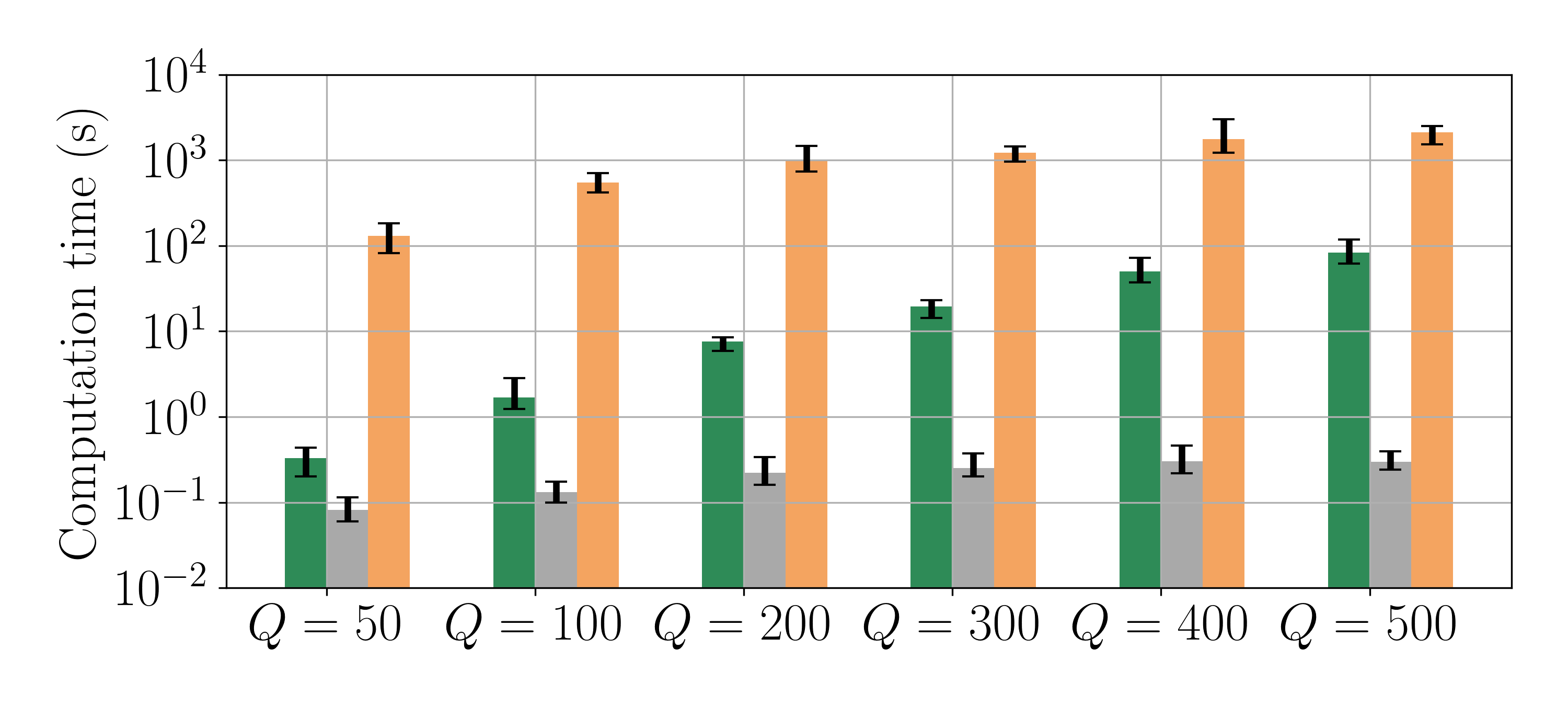}
	\end{subfigure}
	\caption{Induction motor: Computation time of the different \glspl{pce} for training data set size $Q=50,\dots,500$. The \gls{mvsa} and \gls{td} \glspl{pce} are applied for the full response ($M=1201$). The $p/q$-adaptive \gls{lar} \gls{pce} is applied for a reduced dimension ($M' = 24$ for $Q=50$ and $M' \in \left[48, 51\right]$ for $Q \geq 100$) based on \gls{pod}. The colored bars show the average computation time over $10$ different training data sets. The black error bars show the difference between minimum and maximum computation time over these $10$ data sets.}
	\label{fig:scim_computation_time}
\end{figure}

The computation time of each \gls{pce} for different training data set sizes is shown in
Figure~\ref{fig:scim_computation_time}.
As expected, the \gls{td} \gls{pce} is computed almost instantaneously due to its fixed basis. 
However, for the same reason, it is comparatively less accurate in terms of surrogate modeling and uncertainty estimation accuracy.
The \gls{mvsa} \gls{pce} is computed, on average, $\times400 - \times25$ faster than the \gls{pod}-based $p/q$-adaptive \gls{lar} \gls{pce}, despite the latter being applied for a $\times23$ reduced response dimension at minimum.
Notably, even for a reduced dimension of maximum size $M'=51$, the computation of the \gls{pod}-based $p/q$-adaptive \gls{lar} \gls{pce} still requires a substantial amount of time, especially for larger training data sets, e.g., $35$ minutes (on average) for $Q=500$. 
Comparatively, for the same training data set size, the \gls{mvsa} \gls{pce} applied to the full model response ($M=1201$) is computed in less than $2$ minutes in the worst case.

\begin{table}[t!]
	\small
	\caption{Induction motor: Maximum total and univariate degrees of the \gls{mvsa} and the $p/q$-adaptive \gls{lar} \glspl{pce}.}
	\centering
	\begin{tabular}{c c c c c}
		\toprule
		\multirow{2}{6em}{Training data set size $Q$}  & \multicolumn{2}{c}{Maximum total degree} & \multicolumn{2}{c}{Maximum univariate degree}  \\[1ex] 
		\cmidrule(lr){2-3}
		\cmidrule(lr){4-5}
		&  \gls{mvsa} & $p/q$ \gls{lar}  & \gls{mvsa}  & $p/q$ \gls{lar} \\[1em]
		\toprule
		50  & 4 & 8 & 4 & 7\\[0.5ex]
		100 & 4 & 9 & 4 & 7\\[0.5ex]
		200 & 4 & 9 & 4 & 8\\[0.5ex]
		300 & 5 & 9 & 4 & 8\\[0.5ex]
		400 & 7 & 9 & 7 & 9\\[0.5ex]
		500 & 7 & 9 & 7 & 9\\[0.5ex]
		\bottomrule
	\end{tabular}
	\label{tab:scim-max-degrees}
\end{table}

Last, Table~\ref{tab:scim-max-degrees} shows the maximum total degree, i.e., $\max_{\mathbf{k} \in \Lambda}\left|\mathbf{k}\right|_1$, and the maximum univariate degree of the polynomial bases resulting from the \gls{mvsa} and the $p/q$-adaptive \gls{lar} \gls{pce} methods. 
Note that in the case of the  $p/q$-adaptive \gls{lar} \gls{pce}, the maximum degrees over all polynomial bases (i.e., one basis per output) are presented. 
Unlike Section~\ref{sec:beam}, in this test case, the $p/q$-adaptive \gls{lar} \gls{pce} yields higher total and univariate degrees than the \gls{mvsa} \gls{pce}. 
This discrepancy could also explain the difference in their approximation results, especially for the smaller training data set sizes. 
Also note that a \gls{td} \gls{pce} with maximum polynomial degree equal to $7$, i.e., the maximum total degree of the \gls{mvsa} \gls{pce} method, would result in $77520$ coefficients, rendering the method usable only a big data training regime.

\subsection{Power flow in electrical grid}
\begin{figure}[t!]
	\begin{subfigure}[t]{0.49\textwidth}
		\centering
		\includegraphics[scale=0.36]{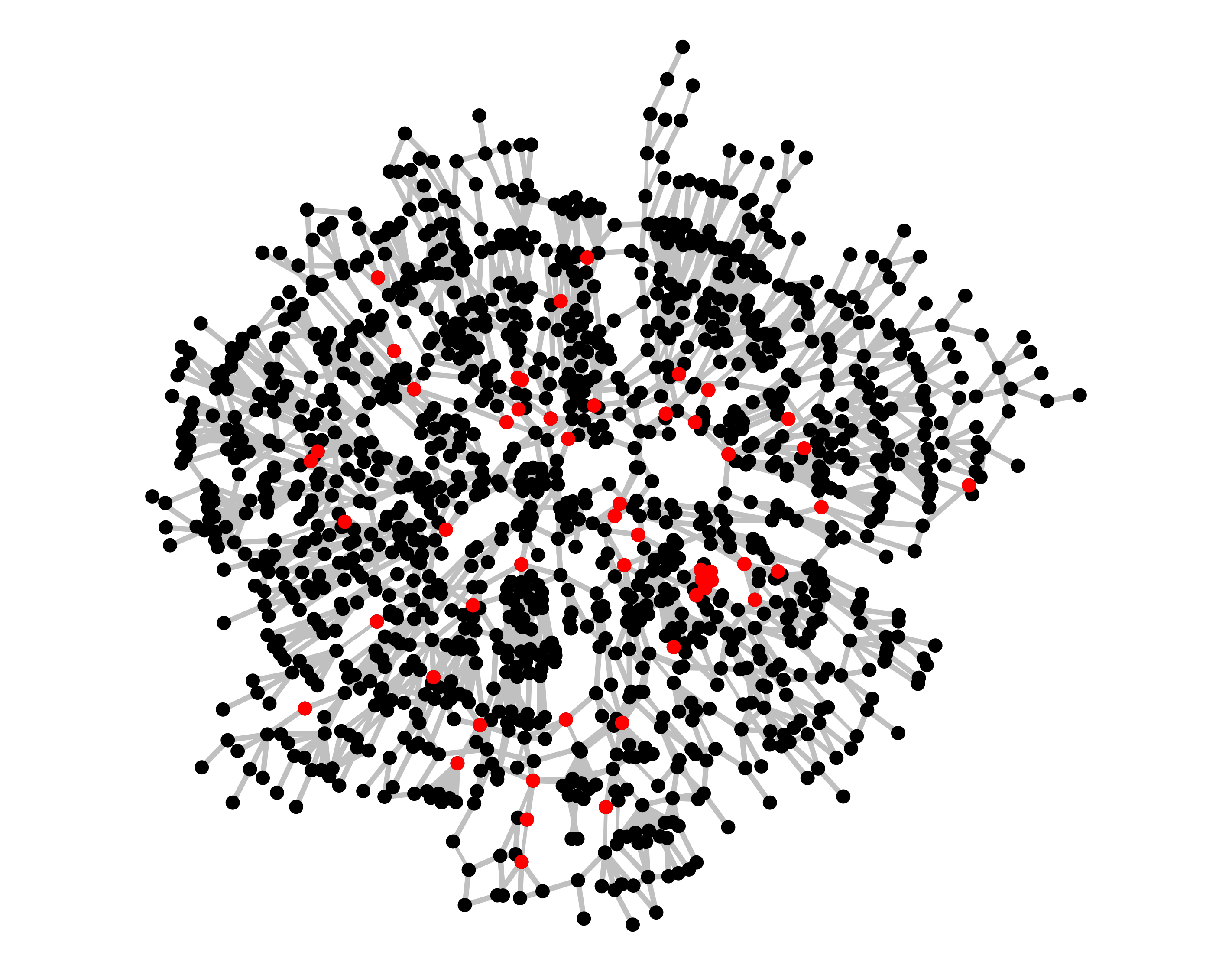}
		\caption{Power grid 1354-PEGASE.}
		\label{fig:grid_sketch}
	\end{subfigure}
	\hfill
	\begin{subfigure}[t]{0.49\textwidth}
		\centering
		\includegraphics[width=1.0\textwidth]{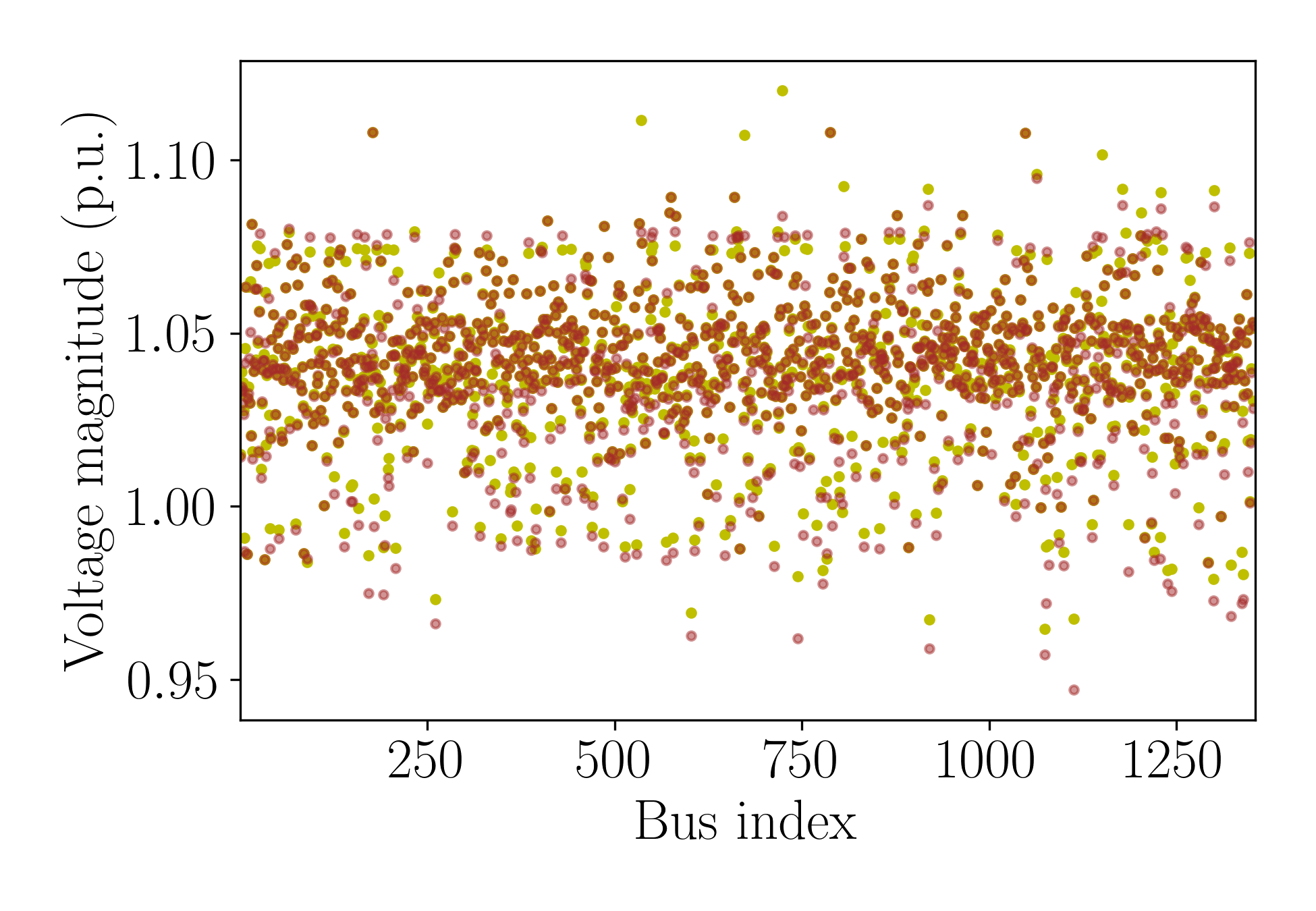}
		\caption{Voltage magnitude per network bus.}
		\label{fig:grid_responses}
	\end{subfigure}
	\caption{(a) The 1354-PEGASE power grid. Network buses are shown in black, transmission lines in gray, and the $52$ random loads in red. (b) Voltage magnitude per network bus in the per-unit (p.u.) system for two realizations of the $52$ random loads, respectively shown in yellow and brown.}
	\label{fig:grid_illustration}
\end{figure}

As third and final test case, we consider a power flow (also referred to as load flow) study based on a power grid model that represents part of the European high-voltage transmission network. 
This network is a standard power system test case known as the 1354-PEGASE grid \cite{fliscounakis2013contingency}, where the number $1354$ refers to the number of buses, i.e., network nodes upon which other elements, e.g., generators, loads, transformers, transmission lines, etc., are connected.
An implementation of this grid is available in the \texttt{pandapower} open-source software \cite{thurner2018pandapower}.
The network is shown in Figure~\ref{fig:grid_sketch}.

The power flow problem is formulated using the so-called power balance equations
\begin{subequations}
	\label{eq:power_balance}
	\begin{align}
		0 &= -\mathrm{Re}\left\{S_i\right\} + \sum_{j=1}^J \left|V_i\right| \left|V_j\right| \left(G_{ij} \cos\theta_{ij} + B_{ij} \sin\theta_{ij}\right), \label{eq:active_power_balance}\\
		0 &= -\mathrm{Im}\left\{S_i\right\} + \sum_{j=1}^J \left|V_i\right| \left|V_j\right| \left(G_{ij} \sin\theta_{ij} + B_{ij} \cos\theta_{ij}\right), \label{eq:reactive_power_balance}
	\end{align}
\end{subequations}
where $S_i$ denotes the apparent power injected at the $i$-th bus, $\mathrm{Re}\left\{S_i\right\} = \left|S_i\right| \cos{\phi_i}$ and $\mathrm{Im}\left\{S_i\right\} = \left|S_i\right| \sin{\phi_i}$ are the active and reactive power for power factor $\phi_i$, $G_{ij}$ and $B_{ij}$ denote the real and imaginary parts of the elements of the network's nodal admittance matrix, $\left|V_i\right|$ and  $\left|V_j\right|$ denote the voltage magnitude at buses $i$ and $j$, and $\theta_{ij} = \theta_{i} - \theta_{j}$ is the voltage angle difference between buses $i$ and $j$.
The known and unknown variables in the balance equations \eqref{eq:power_balance} are determined based on the type of the $i$-th bus. 
That is, the active and reactive power are known for so-called load buses, while the active power and the voltage magnitude are known for so-called generator buses. 
One bus is selected as angular reference, known as the slack or swing bus.
The resulting nonlinear system of equations is solved numerically, using variations of the Newton-Raphson method \cite{milano2010power}.

Following prior works on uncertainty modeling for power grids \cite{jordehi2018deal}, the loads of $52$ buses are considered to be uncertain. 
In particular, we introduce the random variables $X_n$  and the corresponding Beta distributions $B_{n}\left(\alpha=3, \beta=2, l=0, u=1\right)$, $n=1,\dots,N=52$, where $\alpha$, $\beta$ are shape parameters, and $l$, $u$ are the lower and upper limit of the random variable's support. 
The random loads are then given as $S_n = S_n^{\max} X_n$, where $S_n^{\max}$ denotes the maximum allowed load value. 
Note that the maximum loads vary significantly, i.e.,  $S_n^{\max} \in \left[2.8, 1175.0\right]$~MVA.
The power factors $\phi_n$ remain unaffected.
The network buses corresponding to the random loads are marked with red color in Figure~\ref{fig:grid_sketch}.
The quantity of interest is chosen to be the voltage magnitude $\left|V_i\right|$ on each of the $M=1354$ buses, as this is the most critical size regarding network stability.
Each bus is affected much differently by the random loads due to their position on the grid and the different maximum load values.
Figure~\ref{fig:grid_responses} shows the voltage magnitude per network bus for two realizations of the random loads.
Note that the voltage magnitude is computed in the per-unit (p.u) system, that is, relative to the nominal rated voltage of each bus, as is commonly done in power flow studies.
The rated voltage is $220$~kV for $1113$ buses and $380$~kV for $241$ buses.

In this test case, the \gls{mvsa} and \gls{td} \glspl{pce} are employed for training data sets of size $Q=100,200,\dots,1000$. 
The \gls{td} \gls{pce} has maximum polynomial degree $p=2$, however, as will be observed, even the largest training datasets are not sufficient for its training, due to the curse of (input) dimensionality.
The $p/q$-adaptive \gls{lar} \gls{pce} is here omitted, due to the fact that the output dimension is too large for its element-wise application and \gls{pod} does not result in sufficient dimension reduction. 

\begin{figure}[t!]
	\centering
	\begin{subfigure}[b]{0.65\textwidth}
		\centering
		\fbox{\includegraphics[width=0.6\textwidth]{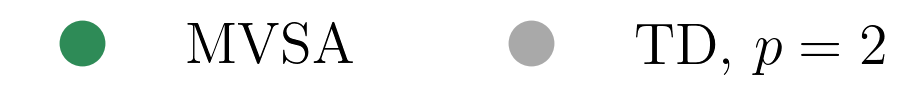}}
	\end{subfigure}
	\\
	\begin{subfigure}[b]{0.49\textwidth}
		\centering
		\includegraphics[width=\textwidth]{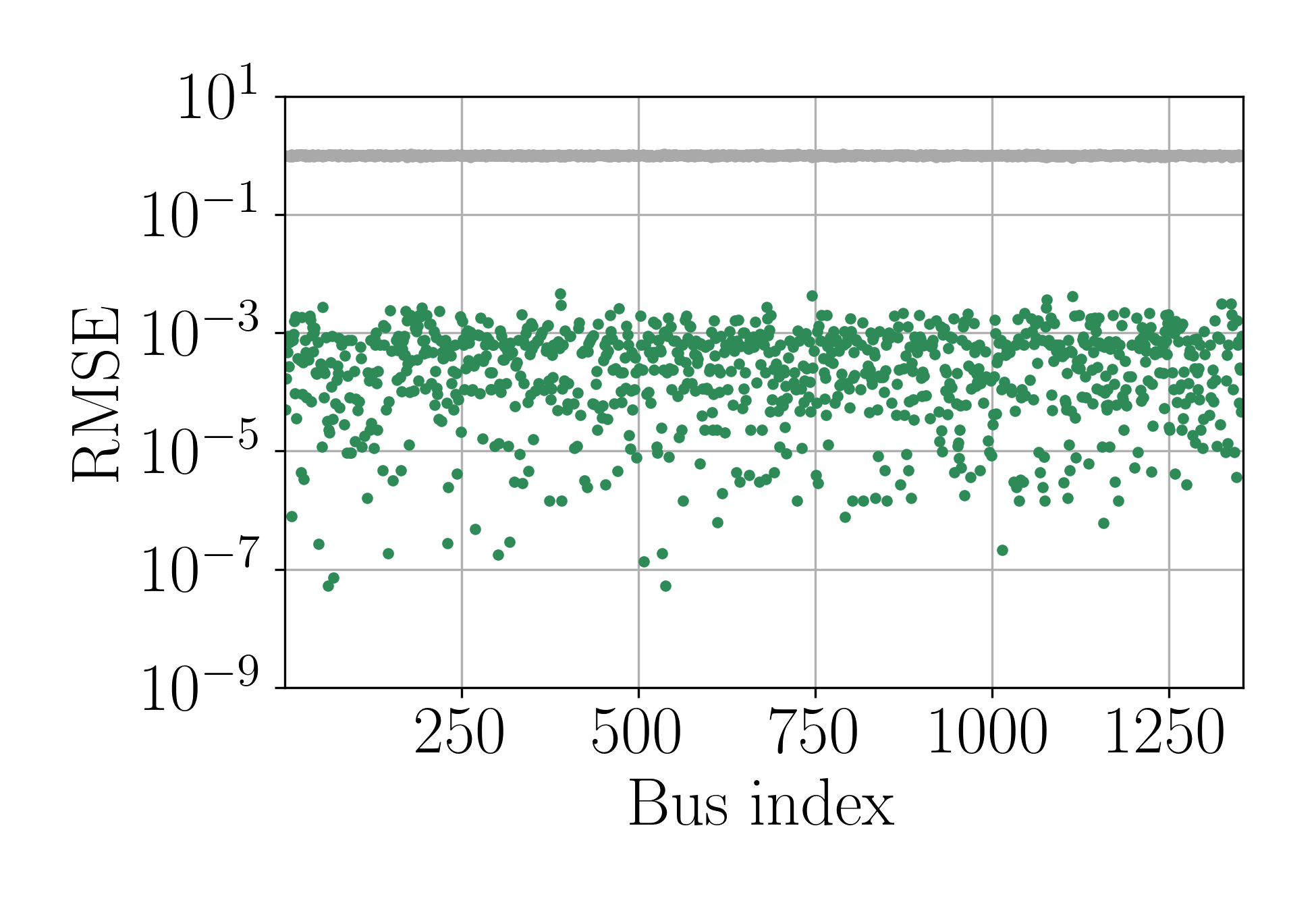}
		\caption{\gls{rmse}, $Q=100$.}
		\label{fig:grid_vector_rmse_Q100}
	\end{subfigure}
	\hfill
	\begin{subfigure}[b]{0.49\textwidth}
		\centering
		\includegraphics[width=\textwidth]{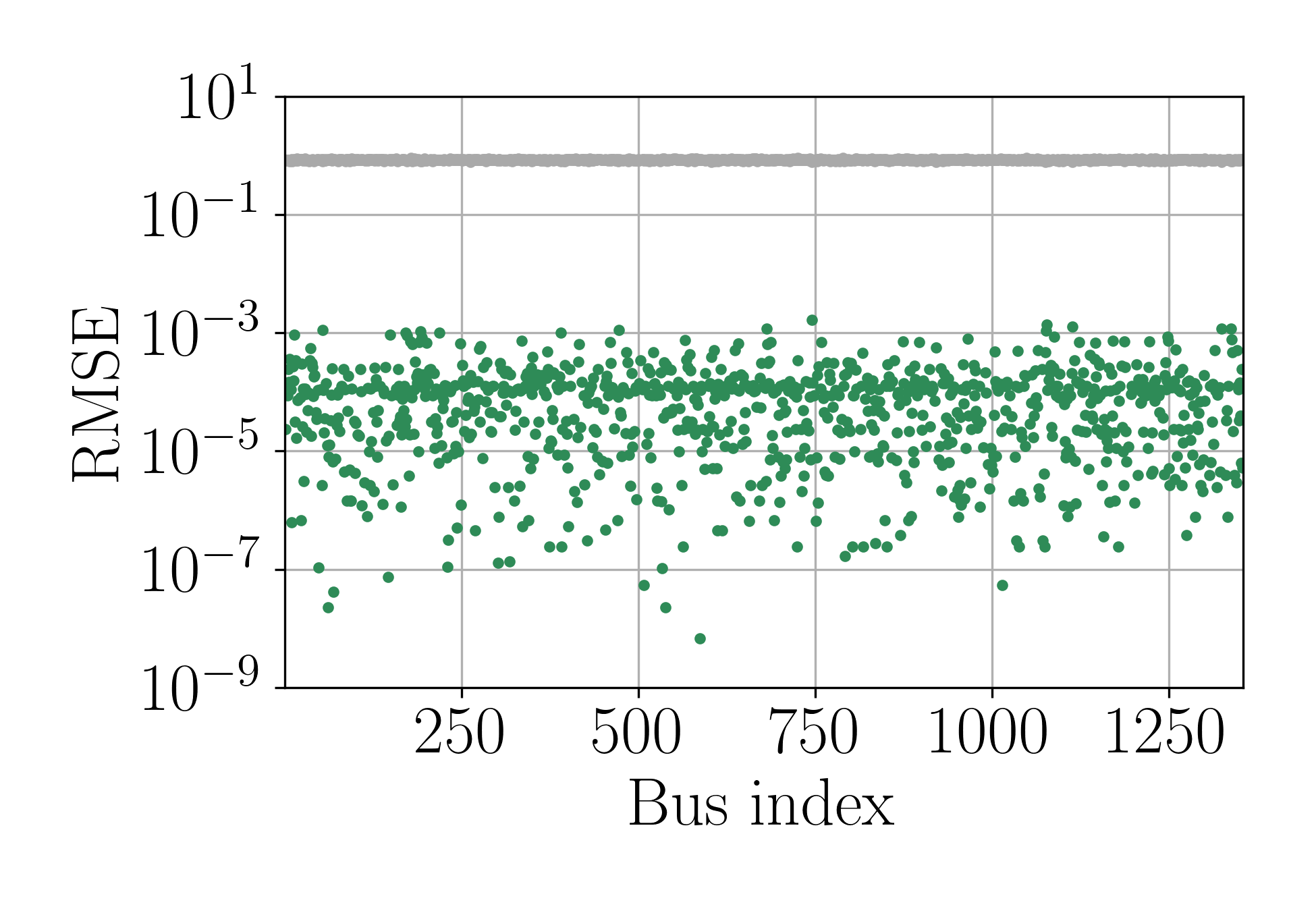}
		\caption{\gls{rmse}, $Q=500$.}
		\label{fig:grid_vector_rmse_Q500}
	\end{subfigure}
	\\
	\begin{subfigure}[b]{0.49\textwidth}
		\centering
		\includegraphics[width=\textwidth]{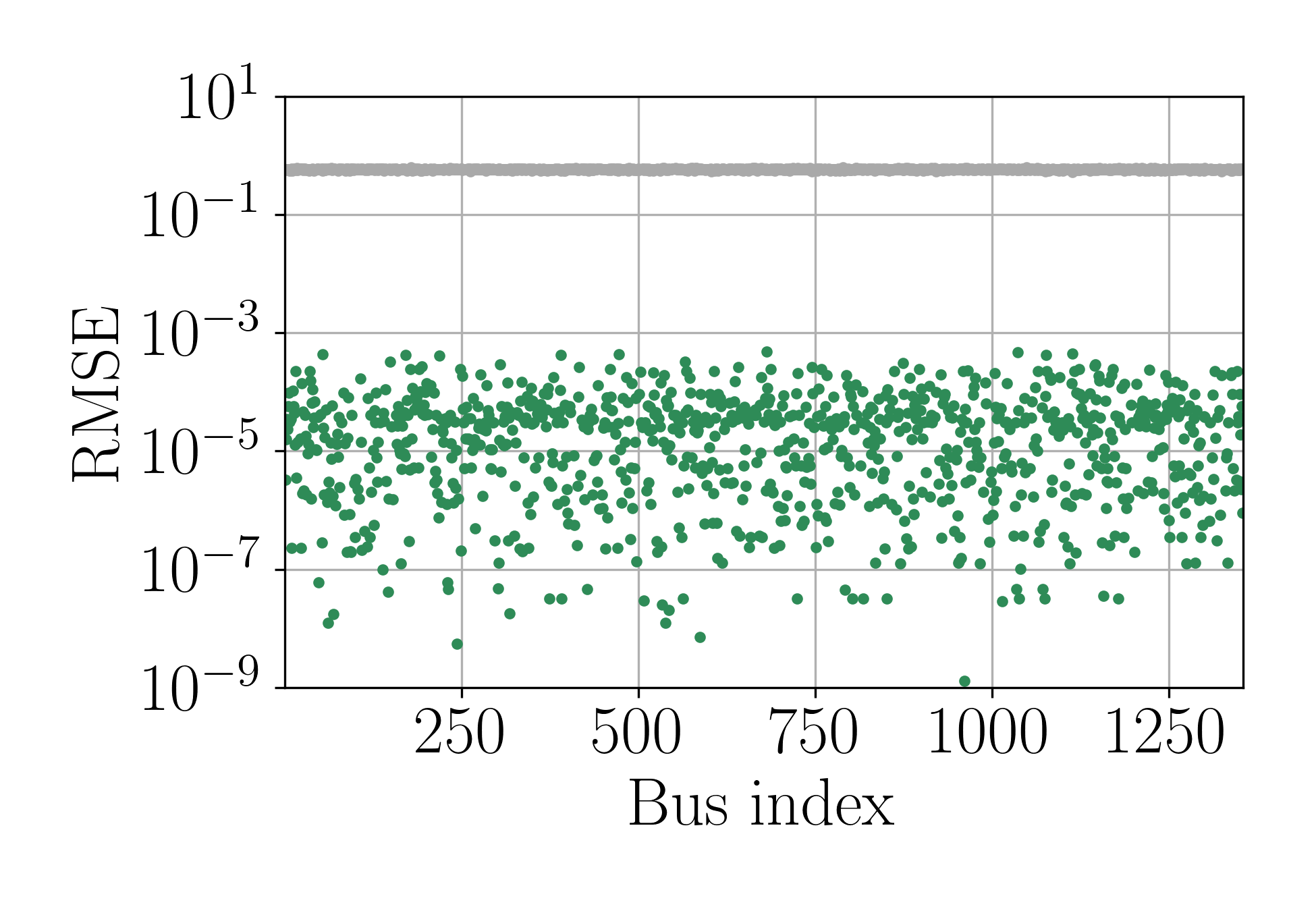}
		\caption{\gls{rmse}, $Q=1000$.}
		\label{fig:grid_vector_rmse_Q1000}
	\end{subfigure}
	\hfill
	\begin{subfigure}[b]{0.49\textwidth}
		\centering
		\includegraphics[width=\textwidth]{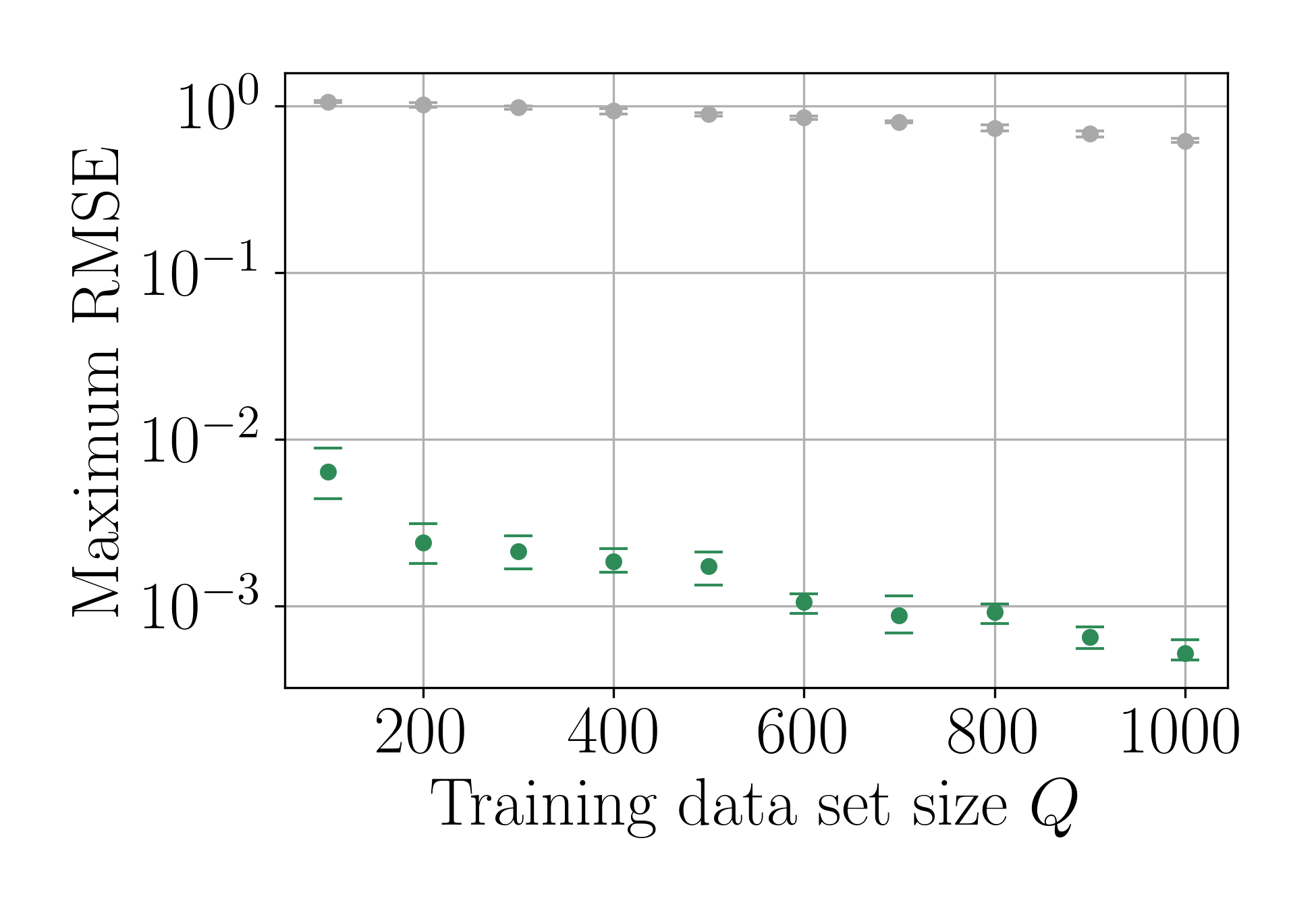}
		\caption{Maximum \gls{rmse}.}
		\label{fig:grid_max_rmse}
	\end{subfigure}
	\caption{Power grid: \Gls{rmse} in voltage magnitude per network bus and maximum \gls{rmse} over all buses for the \gls{mvsa} and \gls{td} \glspl{pce} trained with data sets of increasing size. (a)-(c) show average error values over 10 different training and test data sets. Figure (d) shows the average (circles) and the minimum/maximum (horizontal bars) values of the maximum \gls{rmse} over these $10$ data sets.}
	\label{fig:grid_rmse}
\end{figure}

Surrogate model accuracy results are shown in Figure~\ref{fig:grid_rmse}, where \ref{fig:grid_vector_rmse_Q100}, \ref{fig:grid_vector_rmse_Q500}, and \ref{fig:grid_vector_rmse_Q1000} show the \gls{rmse} per network bus for three training data set sizes, and \ref{fig:grid_max_rmse} shows the maximum \gls{rmse} over all buses for all training data set sizes.
The large input dimensionality ($N=52$) is prohibitive for training the \gls{td} \gls{pce} with the available data sets and only minor improvement in accuracy is observed, despite the tenfold increase in training data set size.
In comparison, the maximum \gls{rmse} of the \gls{mvsa} \gls{pce} is consistently more than two orders of magnitude smaller and is reduced by more than one order of magnitude as the training data set size increases. 
However, for most buses, the \gls{rmse} is several orders of magnitude smaller than the maximum one. 
Note that a number of buses have been omitted in Figures~ \ref{fig:grid_vector_rmse_Q100}, \ref{fig:grid_vector_rmse_Q500}, and \ref{fig:grid_vector_rmse_Q1000}, as the voltage magnitude of said buses is only minimally or not at all affected by the random load variations, hence, the corresponding \glspl{rmse} are almost zero.

\begin{figure}[t!]
	\centering
	\begin{subfigure}[b]{0.55\textwidth}
		\centering
		\fbox{\includegraphics[width=0.6\textwidth]{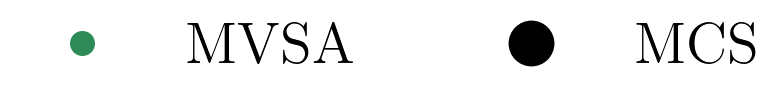}}
	\end{subfigure}
	\\
	\begin{subfigure}[b]{0.49\textwidth}
		\centering
		\includegraphics[width=\textwidth]{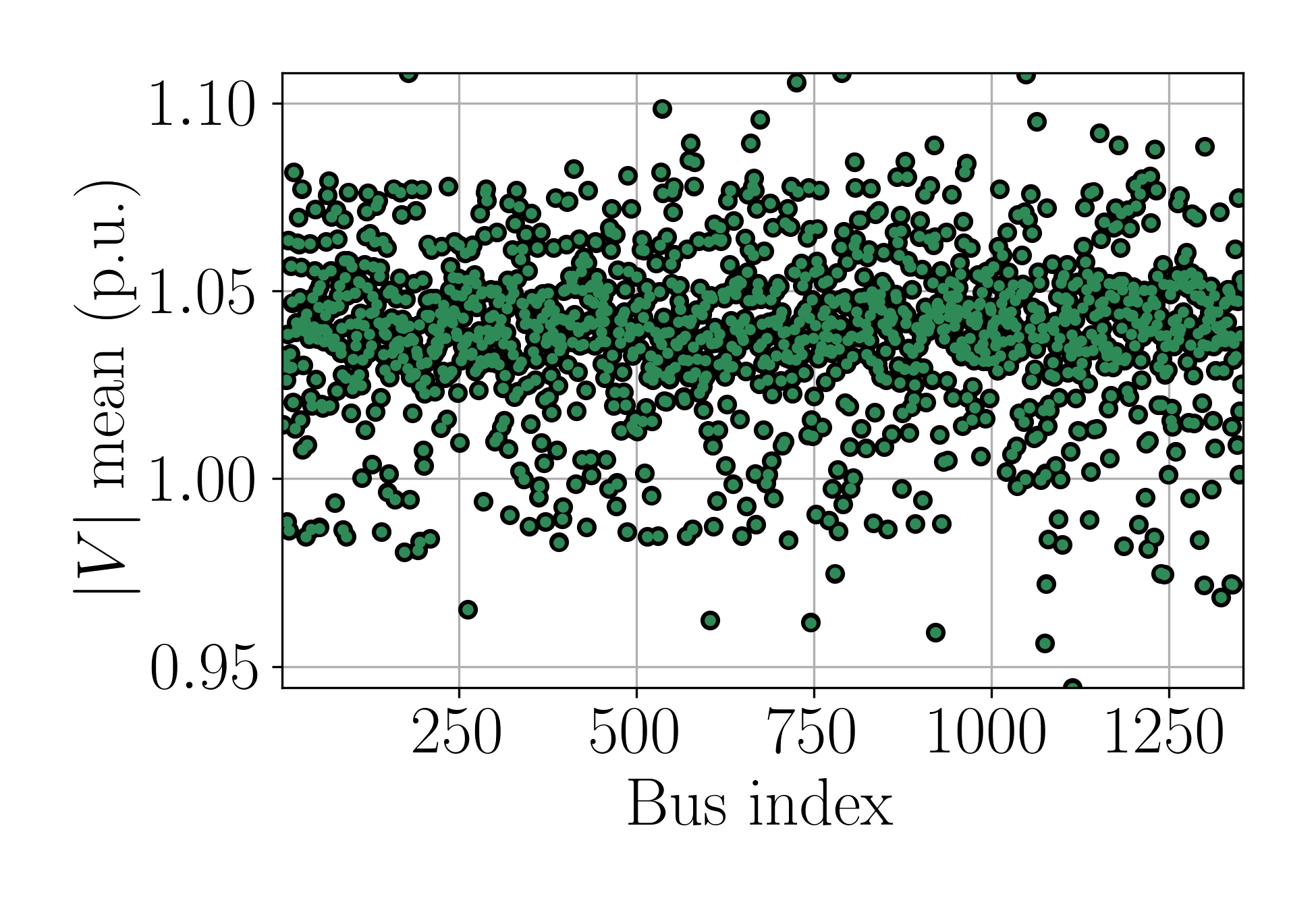}
		\caption{Mean, $Q=100$.}
		\label{fig:grid_mean_Q100}
	\end{subfigure}
	\hfill
	\begin{subfigure}[b]{0.49\textwidth}
		\centering
		\includegraphics[width=\textwidth]{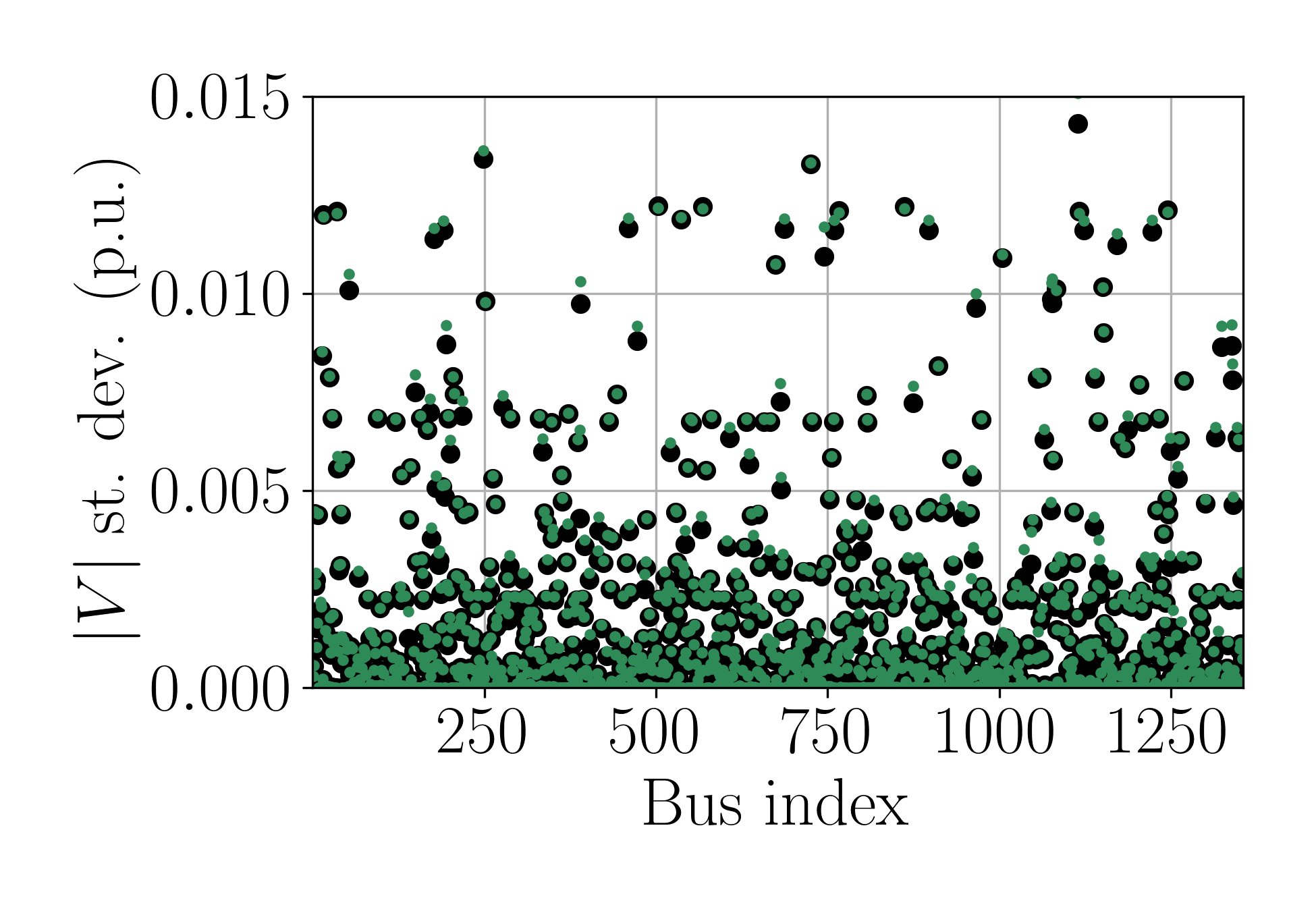}
		\caption{Standard deviation, $Q=100$.}
		\label{fig:grid_std_Q100}
	\end{subfigure}
	\\
	\begin{subfigure}[b]{0.49\textwidth}
		\centering
		\includegraphics[width=\textwidth]{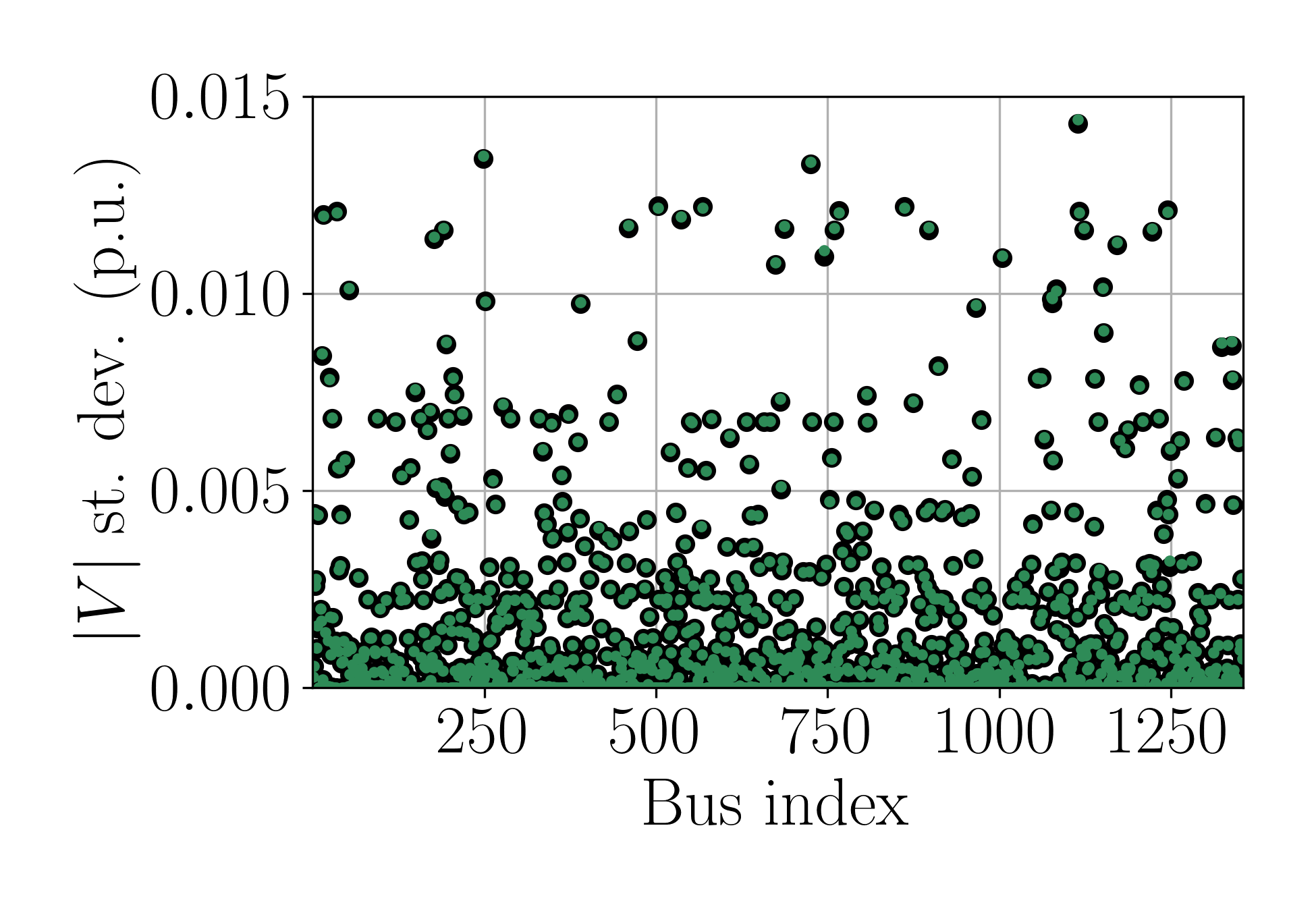}
		\caption{Standard deviation, $Q=500$.}
		\label{fig:grid_std_Q500}
	\end{subfigure}
	\hfill
	\begin{subfigure}[b]{0.49\textwidth}
		\centering
		\includegraphics[width=\textwidth]{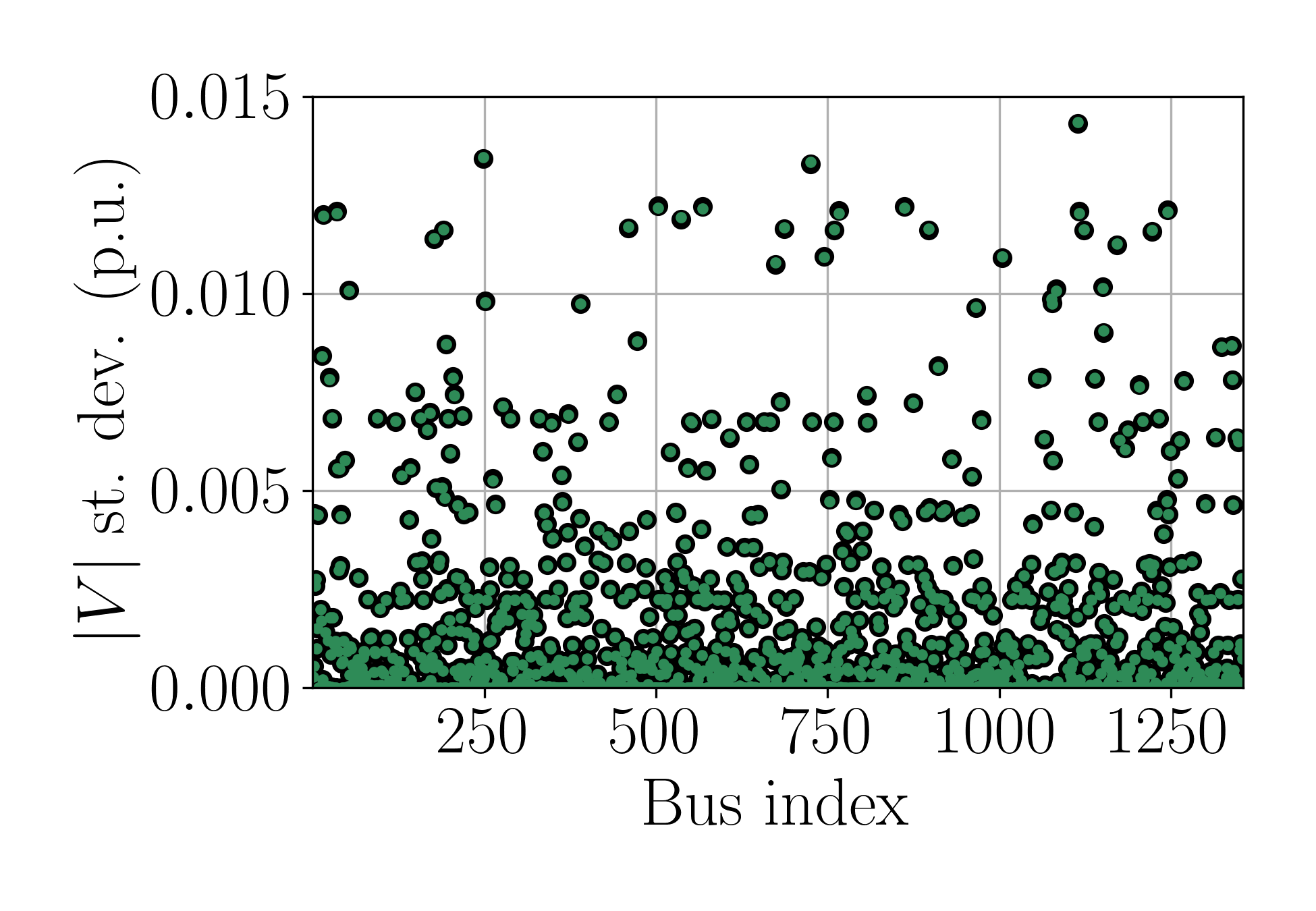}
		\caption{Standard deviation, $Q=1000$.}
		\label{fig:grid_std_Q1000}
	\end{subfigure}
	\caption{Power grid: Mean and standard deviation of the voltage magnitude per network bus, estimated with the \gls{mvsa} \gls{pce} for training data set size $Q \in \left\{100, 500, 1000\right\}$. The reference mean and standard deviation are computed via \gls{mcs} with $15\cdot10^3$ random samples.}
	\label{fig:grid_moments}
\end{figure} 

The differences in impact of the random load variations upon the network bus voltages becomes even more evident by looking at the mean and standard deviation results presented in Figure~\ref{fig:grid_moments}.
That is, for several buses, a very small to negligible standard deviation can be observed, while the standard deviation is quite significant for other buses.
The \gls{mvsa} \gls{pce} approximates the mean very accurately with only $Q=100$ training data points. 
Contrarily, the same training data set size leads to a poor standard deviation estimate for many network buses. 
A significant improvement can be observed for $Q=500$ and the estimate becomes almost identical to the \gls{mcs}-based reference for $Q=1000$.
The results of the \gls{td} \gls{pce} are here omitted, as both mean and standard deviation estimates are very inaccurate. 

\begin{figure}[t!]
	\centering
	\begin{subfigure}[b]{0.4\textwidth}
		\centering
		\fbox{\includegraphics[width=1\textwidth]{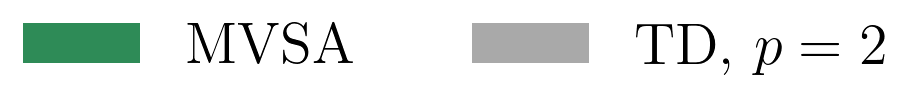}}
	\end{subfigure}
	\\
	\begin{subfigure}[b]{0.8\textwidth}
		\centering
		\includegraphics[width=\textwidth]{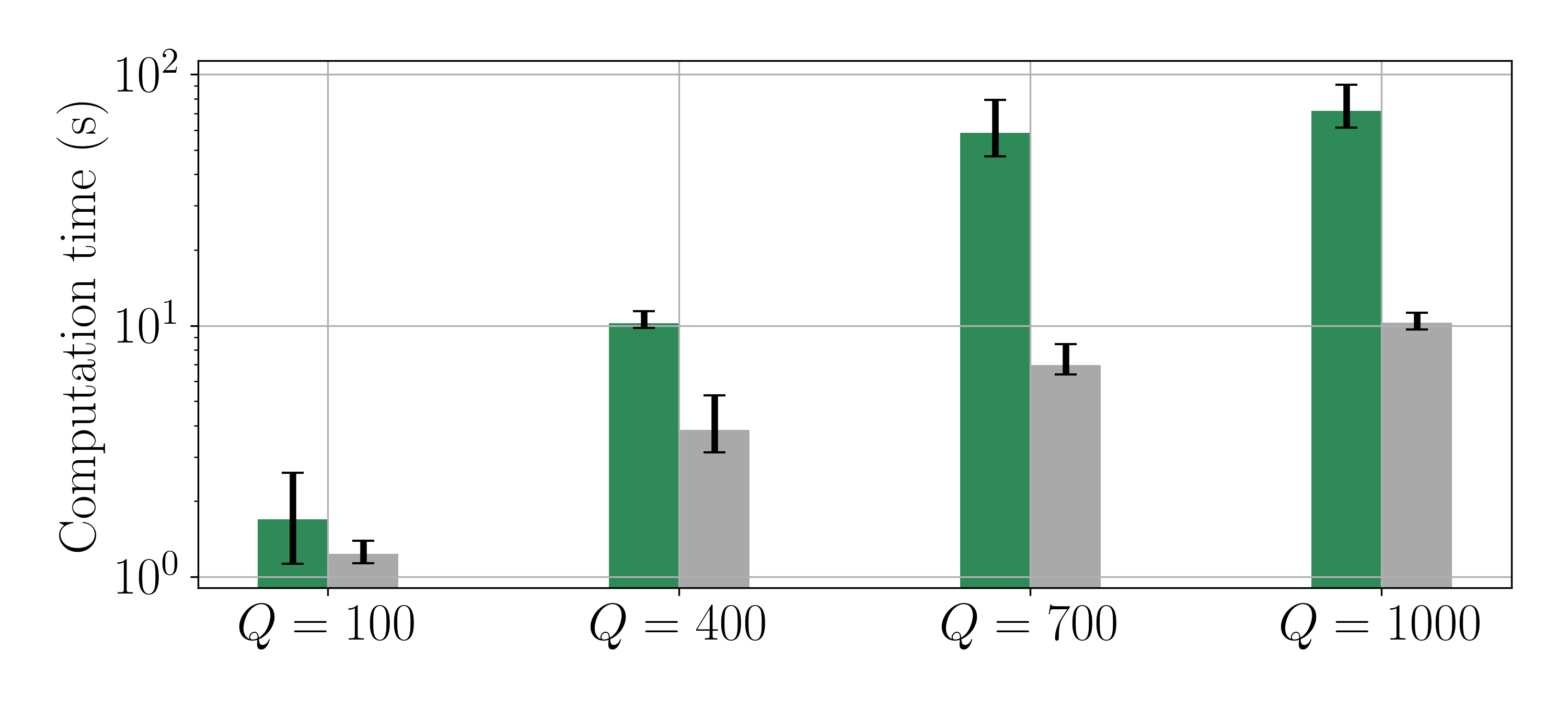}
	\end{subfigure}
	\caption{Power grid: Computation time of the \gls{mvsa} and \gls{td} \glspl{pce} for training data set size $Q \in \left\{100, 400, 700, 1000\right\}$. The colored bars show the average computation time over $10$ different training data sets. The black error bars show the difference between minimum and maximum computation time over these $10$ data sets.}
	\label{fig:grid_computation_time}
\end{figure}

Figure~\ref{fig:grid_computation_time} presents the computation times of the \gls{mvsa} and \gls{td} \glspl{pce} for training data set sizes ranging from $Q=100$ to $Q=1000$ training data points.
Similar to the previous use cases, the \gls{td} \gls{pce} is computed much faster due to its a priori fixed basis, however, for the same reason results in poor approximation and uncertainty estimation results.
In comparison, the computation time of the \gls{mvsa} \gls{pce} remains acceptable, i.e., $91$~s in the worst case, while also yielding accurate surrogate models and uncertainty estimates.

\begin{table}[t!]
	\small
	\caption{Power grid: Maximum total and univariate degrees of the \gls{mvsa} \gls{pce} method.}
	\centering
	\begin{tabular}{c c c}
		\toprule
		Training data set size $Q$  & Maximum total degree & Maximum univariate degree  \\[1ex] 
		\toprule
		100 & 2 & 2 \\[0.5ex]
		500 & 3 & 3 \\[0.5ex]
		100 & 3 & 3 \\
		\bottomrule
	\end{tabular}
	\label{tab:grid-max-degrees}
\end{table}

Last, Table~\ref{tab:grid-max-degrees} shows the maximum total degree, i.e., $\max_{\mathbf{k} \in \Lambda}\left|\mathbf{k}\right|_1$, and the maximum univariate degree of the polynomial bases resulting from the \gls{mvsa} \gls{pce} method.
Note that a \gls{td} \gls{pce} with maximum polynomial degree equal to $3$, i.e., the maximum total degree of the \gls{mvsa} \gls{pce} method, would in this case result in $26235$ coefficients, rendering the method usable only a big data training regime.

\section{Discussion, conclusion, and outlook}
\label{sec:conclusion}

\glsreset{pce}
\glsreset{mvsa}
\glsreset{td}
\glsreset{lar}
\glsreset{uq}

This work presented a novel basis-adaptive method for the construction of \glspl{pce} of high-dimensional, vector-valued model responses. 
The corresponding algorithm is called \gls{mvsa} due to the fact that it employs a multivariate sensitivity analysis metric as criterion for the adaptive expansion of the polynomial basis. 
The method has the advantage of being able to address problems with moderately high-dimensional model inputs, i.e., in the orders of tens, and very high-dimensional model responses i.e., up to the order of thousands, at the same time. 

The \gls{mvsa} \gls{pce} method has been applied to three numerical test cases, each featuring moderately high-dimensional model inputs and very high-dimensional model responses. 
For all three test cases, the \gls{mvsa} \gls{pce} has been evaluated in terms of computational efficiency, surrogate modeling accuracy, and uncertainty estimation accuracy.
Computational efficiency refers to training data demand versus approximation and uncertainty estimation accuracy, as well as to the time needed to compute the \gls{pce}.
The method was compared against standard \gls{td} \glspl{pce} with an a priori fixed polynomial basis, and against a state-of-the-art degree- and $q$-norm-adaptive ($p/q$-adaptive) \gls{pce} method based on \gls{lar}.
Due to the fixed basis, \gls{td} \glspl{pce} are computed faster than the \gls{mvsa} \gls{pce}. However, the latter is typically much more accurate for the same size of training data set.
Additionally, the \gls{mvsa} \gls{pce} remains applicable in cases with comparatively high-dimensional model inputs, where \gls{td} \glspl{pce} become unusable due to the curse of (input) dimensionality.
Compared to the $p/q$-adaptive \gls{lar} \gls{pce}, the \gls{mvsa} \gls{pce} was found to offer either better or at least comparable accuracy for the same training data set sizes.
Additionally, the \gls{mvsa} \gls{pce} is computed several times faster, even for cases where the $p/q$-adaptive \gls{lar} \gls{pce} is combined with a dimension reduction method.
Last, the \gls{mvsa} \gls{pce} remains applicable in cases where the $p/q$-adaptive \gls{lar} \gls{pce} cannot be used at all, e.g., for high-dimensional model responses without effective dimension reduction.

The \gls{mvsa} \gls{pce} method developed in this work can be considered as a useful surrogate modeling and \gls{uq} tool already in its current form. 
Nonetheless, a number of possible developments could further enhance its capabilities. 
First, problems with reduced regularity could be addressed by integrating multi-element approaches \cite{galetzka2023hp}.
Second, the method could be combined with dedicated active learning techniques in an effort to further reduced training data demand \cite{novak2021variance}.
Last, recent developments on physics-informed \glspl{pce} \cite{novak2024physics} offer the possibility of physically conforming \gls{pce}-based surrogate models, rather than pure data-driven ones.
Such developments will be investigated in dedicated follow-up works.

\section*{Acknowledgments}
\noindent The first and third authors are supported by the Deutsche Forschungsgemeinschaft (DFG, German Research Foundation), Project-ID 492661287 -- TRR 361.
The authors would like to thank the reviewers and editors for their constructive feedback.

\section*{Author contributions}
\noindent Conceptualization: DL; Funding acquisition: DL, HDG; Investigation: DL, ED; Methodology: DL; Software: DL, ED; Validation: DL, ED; Writing - original draft: DL; Writing - review \& editing: ED, HDG.





%
%
%
\bibliographystyle{elsarticle-num}
\biboptions{sort&compress}
\bibliography{mvsa-pce-refs}

\end{document}